\documentclass[a4paper,11pt]{JHEP3}

\usepackage{graphicx}
\usepackage{amsmath}
\usepackage{amssymb}
\immediate\write18{../../scripts/bibgen \jobname.tex}
\def\bC{\mathbb{C}}
\def\bR{\mathbb{R}}
\def\bZ{\mathbb{Z}}
\def\cN{\mathcal{N}}
\def\cA{\mathcal{A}}
\def\cB{\mathcal{B}}
\def\ccA{\mathcal{A}}
\def\ccB{\mathcal{B}}
\def\cO{\mathcal{O}}
\def\cT{\mathcal{T}}
\def\cI{\mathcal{I}}
\def\cD{\mathcal{D}}
\def\cP{\mathcal{P}}
\def\cQ{\mathcal{Q}}
\def\cR{\mathcal{R}}
\def\cU{\mathcal{U}}
\def\cV{\mathcal{V}}
\def\cW{\mathcal{W}}
\def\cZ{\mathcal{Z}}
\def\frakg{\mathfrak{g}}
\def\CP{\mathbb{CP}}
\def\tr{\mathop{\mathrm{tr}}}
\def\Vol{\mathop{\mathrm{Vol}}}
\def\two{{\mathbf{2}}}
\def\dim{\mathop{\mathrm{dim}}}
\def\diag{\mathop{\mathrm{diag}}}
\def\id{\mathbf{1}}
\def\stack#1#2{\genfrac{}{}{0pt}{0}{#1}{#2}}
\def\stacks#1#2{\genfrac{}{}{0pt}{1}{#1}{#2}}
\let\epsilon\varepsilon
\let\hat\widehat

\def\skipper{\hskip.5em\relax}
\def\Dn#1#2#3#4#5#6#7{%
{\text{\small$\begin{array}{c@{\skipper}c@{\skipper}c@{\skipper}c@{\skipper}c@{\skipper}c}
&#1&&#5&\\
#2 &  #3 & #4 & #6 & #7
\end{array}$}}}
\def\Esix#1#2#3#4#5#6#7{%
{\text{\small$\begin{array}{c@{\skipper}c@{\skipper}c@{\skipper}c@{\skipper}c@{\skipper}c}
&&#1&& \\
&&#3&&\\
#2 &  #4 & #5 & #6 & #7
\end{array}$}}}

\def\Eseven#1#2#3#4#5#6#7#8{%
{\text{\small$\begin{array}{c@{\skipper}c@{\skipper}c@{\skipper}c@{\skipper}c@{\skipper}c@{\skipper}c}
&&&#3&&\\
#1&#2 &  #4 & #5 & #6 & #7 & #8
\end{array}$}}}

\def\Ee#1#2#3#4#5#6#7#8#9{%
{\text{\small$\begin{array}{c@{\skipper}c@{\skipper}c@{\skipper}c@{\skipper}c@{\skipper}c@{\skipper}c@{\skipper}c}
&&&&&#3\\
#1&  #9&  #8 & #7 & #6 & #5 & #4 & #2
\end{array}$}}}

\preprint{
\hbox{}\hfill UT-07-25\\ 
\hbox{}\hfill arXiv:0709.0348}

\title{A-D-E Quivers and Baryonic Operators}

\author{
Yuji Tachikawa$^1$ and Futoshi Yagi$^2$\\

\bigskip

$^1$ School of Natural Sciences, Institute for Advanced Study,\\
 Princeton,  New Jersey 08540, USA

\medskip

$^2$ Department of Physics, University of Tokyo \\
7-3-1 Hongo, Bunkyo-ku, Tokyo , 113-0033 Japan 
}

\abstract{
We study baryonic operators of the gauge theory on multiple D3-branes
at the tip of the conifold orbifolded by a discrete subgroup $\Gamma$ of $SU(2)$. 
The string theory analysis predicts that the number and the order of the fixed points 
of $\Gamma$ acting on $S^2$ are directly reflected in 
the spectrum of baryonic operators on the 
corresponding quiver gauge theory constructed from
 two Dynkin diagrams of the corresponding type.
We confirm the prediction by utilizing  techniques to enumerate
baryonic operators of the quiver gauge theory which includes the gauge groups
with different ranks.
We also find that the Seiberg dualities act on the baryonic operators in a non-Abelian
fashion.
}

\keywords{AdS/CFT correspondence, McKay correspondence, baryonic operators}

\begin{document}

\section{Introduction}

\subsection*{AdS/CFT and baryons}

$\cN=4$ super Yang-Mills theory with gauge group $SU(N)$
is now  believed to be equivalent to
 Type IIB string theory on $AdS_5\times S^5$
with $N$ units of the five-form flux  \cite{Maldacena},
which is the prototypical example of
the Anti-de Sitter/Conformal Field Theory (AdS/CFT) correspondence.
This amazing correspondence  relates  a theory with gravity  and a genuine gauge theory.
Moreover, it requires the full non-perturbative spectrum
of the superstring theory in the gravity side. 
For example, the baryons,
i.e.~operators which involves the epsilon symbols of the gauge group,
correspond to various wrapped D-branes \cite{BaryonVertex,GGinCFT}.

The duality can be generalized by replacing $S^5$ by other five-dimensional
Einstein manifolds $X^5$. The corresponding  gauge theory is the low energy
limit of the theory
on $N$ D3-branes probing the six-dimensional cone $C(X^5)$ over $X^5$,
and there should be a mapping between D-branes wrapped on $X^5$
and baryonic operators of the gauge theory.  

\FIGURE{
\parbox{.4\textwidth}{
\centerline{\includegraphics[width=.15\textwidth]{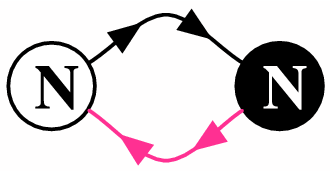}}
}
\caption{Conifold gauge theory \label{KWquiver}}
}
One well-studied example\cite{KlebanovWitten} is to take $X^5=T^{1,1}$ which is
an $S^1$ bundle over $S^2{}_1\times S^2{}_2$.
The dual is an $\cN=1$ supersymmetric
gauge  theory with the group $SU(N)_1\times SU(N)_2$ 
and four bifundamental chiral superfields  $A^i$, $B^j$ ($i,j=1,2$),
which we call the conifold gauge theory. 
We use the quiver diagram to summarize the matter content,
see Fig.~\ref{KWquiver}.
There, each of the nodes signifies an $SU(N)$ gauge group,
and an arrow between the two nodes stands for a bifundamental
chiral superfield, i.e.\ a chiral superfield transforming in the 
fundamental($\square$)/anti-fundamental($\bar\square$)
 representation  under the gauge groups at the head/tail
of the arrow, respectively. We say such a bifundamental field
connects the gauge group at the tail and the one at the head.
Then $A^i$ and $B^j$ transform as the
representations $(\bar \square,\square)$ and $(\square,\bar\square)$
under $SU(N)_1\times SU(N)_2$, respectively.
We often abuse the terminology and just used the word the quiver to mean either 
the gauge theory or the diagram. 
There are two $SU(2)$ symmetries acting on the indices $i$ of $A^i$
and $j$ of $B^j$ respectively, and they correspond to the rotation
of $S^2{}_1$ and $S^2{}_2$ in the geometry.

\FIGURE{
\includegraphics[width=.3\textwidth]{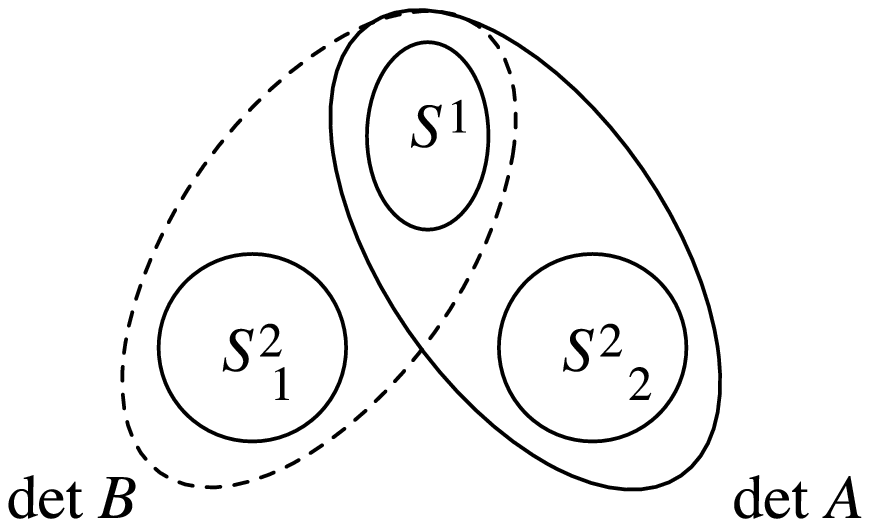}
\caption{Dibaryons on $T^{1,1}$.\label{t11baryon}}
}
D3-branes wrapped on the $S^3$ part of $T^{1,1}\sim S^2\times S^3$
have been successfully identified \cite{GubserKlebanov} 
with determinants of the bifundamentals $A^i$ and $B^j$, 
 see Fig.~\ref{t11baryon}.
These operators are called dibaryons, since we used two epsilon symbols
to construct them.

Our focus in this paper is the orbifold of $T^{1,1}$ 
by a discrete subgroup $\Gamma\subset SU(2)_1$ of the isometry. 
 $T^{1,1}/\Gamma$ has various types of three-cycles,
of the form $S^3/\bZ_n$ or $S^3/\Gamma$.  Since the volume of the
three-cycle is proportional to the dimension of the dual operator,
it translates to a rich and intricate spectrum of baryonic operators
of the dual gauge theory.  
Our objective in this article is then to establish the one-to-one
mapping between them.

People have studied the mapping of  baryons and  wrapped 
D-branes in various dual pairs, e.g.~for $S^5/\bZ_3$ in \cite{GRW},
for generalized conifolds and del Pezzos in \cite{BeasleyPlesser,HerzogMc,IWbaryon,HerzogWalcher}.  Quite recently
people started to enumerate baryonic operators  systematically 
for $X^5$ with $U(1)^3$ 
isometry \cite{baryonic-counting0,baryonic-counting1,baryonic-counting2}.
Our setup is arguably simpler than these previous works in the gravity side. Indeed,
the action of $\Gamma$ on the $S^2$ of $T^{1,1}$ 
is exactly as the symmetry group
of a regular polyhedron,  which has been known to us since the days of
classical Greek natural philosophers. It allows us to concentrate on
and uncover  the dual
phenomena on the gauge theory side, which will be the main topic of this article.

\subsection*{McKay correspondence and baryons}
%
%
%
As is well known, discrete subgroups $\Gamma$ of $SU(2)$ are exhausted by
cyclic groups $\bZ_n$,  binary dihedral groups $\hat\cD_n$ and 
binary tetra-, octa- and icosahedral groups $\hat\cT$, $\hat\cO$, $\hat\cI$. 
It follows the
pattern $A_n$, $D_n$ and $E_{6,7,8}$, which
can be understood following the observation of McKay \cite{McKay}.
There, one associates a node of the extended Dynkin diagram
to each irreducible representation of the corresponding group, 
and the edges encode the decomposition of the tensor product
of the irreducible representation with the standard two-dimensional representation.
The McKay correspondence has a physical
realization using D3-branes probing $\bC^2/\Gamma$ \cite{DouglasMoore},
where the nodes of the Dynkin diagram 
correspond to the fractional branes at the origin, 
and the edges to the open strings stretching between two fractional branes.

\FIGURE{
\includegraphics[width=.45\textwidth]{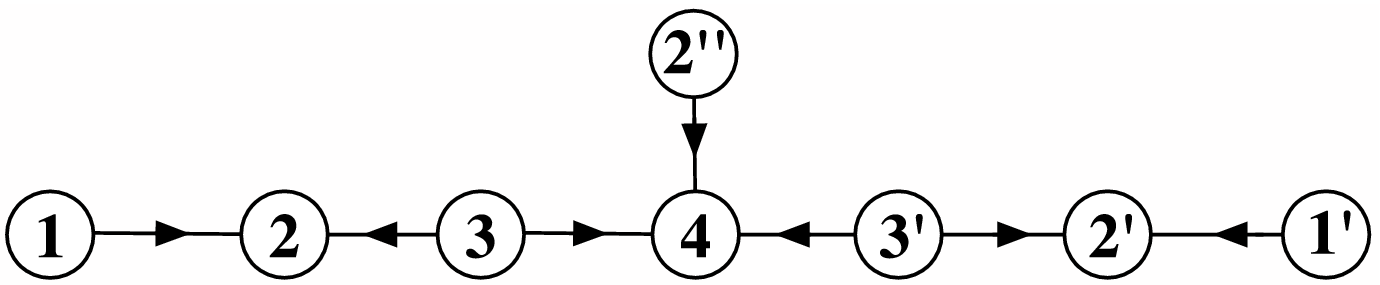}
\caption{Subquiver of the octahedral theory.\label{example-of-alternating}}
}
Application of the procedure to our case produces a quiver gauge theory 
which includes the alternating extended Dynkin diagram as a subquiver,
which we will see in more detail in Sec.~\ref{DouglasMoore}.
We call them alternating in the sense that
 the arrows are all incoming or all outgoing at each node.
For example, we get the subquiver in 
Fig.~\ref{example-of-alternating} if we take the binary
octahedral group $\hat\cO$ as the orbifold group $\Gamma$.
A number $k$ in the circle stands for an $SU(Nk)$  gauge group,
and primes are used to distinguish different gauge groups with the same ranks.

What will be the spectrum of the baryonic operators of this quiver?
From the AdS/CFT correspondence,
it should reproduce the structure of three-cycles
of $T^{1,1}/\Gamma$, and  we will see shortly that it is dictated by
the action of  $\Gamma$ on $S^2$.  In other words, we can expect that
the baryonic spectrum of the Dynkin quiver `knows' the action of $\Gamma$.

\FIGURE{
\includegraphics[width=.25\textwidth]{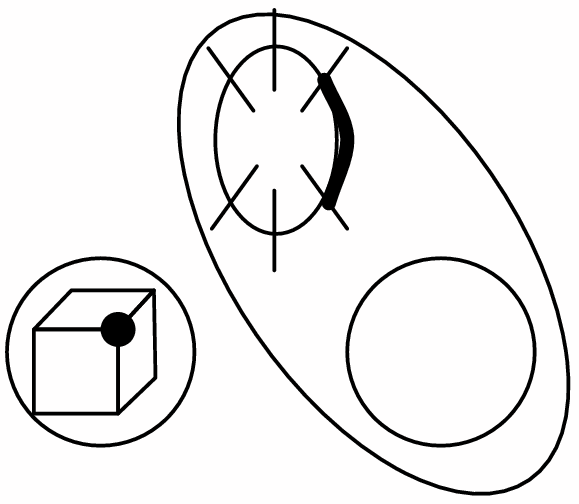}\vspace*{-1em}
\caption{D3-brane on $T^{1,1}/\Gamma$.\label{orbifoldbaryon}}
}
Let us take $\Gamma=\hat\cO$ again as the example, and 
consider
D3-branes wrapped at $S^3{}_2$ which is an $S^1$ bundle 
over $S^2{}_2$, see Fig.~\ref{orbifoldbaryon}. We have the choice 
on which point of $S^2{}_1$ to put the D3 brane.
When we put the D3-brane at a vertex of the cube, $\bZ_6\subset \hat\cO$
acts on the D3-brane worldvolume as the $1/6$-rotation of the $S^1$ fiber.
Therefore, it is $1/6$ times as heavy as the dibaryon of the unorbifolded theory.
The worldvolume is $S^3/\bZ_6$, and thus we have a choice of the Wilson line
$\alpha^6=1$, which leads to six operators of the same dimension of
 the gauge theory \cite{GV}.
Similarly, by putting the D3-brane at the center of a face or at the midpoint of an edge,
we have a wrapped D3-brane which is $1/8$ and $1/4$ as heavy as the original dibaryon.
The D3-brane at generic points of $S^2{}_1$ is half as heavy as the baryon of 
the unorbifolded theory, but it has the moduli space and we need to quantize it.

We will carry out the analysis of the geometry in detail in Sec.~\ref{baryons},
and the general statement about the spectrum of the baryons
 for non-Abelian $\Gamma$ is the following:
Let $\Gamma$ be a 
discrete subgroup of $SU(2)$, generated by elements $a$, $b$, $c$ and $z$
by the relation $a^p=b^q=c^r=z$, and $z^2=1$.
Consider an alternating extended Dynkin quiver of the same
type $\Gamma$.
Then, the baryonic operators of the quiver are
 generated by the following sets of operators\\[1ex]
\hbox{}\qquad\qquad\begin{tabular}{ccll}
$\cP_1\ldots,\cP_p$ &:&  of weight $|\Gamma|/(2p)$,\\
$\cQ_1\ldots,\cQ_q$ &:& of weight $|\Gamma|/(2q)$,\\
$\cR_1\ldots,\cR_r$ &:& of weight $|\Gamma|/(2r)$ ,\\
 $\cO(N)$ operators &: & of weight $|\Gamma|/2$ .
\end{tabular}\\[1ex]
Here the weight of the operator is defined as the number of bifundamental
fields in it, divided by $N$.

A large part of our paper is devoted to check this prediction 
in the gauge theory. In fact,
this mathematical statement about the alternating Dynkin quiver was 
proved in \cite{SkowronskiWeyman},
and  the AdS/CFT correspondence with $T^{1,1}/\Gamma$ gives a 
string-theoretic reason of existence of such a theorem. 
The proof in \cite{SkowronskiWeyman} 
was done without reference to the discrete group $\Gamma$,
whereas  we analyze the problem emphasizing its relation to
the McKay correspondence.


\subsection*{Non-toric/non-conformal quiver and baryons}
One distinguishing feature of the space $T^{1,1}/\Gamma$ for 
non-Abelian $\Gamma$ is that the isometry is reduced to $SU(2)\times U(1)$
of rank 2. It is a non-toric Einstein manifold and correspondingly we have 
$SU$ gauge groups of different ranks  in the quiver gauge theory. 
Baryonic operators on such a theory is much subtler compared to the
baryons of toric quiver gauge theory.

The problem is that the bifundamentals are no longer  square matrices, therefore
we can no longer form simple determinants from them.
Let us again consider the alternating quiver for $\Gamma=\hat\cO$, 
Fig.~\ref{orbifoldbaryon}. 
One gauge-invariant operator is of the form 
\begin{equation}
\epsilon_{(1)} (A_{1\to2})^N 
\epsilon^{(2)} (A_{3\to2})^N
\epsilon_{(3)} (A_{3\to 4})^{2N} 
\epsilon^{(4)} (A_{2''\to 4})^{2N}
\epsilon_{(2'')} 
\end{equation} 
where $A_{a\to b}$ is the bifundamental 
between $SU(aN)$ and $SU(bN)$ gauge groups, and $\epsilon_{(k)}$ is the
epsilon symbol for $SU(kN)$. 
It is a product of $6N$ fields, i.e. of weight six.

Now $|\hat\cO|=48$ and $(p,q,r)=(4,3,2)$, and we saw above
that the AdS/CFT correspondence predicted the lowest weight 
of the baryonic operator is $|\hat\cO|/(2p)=6$.
We need to construct three others, 
and show that they exhaust gauge-invariant fields
made of $6N$ bifundamentals.

One might be able to construct other baryonic operators by inspection, but
it requires systematic techniques to enumerate  and classify them.
We develop two such techniques in this article. One is the untangling procedure
which is based on the relation \begin{equation}
\epsilon^{i_1\ldots i_k a_1\ldots a_{N-k}}
\overline\epsilon_{j_1\ldots j_k a_1\ldots a_{N-k}}\propto
\delta^{[i_1}_{j_1}\cdots \delta^{i_k]}_{j_k}.
\end{equation}
Another is the application of the theory of quiver representations,
where we will find a baryonic operator can be naturally associated to an indecomposable
representation of the quiver.
In the previous application of the quiver representations in string theory,
the dimensions of the vector spaces associated to the nodes are identified with
the ranks of the gauge groups. A curious feature of our case is that
the dimensions correspond to the number of the epsilon symbols used 
in the baryonic operator.
We will also see that the action of Seiberg dualities on the baryons 
is quite non-trivial, and that it can be utilized in the classification.

The appearance of gauge groups of different ranks is also a prominent feature
of non-conformal deformations of toric quiver gauge theory, and the techniques
we develop will hopefully have some utility in studying baryons and baryonic
branches of these non-conformal theories.


\subsection*{Organization of the paper}

The rest of the article is structured as follows:
we start the discussion in Sec.~\ref{T11}
by reviewing the known correspondence
of the conifold gauge theory and Type IIB string on 
$AdS_5\times T^{1,1}$.
In Sec.~\ref{DouglasMoore}, we review the McKay correspondence
and the construction
of Douglas and Moore, and apply them
to obtain the quiver gauge theory dual to Type IIB string
on $AdS_5\times T^{1,1}/\Gamma$. We also perform some elementary
check of the AdS/CFT correspondence in our cases.
Then in Sec.~\ref{baryons}, we study the spectrum of D3 branes
wrapping three-cycles of $T^{1,1}/\Gamma$, and we construct
the corresponding baryonic operators in the quiver gauge theory.
In Sections \ref{direct} and \ref{indirect} we proceed to 
show that the operators constructed up to that point
exhaust the baryonic operators of the quiver gauge theory, which requires
a quite lengthy analysis.  We use in Sec.~\ref{direct}
a direct approach which analyze
the structure of the contraction of the indices of the epsilon tensors,
whereas in Sec.~\ref{indirect} we employ the mathematical theory
of the quiver representations to classify the baryonic operators.
Each approach has its own virtues and we think they provide
a valuable tool to analyze baryonic operators of non-toric and/or non-conformal
quiver gauge theories.
Finally in Sec.~\ref{Abranch} we show
 that the generators of the baryonic
operators have no non-linear constraint among them if $N>1$. 
Therefore the baryonic chiral ring of the alternating Dynkin quiver 
is just a polynomial ring with generators determined by the geometry
of $S^2/\Gamma$,  agreeing  with  the mathematical result in \cite{SkowronskiWeyman}.
We conclude our article by a discussion about future prospects in Sec.~\ref{summary}.
We have several appendices which complement the main discussion.

\subsection*{Note added in version 2}
The authors were informed after submitting the version 1 of the paper on the arXiv
that the result about the baryons of the alternating Dynkin quiver was
 proved in a mathematical paper \cite{SkowronskiWeyman} in 2000
in a different method.
The authors would like to thank Yoshiyuki Kimura for finding out
the reference \cite{SkowronskiWeyman}.
The wording of the version 2 of the paper was modified accordingly.
The readers can think of the work either as a further check of the 
AdS/CFT correspondence using the known mathematical result,
or as a `postdiction' of it from
the application of the correspondence to $T^{1,1}/\Gamma$.

\section{Review of the correspondence for $T^{1,1}$}\label{T11}
Let us begin by  recalling the AdS/CFT correspondence 
for the case of Type IIB string theory on $T^{1,1}\times AdS_5$.
The metric for the five-dimensional space $T^{1,1}$ is given by 
\begin{equation}
ds_{T^{1,1}}^2=\frac16\sum_{i=1,2}(d\theta_i^2+\sin^2\theta_i d\phi_i^2) 
+\frac19(d\psi + \cos\theta_1 d\phi_1+\cos\theta_2d\phi_2)^2
\label{T11metric}
\end{equation} where $0\le\theta_i<\pi$ and $0\le \phi_i<2\pi$
parametrize $S^2{}_1\times S^2{}_2$ in an obvious way, 
and $0\le\psi <4\pi$ is the coordinate
of the $S^1$ fiber over it.   The set $(\theta_i,\phi_i,\psi)$ for each $i$
describes the three-manifold which is topologically $S^3$ in the form of
the Hopf fibration, although the fiber direction is squashed compared to the
round sphere. Still it has  $SU(2)_i$ isometries for $i=1,2$ acting on
each set of coordinates $(\theta_i,\phi_i,\psi)$. Combined with the shift of $\psi$,
$T^{1,1}$ has the isometry group $SU(2)_1\times SU(2)_2\times U(1)$.

The metric cone $C$ over $T^{1,1}$ is the well-known conifold, which
 has an alternative description as a 
hypersurface determined by $xw=yz$ in $(x,y,z,w)\in \bC^4$.
Now, let us consider Type IIB string theory on $\bR^{3,1}\times C$,
and introduce $N\gg 1$ of D3 branes which fill $\bR^{3,1}$ direction.
Since the conifold $C$ is Calabi-Yau, there remains $\cN=1$ supersymmetry
in four dimensions. 
The gauge theory on the stack of D3 branes should have $N$-th symmetric
product of the conifold as a branch of the moduli space, 
and is known to be described
by the gauge theory whose matter content is summarized in Fig.~\ref{KWquiver}.

To specify an $\cN=1$ supersymmetric gauge theory we need to give the
superpotential, which is 
\begin{equation}
W=\epsilon_{ij}\epsilon_{kl}\tr A^i B^k A^j B^l
\end{equation}
for the conifold theory. Then, F-term and D-term conditions can be satisfied
by taking all of $A^i$ and $B^k$ diagonal matrices, and thus
a branch of the moduli space is given by the $N$-th symmetric
product of the space described by the first entries of $A^i$ and $B^j$.
Let us call them  $a^i$ and $b^j$. The gauge-invariant
combinations of them are \begin{equation}
x=a^1 b^1, \quad y=a^1 b^2, \quad
z=a^2 b^1, \quad w=a^2 b^2,
\end{equation} which satisfy $xw=yz$  by construction. 
Thus we confirmed the gauge theory has the $N$-th
symmetric product of the conifold as a branch in the moduli space.

Let us now put all of the D3 branes at the tip of the conifold.
For large $N\gg 1$, the tension of the branes bends the spacetime,
and the low-energy dynamics is captured by the near-horizon geometry
which is $T^{1,1}\times AdS_5$ with $N$ units of the 5-form flux through $T^{1,1}$.
$AdS_5$ has the isometry $SU(2,2)$, which is isomorphic to the conformal group
of $\bR^{3,1}$. The proposal by Maldacena \cite{Maldacena}, 
applied in this context, means that
the dynamics of the near-horizon region is dual to 
that of the low-energy limit of the conifold gauge theory, which should be
 a non-trivial conformal field theory.
 
 There are various tests of this AdS/CFT correspondence.  Firstly, 
 the low-energy limit of the gauge theory should be superconformal.
Let us suppose it is so; then the scaling dimension of the superpotential should be three. 
It is most natural to assign the scaling dimension $3/4$ to $A^i$ and $B^j$ by the 
consideration of the symmetry.  Now the Novikov-Shifman-Vainshtein-Zakhalov
(NSVZ) beta functions for the gauge coupling constants of two $SU(N)$
groups vanish
with this scaling dimension for the bifundamental fields, which is as it should
be for a superconformal field theory.   Internal global  symmetries also agree.
Indeed, two $SU(2)$ symmetries acting on $A^i$ and $B^j$ and the $U(1)$
R-symmetry of the gauge theory nicely account for the isometry $SU(2)_1\times
SU(2)_2\times U(1)$ of $T^{1,1}$. There is one additional global symmetry
which we call the baryonic charge, which assigns charge $+1$ to $A^i$ and $-1$ to $B^j$.
It comes from the 4-form  potential reduced
along $S^3$ in $T^{1,1}$.

Second is the matching
of the  central charges $a$ and $c$ calculated in
the gravity and the gauge theory descriptions\cite{Gub}.
In the gauge theory side,
they can be determined from the 't Hooft anomaly of the $R$-symmetry, using the formula
\begin{equation}
a=\frac{3}{32}\left(3\tr R^3-\tr R\right),\qquad
c=\frac{1}{32}\left(9\tr R^3-5\tr R\right),
\end{equation} where the trace runs over  the Weyl fermions of the ultraviolet theory.
The $R$-charges of a chiral superfield
is fixed to be two thirds of its scaling dimension, 
so in this case we have \begin{equation}
a=c=\frac{27}{64}N^2\label{T11a}
\end{equation} in the large $N$ limit. 
In the gravity side, the central charges are determined \cite{HS} by the
response of the bulk metric to the boundary perturbation via the prescription
of \cite{GKP,Witten}, with the result \begin{equation}
a=c=\frac{N^2\pi^3}{4\Vol X}\label{holographicA}
\end{equation} where $\Vol X$ is the volume of the internal Einstein manifold
normalized to have $R_{mn}=4g_{mn}$.
From the explicit metric \eqref{T11metric} of $T^{1,1}$, we find \begin{equation}
\Vol T^{1,1}=\frac{16}{27}\pi^3,
\end{equation}which reproduces \eqref{T11a} after substituting in  \eqref{holographicA}.

Third is the correspondence of the dibaryon operators \cite{GubserKlebanov}
in the gauge theory
and the wrapped D3-branes in $T^{1,1}$.   
Dibaryon operators are the following gauge-invariant operators \begin{equation}
\det A^{i_1,\ldots,i_N} \equiv \epsilon_1\epsilon_2 A^{i_1} \cdots A^{i_N} ,\qquad
\det B^{j_1,\ldots,j_N} \equiv \epsilon_1\epsilon_2 B^{j_1} \cdots B^{j_N},
\end{equation} which are  called as such because two epsilon symbols are necessary to
construct them. Here $\epsilon_{1,2}$ are the epsilon symbols for $SU(N)_{1,2}$
respectively, and  we omitted the gauge indices for simplicity.
The $SU(2)_1$ indices $i_1,\ldots,i_N$ and the $SU(2)_2$ indices $j_1,\ldots,j_N$
are automatically symmetrized because of the presence of the two epsilon
symbols, and thus they come in the spin $N/2$ representation
of the $SU(2)_{1,2}$ global symmetry groups. They have scaling dimensions
$3N/4$, and should correspond to heavy objects in the bulk geometry.

In fact, they are known to be represented by D3-branes wrapping $S^3$ of $T^{1,1}$
\cite{BHK}. One family of $S^3$, which we call the $\ccA$-family,
 is given by fixing $(\theta_1,\phi_1)$,
and the worldvolume is parametrized by $(\theta_2,\phi_2,\psi)$.
Another family, which we call the $\ccB$-family,
is given by exchanging $(\theta_1,\phi_1)$ and $(\theta_2,\phi_2)$ 
in the description above.  These $S^3$ are known to be
supersymmetric cycles, which corresponds to the fact that the dibaryons 
preserves half of the supersymmetries.
Their mass, calculated from the tension
of the D3 brane, matches with the dimension of the dibaryon operators \cite{BHK}.
As they wrap non-trivial homological cycles, they carry associated conserved 
charges.  The fact that $S^3$ of the $\ccA$- and $\ccB$- family are homologically 
opposite to each other fits nicely to the fact
that  the baryonic charges of the dibaryons $\det A$ and $\det B$ are 
opposite to each other.
Finally one can calculate the $SU(2)_{1,2}$ spin of each of the family.
For the $\ccA$-family, the low-energy dynamics of the brane are 
given by the supersymmetric quantum mechanics of
the motion of its center-of-mass coordinates on $S^2$ parametrized by
$(\theta_1,\phi_1)$.
As detailed in \cite{BHK}, the Chern-Simons coupling
of the D3 brane to the $N$ units of the 5-form flux in the geometry
leads to the presence of $N$ units of the magnetic flux  through $S^2$.
It leads to $N+1$  zero-modes of the Hamiltonian of the motion
of the brane, which transform in spin $N/2$ representation of $SU(2)_1$
acting on the coordinates $(\theta_1,\phi_1)$.
Each zero-mode represents a distinct particle if viewed from
the $AdS$ space, which then should give rise to a distinct operator in the gauge theory.
We can repeat exactly the same analysis for the $\ccB$-family,
by exchanging $(\theta_1,\phi_1)$ and $(\theta_2,\phi_2)$.

\section{Quivers for  $T^{1,1}/\Gamma$}\label{DouglasMoore}
Now we move on to the construction of the quiver gauge theory
corresponding to the orbifold of $T^{1,1}$ by a 
discrete subgroup of $SU(2)_1$. We begin by the review of the
property of the discrete subgroups of $SU(2)$.

\subsection{A-D-E classification of $SU(2)$ subgroups}\label{classification}
There are diverse mathematical objects
which are classified by the pattern\footnote{See e.g.\ 
the list in Section 2.2 of  \cite{Zuber}.} 
A-D-E, and the earliest in history is 
the classification of the Platonic solids, or more precisely their 
symmetry groups.  They form discrete subgroups $\Gamma_0$ of $SO(3)$. 
There are
two infinite families of cyclic and  dihedral groups,
and three exceptional cases of tetra-, octa- and icosahedral groups,
which we denote by $\bZ_n$, $\cD_n$, $\cT$, $\cO$ and $\cI$, respectively.
Their properties are summarized in Table \ref{grouptable}.

\TABLE{
\centerline{\def\arraystretch{1.2}
\begin{tabular}{l|l|cc}
&& $\Gamma_0$ & $ \Gamma$\\
\hline
cyclic&$A_{2n}$  & n/a & $\bZ_{2n+1}$\\
cyclic&$A_{2n-1} $ & $\bZ_n$ & $\bZ_{2n}$
\end{tabular}\qquad
\begin{tabular}{c|l|ccc|ccc}
& &$\Gamma_0$ & $ \Gamma$  & $|\Gamma|$   &  $p$ & $q$ & $r$\\
\hline
dihedral & $D_{n+2}$  & $\cD_n$ & $ \hat\cD_n$& $ 4n $ & $n$ & $2$ & $2$\\
tetrahedral &$E_6$ & $\cT$ & $\hat\cT$&$ 24 $& $3$ & $3$ & $2$  \\
octahedral &$E_7$  & $\cO$ &$\hat\cO$& $ 48 $& $4$ & $3$ & $2$ \\
icosahedral&$E_8$  & $\cI$ & $\hat\cI$&$ 120 $ & $5$ & $3$ & $2$\\
\end{tabular}}
\caption{Data of discrete subgroups of $SO(3)$ and $SU(2)$\label{grouptable}}
}

As an abstract group, each of the non-Abelian subgroup $\Gamma_0$
can be defined by the 
following relations about the generators $a$, $b$, $c$: \begin{equation}
a^p=b^q=c^r=abc=1,
\end{equation} where $(p,q,r)$ is a triple of positive integers satisfying \begin{equation}
\frac1p +\frac 1q + \frac1r>1.
\end{equation} Such a triple is called a Platonic triple, and there are one-to-one 
correspondence with  Platonic triples and non-Abelian discrete subgroups
of $SO(3)$.  
Furthermore, the  fundamental domain $S^2/\Gamma_0$ 
has three orbifold singularities of the form $\bC/\bZ_p$, $\bC/\bZ_q$, $\bC/\bZ_r$;
see Fig.~\ref{fundamental-region}.  More about classical aspects 
of these groups can be found in the beautiful textbook by Coxeter \cite{Coxeter}.

\FIGURE{\centerline{
\begin{tabular}{c|ccccccccccccc}
$\Gamma_0$& $\bZ_n$ & $\cD_{n}$ & $\cT$ & $\cO$ & $\cI$\\[1ex]
\raise6ex\hbox{$S^2/\Gamma_0$} 
& \includegraphics[width=.15\textwidth]{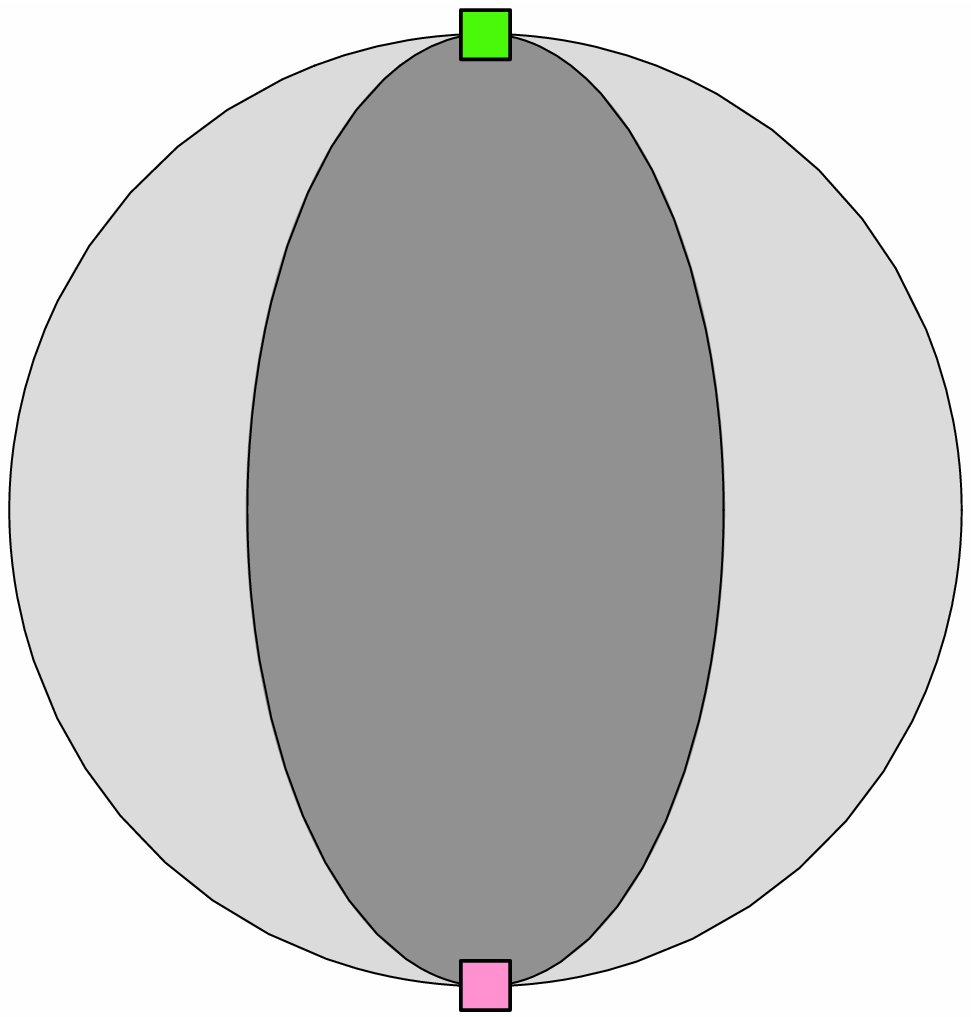}&
\includegraphics[width=.15\textwidth]{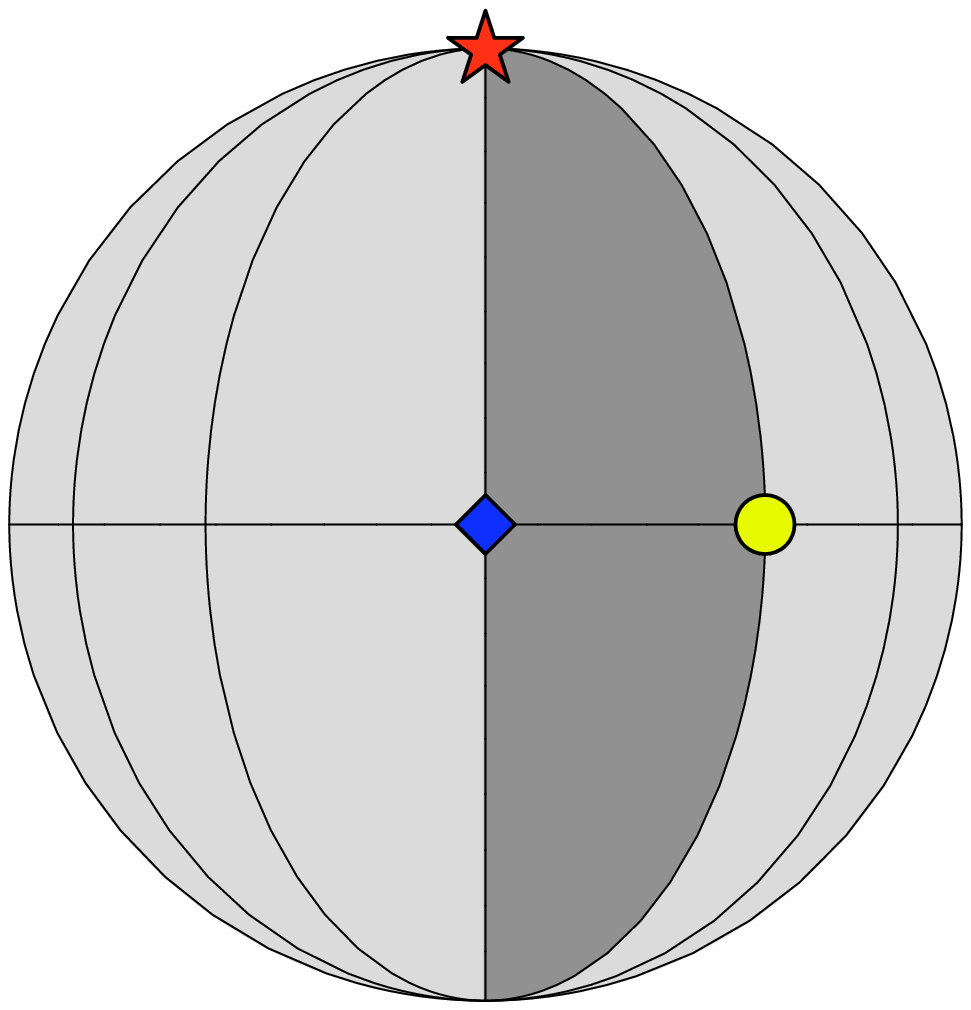}&
\includegraphics[width=.15\textwidth]{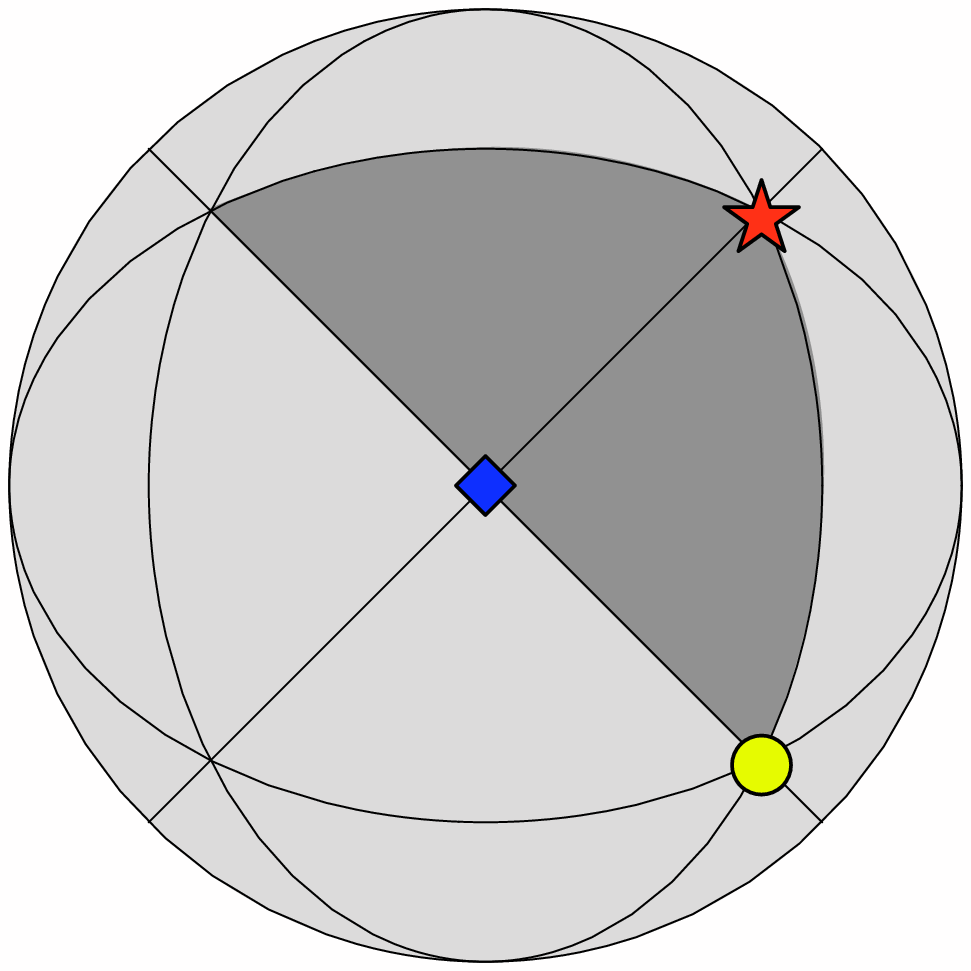}&
\includegraphics[width=.15\textwidth]{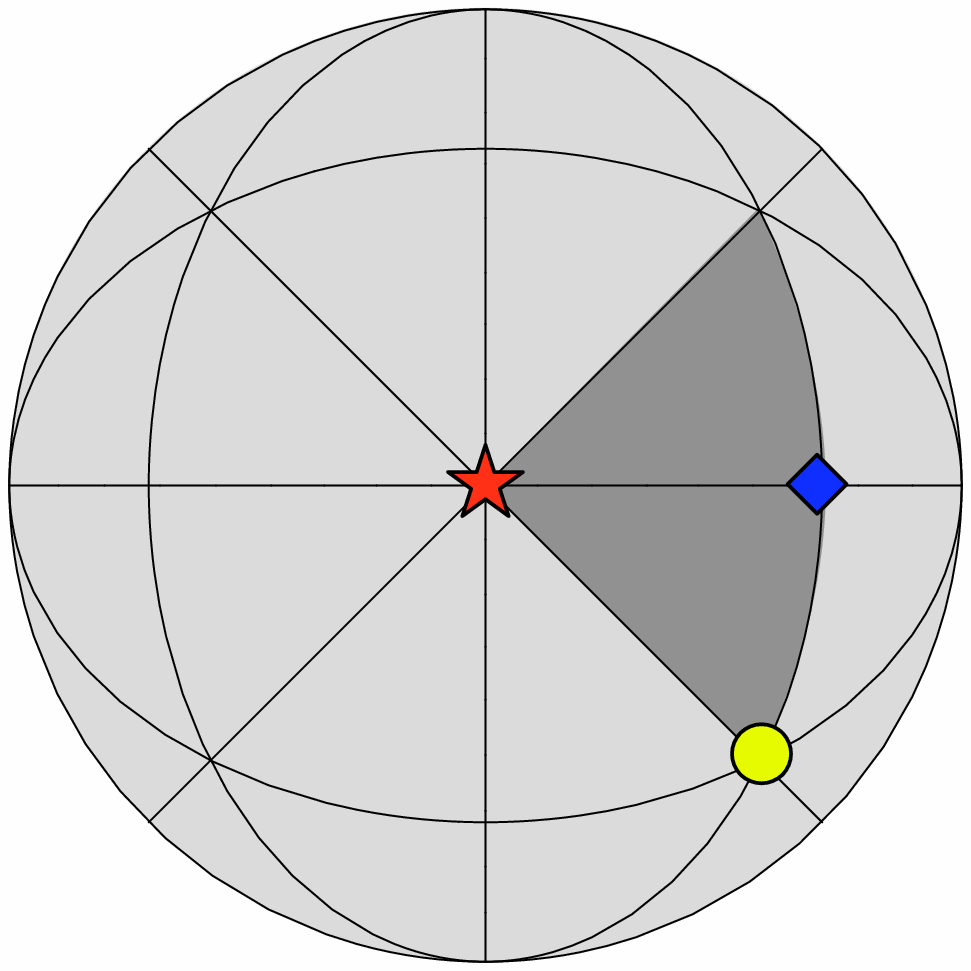}&
\includegraphics[width=.15\textwidth]{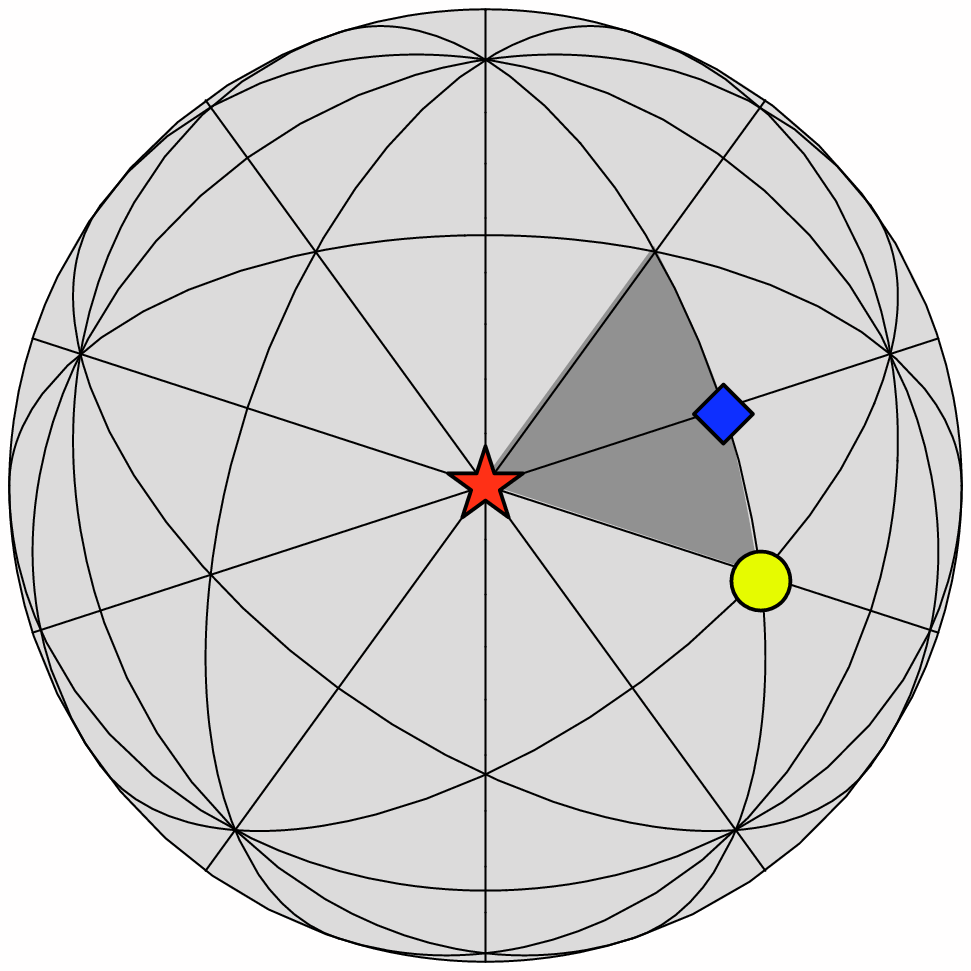}
\end{tabular}
}
\caption{Form of $S^2/\Gamma_0$. The fundamental region is shaded in dark grey. 
We used $\bZ_{6}$ and $\cD_6$ for illustration.
 For non-Abelian groups,
the red star, the yellow circle, and the blue rhombus
are the rotation axes of the generators $a$, $b$, $c$, respectively.
\label{fundamental-region}}
}

Any discrete subgroup $\Gamma_0$ of $SO(3)$
is  the projection of a subgroup $\Gamma$ of $SU(2)$ with $|\Gamma|=2|\Gamma_0|$,
 which are  called binary dihedral groups, the binary tetrahedral group, etc.   
Every non-Abelian finite subgroup of $SU(2)$ is obtained in this way.
For  Abelian subgroups, 
$\bZ_n\subset SO(3)$ comes from $\bZ_{2n}\subset SU(2)$, while 
$\bZ_n\subset SU(2)$ for odd $n$,
generated by $\diag(e^{2\pi i /n},e^{-2\pi i/n})$, does not
arise from a subgroup of $SO(3)$ in this way.

The groups listed in Table \ref{grouptable} are tagged with 
the types $A$, $D$ and $E$. The
 Dynkin diagram can be assigned to each of the group in various ways,
 but the one most relevant to us is the McKay correspondence.
It goes as follows: 
for a discrete subgroup $\Gamma$ of $SU(2)$, let $\rho_s$ ($s=1,\ldots,n_\Gamma$)
be its irreducible representations, and prepare $n_\Gamma$ nodes associated 
to them. $\Gamma$ has a standard two-dimensional representation $\rho_{\two}$
as a subgroup of $SU(2)$, and it happens that 
the irreducible decomposition of $\rho_s\otimes \rho_\two$ contains
each of the irreducible representation at most once. Thus we can write\begin{equation}
\rho_s \otimes \rho_{\two}=\bigoplus_{t\in S_s}\rho_t \label{decomposition}
\end{equation} using $S_s\subset \{1,\ldots,n_\Gamma\}$.
The fact that $\two$ is the conjugate representation of itself means
$t\in S_s$ if and only if $s\in S_t$.  The surprising correspondence found by McKay
is that the graph of $n_\Gamma$ 
points with edges drawn between $s$ and $t$ if $t\in S_s$ 
is one of the extended Dynkin diagrams of A-D-E type, see Fig.~\ref{mckay}.
For $D_n$ and $E_n$, irreducible representations are labeled
with their dimensions, and we put primes to distinguish different
irreducible representations of the same dimension.
The readers can find the character tables of these groups in Appendix~\ref{bigtable}.

\FIGURE{
\def\dynkinspacer{2em}
\def\get#1{$\vcenter{\hbox{\includegraphics[scale=.5]{#1}}}$}
\begin{tabular}{l|c}
$A_{2n-1}$ &\get{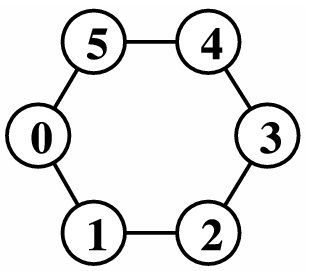} \\[\dynkinspacer]
$D_{n+2}$ &\get{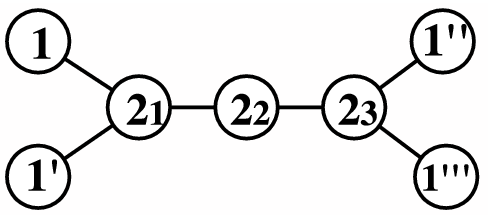} \\[\dynkinspacer]
\end{tabular}\qquad
\begin{tabular}{l|c}
$E_6$ &\get{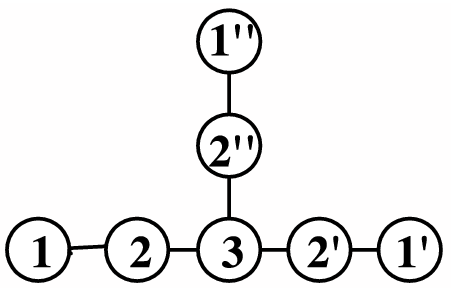} \\[\dynkinspacer]
$E_7$ &\get{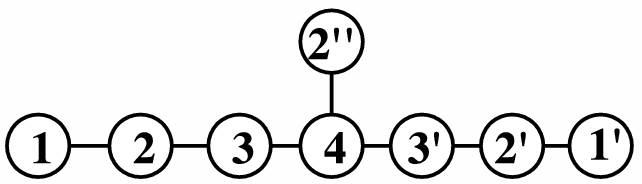} \\[\dynkinspacer]
$E_8$ &\get{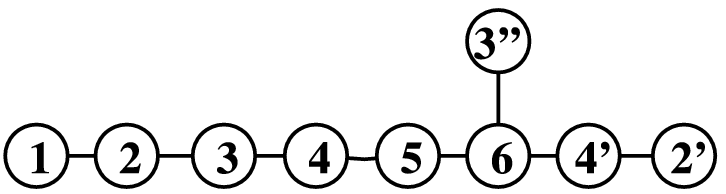} \\[\dynkinspacer]
\end{tabular}
\caption{Diagrams formed by irreducible representations.
We used $A_5$ and $D_6$ for illustration.\label{mckay}}
}

One immediate consequence of \eqref{decomposition}
is that  $d_s=\dim \rho_s$ satisfies \begin{equation}
2 d_s=\sum_{t\in S_s} d_t,\label{eigenvector}
\end{equation}i.e.\ the vector $(d_s)$ is  the eigenvector of the Cartan matrix
with eigenvalue $0$. In a similar way, for any element $g\in \Gamma$,
the vector of the characters $(\tr_{\rho_s} g)$ is an eigenvector
with the eigenvalue $2-\tr_{\two} g$.

Another important representation of $\Gamma$ is the regular representation $\rho_r$
which is $|\Gamma|$ dimensional: its orthonormal basis is given by $e_g$ for
$g\in \Gamma$ and the action of $\Gamma$ is given by \begin{equation}
\rho_r(h) e_g = e_{hg}.
\end{equation} A fundamental theorem of finite group theory states that
the regular representation decomposes as \begin{equation}
\rho_r=\bigoplus_{s=1}^{n_\Gamma} \rho_s ^{\oplus d_s}\label{regulardecomposition}
\end{equation} and thus \begin{equation}
|\Gamma|=\sum_{s=1}^{n_\Gamma} d_s{}^2.\label{sum-of-dim2}
\end{equation}

\subsection{Geometry}
$T^{1,1}/\Gamma$ is a smooth space with no orbifold singularity,
because the action of $\Gamma\subset SU(2)_1$ 
to the coordinates $(\theta_1,\phi_1,\psi)$ is topologically
the group action of $SU(2)$ from the left of $S^3\sim SU(2)$,
and thus it has no fixed points in $S^3$.
One can also understand this fact using the Hopf fibration $S^3\to S^2$.
The action of $SU(2)_1$ to the $S^2$ parametrized by $(\theta_1,\phi_1)$
is the usual rotation of $SO(3)$. Thus, for any element $g\in SU(2)_1$ which is not
$\pm\id$, there are two points on $S^2$ fixed by $g$.  
At these fixed points, $g$ acts as the translation of the coordinate $\psi$ of
the $S^1$ fiber by \begin{equation}
\psi\to\psi+4\pi/n
\end{equation}where $n$ is the smallest nonzero integer such that $g^n=\id$.

As an orbifold without fixed points, the fundamental group is given by 
the orbifolding group itself, that is $\pi_1(T^{1,1}/\Gamma)=\Gamma$.
Utilizing this, let us check that the orbifolding does not reduce
the amount of supersymmetry preserved by the background.
The covariantly constant spinor $\psi$ is determined only up to an
overall phase, and the supersymmetry is broken
if there is  a non-trivial phase  $\alpha(g)\in U(1)$
after the parallel transport along the path $g\in\pi_1$,
\begin{equation}
g_* \psi = \alpha(g) \psi.
\end{equation} Here $g_*\psi$ is the spinor after the parallel transport.
To show $\alpha(g)=1$,  one only needs to realize 
that one can define $\alpha(g)$  for arbitrary $g\in SU(2)$ by the pull-back.
Then  $\alpha: SU(2)\to U(1)$ must be a one-dimensional representation of $SU(2)$,
which is automatically trivial.

\subsection{Construction of the quiver}\label{construction-of-quiver}
Let us now construct the dual quiver gauge theory for the Type IIB string on
$T^{1,1}/\Gamma\times AdS_5$, where $\Gamma$ is one of the
discrete subgroup of $SU(2)_1$.
Its moduli space  should
contain the $N$-th symmetric power of  $C/\Gamma$, i.e. 
the conifold  $C$ orbifolded by $\Gamma$.

We follow the procedure given by Douglas and Moore \cite{DouglasMoore} in the case
of $\cN=2$ orbifold of $\bC^2$:  to realize  $N$ D3-branes moving
on $C/\Gamma$, we consider $\tilde N=N|\Gamma|$ D3-branes
on $C$ so that D3-branes occupy the points related by the action of $\Gamma$.
Therefore, 
we start from the conifold gauge theory with two $SU(\tilde N)$ gauge groups
with four bifundamentals $A^i$ and $B^j$.
We label $\tilde N$ rows and  columns by the pair 
$(k,g)$ where $k=1,\ldots,N$ and $g\in \Gamma$.
The branch which concerns us is the one where $A^i$ and $B^j$ are all diagonal;
we denote the diagonal entries by $a^i_{k,g}$ and $b^i_{k,g}$.
Then we need to impose \begin{equation}
a^i_{k,hg}=\rho_\two(h)^i_j a^j_{k,g},\qquad
b^i_{k,hg}=b^i_{k,g}
\end{equation} for all $h\in\Gamma$, 
so that the $\tilde N$ branes are placed at the points
related by $\Gamma\subset SU(2)_1$.

The conditions above can be enforced by demanding that \begin{equation}
A^i{}_{k}{}^{l}=\rho_\two(h)^i_j \rho_r(h) A^j{}_k{}^{l} \rho_r(h)^{-1},\qquad
B^i{}_{k}{}^{l}= \rho_r(h) B^i{}_k{}^{l} \rho_r(h)^{-1}.\label{AB}
\end{equation} We omitted the indices for the regular representation to reduce
the clutter. To be consistent, the generator $X$ of the gauge transformations 
should also be restricted so that \begin{equation}
X_k{}^l=\rho_r(h) X_k{}^l \rho_r(h)^{-1}. \label{X}
\end{equation}

To analyze further, we change the basis  of the regular representation to the
RHS of \eqref{regulardecomposition}, and replace the index $g\in \Gamma$
by the triple $(s,\alpha,a)$, where $s=1,\ldots, n_\Gamma$ labels irreducible
representations of $\Gamma$, 
$\alpha=1,\ldots,d_s$ labels $d_s$ copies of $\rho_s$ in $\rho_r$,
and $a$ is the index on which $\rho_s$ acts.  The equation \eqref{X} becomes
\begin{equation}
X_{k,s,\alpha}{}^{l,t,\beta}=\rho_s(h) X_{k,s,\alpha}{}^{l,t,\beta}\rho_t(h)^{-1}
\end{equation} and we omitted the indices for $\rho_{s,t}$.
The solution is given by \begin{equation}
X_{k,s,\alpha}{}^{l,t,\beta}=  x_{k,\alpha}^{l,\beta} \delta_s^t\id_s.
\end{equation} via Schur's lemma. 
Here $\id_s$ is the identity matrix of the representation $\rho_s$.
 Thus each of the two $SU(\tilde N)_{1,2}$
gauge groups is projected to the product of $SU(N d_s)_{1,2}$ gauge groups,
$s=1,\ldots,n_\Gamma$. In the same way, the  field $B^i$
is decomposed to  the bifundamental fields $B^i_{s}$ 
connecting $SU(N d_s)_2$ to $SU(N d_s)_1$.   
The $SU(2)_2$ global symmetry acting on $\ccB$-type fields remains unbroken
by  orbifolding.

The condition \eqref{AB} for the field $A$ is slightly more complicated: in the
new basis it becomes \begin{equation}
A^i{}_{k,s,\alpha}{}^{l,t,\beta}=
\rho_\two(h)^i_j\rho_s(h) A^j{}_{k,s,\alpha}{}^{l,t,\beta}\rho_t(h)^{-1}.
\end{equation} Again, Schur's lemma means the solution is given by \begin{equation}
A^i{}_{k,s,\alpha,a}{}^{l,t,\beta,b}=
a_{k,s,\alpha}{}^{l,t,\beta} P^{i,b}_{a}
\end{equation} for $s\in S_t$, and zero otherwise. 
Here $P^{i,b}_{a}$ is the projector from $\rho_\two\otimes \rho_t$
 to the component $\rho_s$ in \eqref{decomposition}.
Thus, the field $A^i$  is projected to the bifundamental fields $A_{s\to t}$
connecting $SU(Nd_s)_1$ and $SU(Nd_t)_2$ whenever the nodes 
$s$ and $t$ are connected in the Dynkin diagram. 

\FIGURE{\def\quiverspacer{3em}
\def\get#1{$\vcenter{\hbox{\includegraphics[scale=.35]{#1}}}$}
\begin{tabular}{l|c}
$A_{2n-1}$ &\get{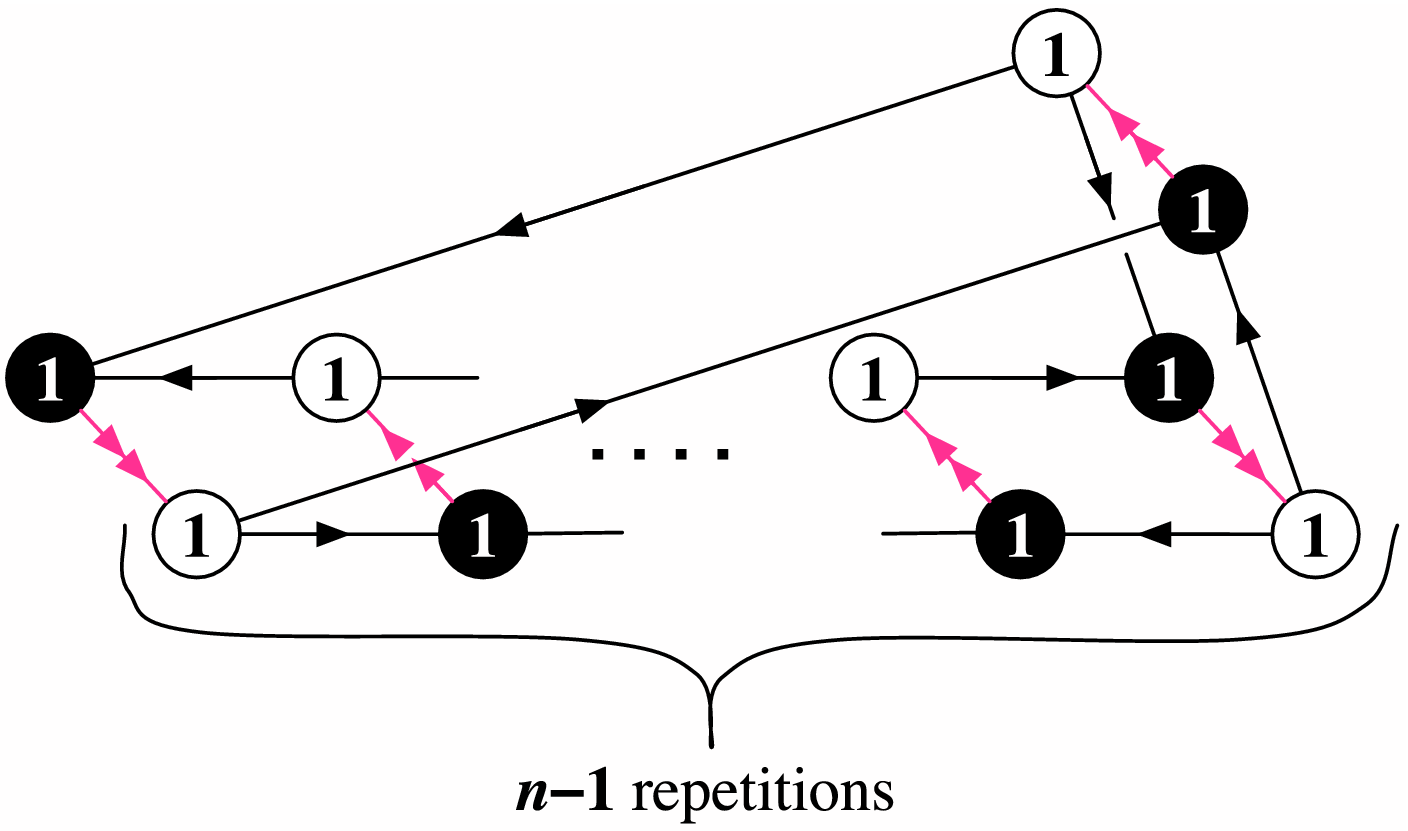} \\[\quiverspacer]
$D_{n+2}$ &\get{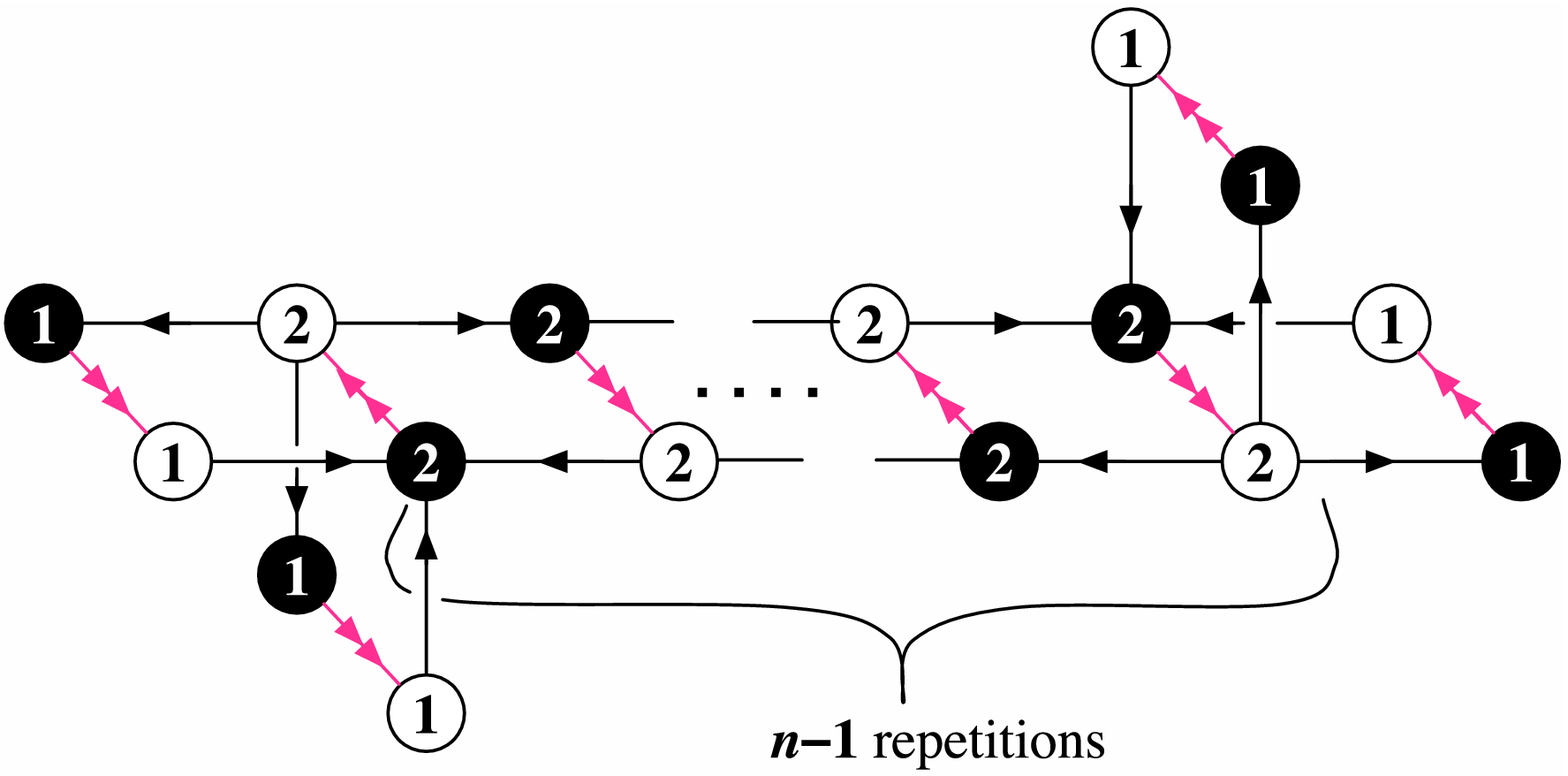} \\[\quiverspacer]
$E_6$ &\get{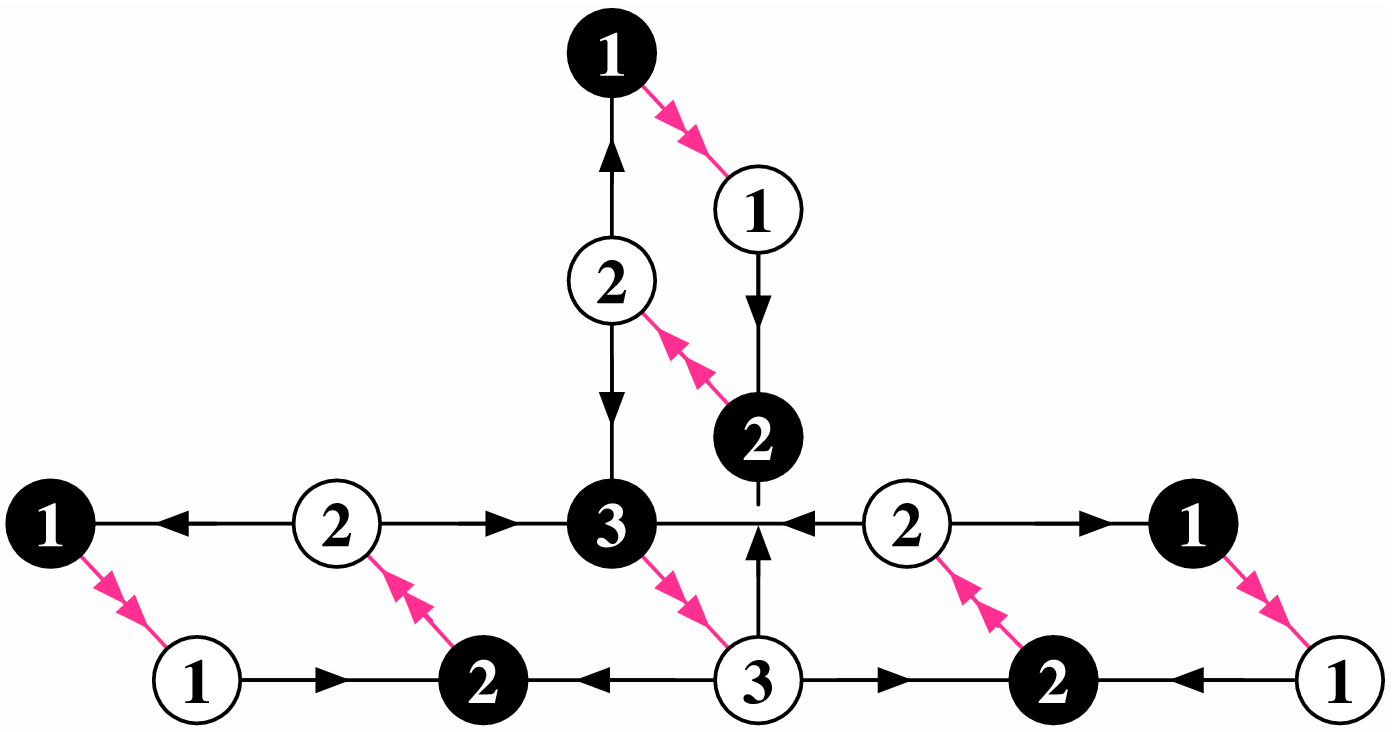} \\[\quiverspacer]
$E_7$ &\get{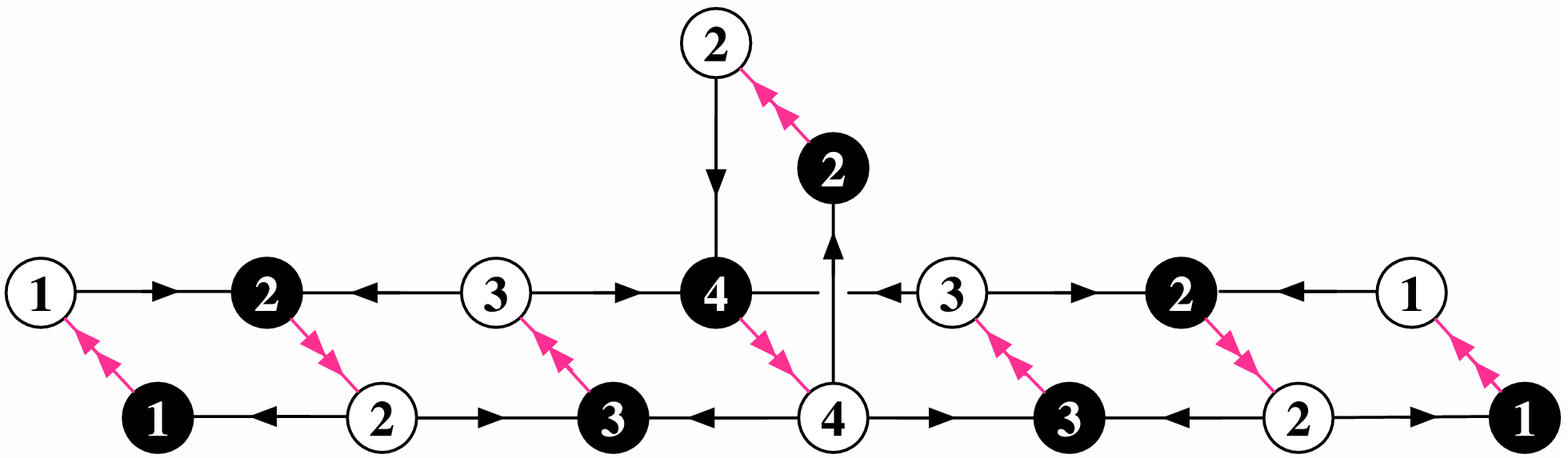} \\[\quiverspacer]
$E_8$ &\get{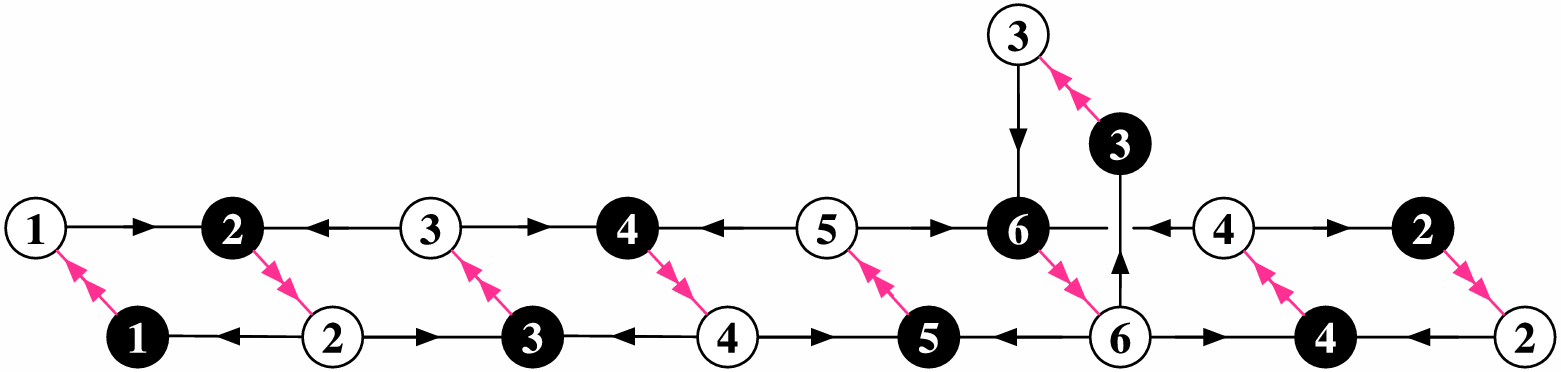} \\[\quiverspacer]
\end{tabular}
\caption{Quivers.\label{quivers}}
}

The resulting quivers  are depicted in Fig.~\ref{quivers}.
There, the $SU(N d_s)_1$ gauge group is represented by a white circle
with $d_s$ inside, the $SU(N d_t)_2$ gauge group by a black circle with $d_t$
inside, $A_{s\to t}$ fields by black single arrows from
a white circle to a black one, and finally $B^i_s$ fields by red double arrows
from a black circle to a white one.
We call the fields $A_{s\to t}$ and $B^i_s$ collectively as
$\ccA$-type and $\ccB$-type fields%
\footnote{We also denote baryonic operators constructed from 
$\ccA$-type and $\ccB$-type fields as $\ccA$-type and $\ccB$-type baryons.
Do not confuse `$\ccA$-type baryons' with `baryons of $A$-type quivers.'}, 
respectively. 
The superpotential of the theory is quartic, coming from 
$\epsilon_{ij}\epsilon_{kl}\tr A^iB^kA^jB^l$ of the unprojected 
$SU(N|\Gamma|)\times SU(N|\Gamma|)$ theory.

The structure of the quiver is quite simple; it consists of two extended Dynkin
diagrams of the A-D-E type of the discrete subgroup used, connected
to a ladder by double arrows of  $\ccB$-type fields.
For $\Gamma=\bZ_n$, this is equivalent to the so-called $Y^{n,0}$ quiver.
An important property of each of the Dynkin sub-quiver formed
by $\ccA$-type fields is that the direction of the arrows are alternating, i.e.\ 
 the single black arrows at each nodes are all incoming or all outgoing.
We call them  alternating Dynkin quivers.
This particular orientation of the arrows is known to be natural
from the point of view of McKay correspondence \cite{Lusztig}.
Indeed, it can be realized by first
 classifying the nodes into two sets depending whether
$-\id \in SU(2)$ is represented as $+1$ or $-1$ in the corresponding representation,
and then by connecting the first set to the second by arrows.

\subsection{Some checks of the correspondence}

We constructed the quiver gauge theory to have $N$-th symmetric
product of $C/\Gamma$ as a branch of the moduli space, so it passes
the first test that it should describe the motion of $N$ D3-branes of $C/\Gamma$.
Also, the quiver we obtained is free from cubic gauge anomalies
of the $SU(Nd_s)_{1,2}$ gauge groups as there are
as many  incoming arrows as 
outgoing ones  at each node because of the relation \eqref{eigenvector}.

Now we take the low-energy limit of the theory when all of the D3-branes
are at the tip of the cone, which leads to the correspondence of the Type IIB
string on $T^{1,1}/\Gamma\times AdS_5$ and the infrared limit
of the quiver gauge theory of the last subsection.
It is natural to assign the scaling dimension  $3/4$ to all of the bifundamentals 
as was the case in the 
un-orbifolded conifold theory. 
Then the NSVZ beta function of each gauge group can be checked to vanish
using the relation \eqref{eigenvector}.

Next, the central charges calculated from the geometry and the gauge theory
agree. Indeed, from the formula \eqref{holographicA} we have \begin{equation}
a=c=\frac{27}{64} N^2 |\Gamma|\label{AofT11/G}
\end{equation} since $\Vol (T^{1,1}/\Gamma)=\Vol T^{1,1}/ |\Gamma|$.
From the point of view of the gauge theory, there are $\sum_s d_s{}^2$ times
as many  vector multiplets, $\ccA$- and $\ccB$-type chiral multiplets 
with the same assignments of $R$-charges.  Then, the central charges
are $\sum_s d_s{}^2$ times those of the conifold theory, and thanks to the 
relation \eqref{sum-of-dim2} it is equal to \eqref{AofT11/G}.

Finally, the internal global symmetry of both sides agree: 
the isometry $SU(2)_1\times SU(2)_2\times U(1)_\psi$ of $T^{1,1}$
is broken to $U(1)\times SU(2)_2\times U(1)_\psi$ or $SU(2)_2\times U(1)_\psi$
depending on whether $\Gamma$ is Abelian or not.
In the gauge theory, $SU(2)_2$ acting on $B^i$ fields is left intact under the
orbifold projection, and the same is true for $U(1)_R = U(1)_\psi$.
As for $SU(2)_1$ symmetry acting on the $A$ fields, 
there remains a $U(1)$ subgroup if $\Gamma=\bZ_n$ 
by assigning the charge $+1$ to $A_{s\to s+1}$ and $-1$ to $A_{s\to s-1}$,
whereas nothing remains as the symmetry if $\Gamma$ is non-Abelian.
The baryonic symmetry of the conifold theory,
with the charge $+1$ for $A$ fields and $-1$ for $B$ fields,
is inherited in the orbifolded theory by the same assignment  of the charges,
and it agrees with the geometry in that
 $\dim H^3(T^{1,1}/\Gamma,\bR)=1$.

\section{Baryons on  $T^{1,1}/\Gamma$}\label{baryons}

We have constructed, \`a la Douglas and Moore,
the quiver gauge theory describing D3-branes
probing the conifold orbifolded by a discrete subgroup $\Gamma$
of its $SU(2)$ isometry. In the large $N$ limit
it should be the dual gauge theory of  Type IIB string theory on 
$T^{1,1}/\Gamma\times AdS_5$,
 and we performed some preliminary checks of the correspondence.
The checks we did so far were satisfied more or less by
construction of the quiver. We now move on to the main topic
of our paper, namely the study of the baryonic operators
and of their realization as wrapped  D3 branes.

We will exclude the subgroup $\bZ_{2n-1}\subset SU(2)$ in the following analysis,
because it does not include $-\id\in SU(2)$  and shows a quite different behavior
compared to other subgroups.  In any case, the orbifold of the conifold
by $\bZ_{2n-1}$ is toric, whose baryonic operators are the subject of intense
study by various groups\cite{baryonic-counting0,baryonic-counting1,baryonic-counting2}, 
and will hopefully be treated elsewhere.

We also study only the baryonic operators 
which are constructed solely from $\ccA$-type bifundamentals,
or those made solely of $\ccB$-type bifundamentals.  It is mainly because of
technical complexity of the analysis of  baryons of mixed types, 
as can be inferred from the analysis in the toric cases.
We will find a quite intricate structure even in the reduced classes  
of operators which we will analyze in the following.

\subsection{$\ccB$-type baryons}
Let us wrap a D3-brane at fixed $(\theta_2,\phi_2)$.
In the un-orbifolded case, 
the brane corresponds to the operator $\det B$ in the conifold gauge theory.
Hence in the orbifolded case, the operator should be constructed solely from the
$\ccB$-type fields in the quiver.

Here, the worldvolume is $S^3/\Gamma$.  
We can wrap multiple, say $k$ of D3-branes
at the same place, then we have the choice of the flat worldvolume
gauge field in $U(k)$ \cite{GRW,GV}. 
Since $\pi_1(S^3/\Gamma)=\Gamma$,
the freedom in the Wilson lines is given by a $k$-dimensional representation 
of $\Gamma$, which decompose to the direct sum of  irreducible representations
$\rho_s$ of $\Gamma$.

Hence there should be an  operator $\cB_s$
in the gauge theory for each irreducible
representation $\rho_s$.  The motion along $(\theta_2,\phi_2)$ needs
to be quantized. For the trivial representation of $\Gamma$,
it gives rise to the baryonic operator which transforms as the
spin $N/2$ representation of $SU(2)_2$.  For $\rho_s$ with $d_s=\dim \rho_s>1$,
the moduli space of the center of mass is still $S^2$, but we have $d_s$ times
as much five-form flux coupling to the worldvolume.
Therefore  it comes
in the spin $Nd_s/2$ representation of $SU(2)_2$ global symmetry.

The scaling dimension of these operators $\cB_s$ can be fixed by the comparison with
the un-orbifolded case. Recall that the scaling dimension is 
given  by the mass
of the wrapped D3-branes  times the curvature radius of the AdS space.
By the equations of motion  of Type IIB string theory,
it is clear that the curvature of the AdS space is the same for
the $T^{1,1}/\Gamma$ theory  with $N$ units of flux
and $T^{1,1}$ theory with $N|\Gamma|$ units of flux. 
In the latter case, the brane wrapped on $S^3$ parametrized by 
$(\theta_1,\phi_1,\psi)$  corresponds to the operator $\det B$ with
$\Delta=3N|\Gamma|/4$.     
In the former, the brane wrapped on $S^3/\Gamma$
has $1/|\Gamma|$ as much mass as that of the latter.
Therefore, the gauge theory operator $\cB_s$ 
should have the dimension \begin{equation}
\Delta(\cB_s)= 3Nd_s /4.
\end{equation}  In the following, the scaling dimension always appears 
with a factor of $3N/4$, so we define the weight $w$ of an operator $\cO$
by \begin{equation}
\Delta(\cO)=(3N/4) w(\cO).
\end{equation} Then $w(\cO)=d_s$.

The gauge theory naturally reproduces this result. Indeed, since $\ccB$-type
bifundamental fields are disconnected in the quiver diagram, any operator
constructed solely out of $\ccB$-type fields are the product of the following operators
\begin{equation}
\det B_s^{i_1\ldots i_{Nd_s}}
 \equiv  \epsilon_{1s} \epsilon_{2s} B_s^{i_1}\cdots B_s^{i_{Nd_s}},
\end{equation} where $\epsilon_{is}$, $i=1,2$ are the epsilon symbols
of $SU(Nd_s)_i$ gauge groups, and we omitted the gauge indices for brevity. 
It is easy to see that 
$\det B_s$ has weight $w(\det B_s)=d_s$, and comes
in  the spin $N d_s/2$ representation of $SU(2)_2$
because two epsilon symbols symmetrize 
the indices $i_1,\ldots,i_{Nd_s}$.

\subsection{$\ccA$-type baryons}
\subsubsection{Geometry}\label{geom-A}
Next, let us consider D3 branes wrapped 
on three-cycles at fixed $(\theta_1,\phi_1)$.
The corresponding operator in the un-orbifolded case
is $\det A$. Therefore the operators for these branes in the orbifolded case
should be constructed out of the $\ccA$-type  bifundamentals only.

If the coordinates $(\theta_1,\phi_1)$ are generic, the only element
in $\Gamma$ which fixes these coordinates is $-\id$, which shifts 
the $\psi$ coordinate halfway, $\psi\to \psi +2\pi$.  Thus we know 
the worldvolume
topology is $S^3/\bZ_2$, and we have the choice of the worldvolume
Wilson line which gives the phase $\pm 1$ when one traverses the $\psi$ coordinate.
The weight of the corresponding operator is determined
by repeating the previous argument, and is \begin{equation}
w= |\Gamma|/2.\label{genericDeltaOfA}
\end{equation} Let us denote  operators of this kind collectively by $\cA$.

The D3-brane can be moved along $(\theta_1,\phi_1)$
preserving  supersymmetry, and hence
we need to quantize the motion along this direction.
The moduli space
of the D3-brane  in this direction is $S^2/\Gamma$,
which we presented in Fig.~\ref{fundamental-region}.
$S^2/\Gamma$ has orbifold singularities,
and the D3-brane put at these points can decay into multiple D3-branes
as we will see shortly.
Therefore the quantization will be a delicate procedure; 
but if we neglect the subtlety, there is $N+1$ zero modes.
It is because its wavefunction should be a holomorphic section of a line bundle
over $S^2/\Gamma$ with $c_1=N$, as was briefly reviewed in Sec.~\ref{T11}.
It is natural to suppose the correction to the number of the zero-modes will be
$O(1)$ in the large $N$ limit. Combined with the choice of the Wilson line
along the $\psi$  direction, we predict the existence of $2N+O(1)$ distinct
operators which we collectively call $\cA$,
with weight given by \eqref{genericDeltaOfA}.

Let us consider what happens if we put the D3-branes at one of the 
orbifold points of $S^2/\Gamma$.  Suppose the subgroup fixing the point is
$\bZ_{2k}\subset \Gamma$.  Then the $S^1$ fiber above is acted by the shift
in the $\psi$ direction \begin{equation}
\psi \to \psi +2\pi/k.
\end{equation}  
Thus, the length of the fiber is $1/k$ of that of the $S^1$ fiber
over generic points of $S^2/\Gamma$, 
and a single D3-brane wrapped at a generic point
of $S^2/\Gamma$ can decay into $k$ D3-branes
when moved to  the orbifold point.

The worldvolume is topologically
$S^3/\bZ_{2k}$, and as always we have $2k$ choices $\alpha$ of the Wilson line phase 
along the $\psi$ direction which should satisfy $\alpha^{2k}=1$ \cite{GRW,GV}.
Then our prediction for the gauge theory operators is that,
 for each of the orbifold points of $S^2/\Gamma$, there are $2k$ distinct 
operators of  weight $w=|\Gamma|/(2k)$
where $2k$ is the order of the orbifold point.
As discussed in Sec.~\ref{classification}, there are
two orbifold points with $k=n$ for $\Gamma=\bZ_{2n}$ and 
there are three orbifold points with $k$ given by one of $(p,q,r)$ of the Platonic triple
if $\Gamma$ is non-Abelian.

\subsubsection{Gauge theory}\label{A-gauge}

A quick inspection of the quiver diagram, Fig.~\ref{quivers}, tells us
that the $\ccA$-type bifundamentals form two disjoint sets of Dynkin diagrams,
and each of the Dynkin diagram has the direction of its arrows alternating in the sense
that each node has all arrows connected to it either all incoming 
or all outgoing.
The direction of arrows of one Dynkin diagram is the reverse of that of 
the other Dynkin
diagram. Since the reversal of the arrows is  a matter of
a change in  convention, it is clear that two Dynkin diagrams gives the same
number of gauge-invariant operators of the same scaling dimension.

Thus, from the analysis of the geometry, we expect that each 
alternating Dynkin quiver gives, for each of the 
orbifold point of $S^2/\Gamma$ of the form $\bC/\bZ_k$, 
$k$ gauge-invariant operators of weight $|\Gamma|/(2k)$.
It is a  non-trivial mathematical prediction
on the inter-relation among objects classified by A-D-E
 coming from the
AdS/CFT correspondence, in that the Dynkin diagram of type $\Gamma$
`knows' how $\Gamma$ acts on $S^2$ as a subgroup of $SU(2)$.
Indeed, this mathematical statement is precisely what was proved 
in \cite{SkowronskiWeyman}.

Let us now construct some baryonic operators:
the quiver gauge theory
was obtained by imposing the condition \eqref{AB} on
the conifold gauge theory with $SU(N|\Gamma|)^2$ gauge group,
and therefore we can embed the $\ccA$-type bifundamental fields
in  the fields $A^{1,2}$ of the unorbifolded gauge theory. 
We can then
 construct an $\ccA$-type baryon by forming the dibaryon \begin{equation}
\det (\lambda_i A^i)\label{bigbaryon}
 \end{equation}
 given a two-dimensional complex vector $\lambda_i$.
The condition \eqref{AB} on the matrices $A^i$ means
that $\lambda_i$, $\lambda_i'$ related by the action of
$g\in\Gamma\subset SU(2)$ defines the same operator,
and obviously a scalar multiplication of $\lambda_i$ does not matter either.
Therefore the moduli space of such operators forms the space $\CP^1/\Gamma$.
 
It is a $N|\Gamma|$-by-$N|\Gamma|$  matrix, and thus gives
a weight-$|\Gamma|$ operator which is twice as much as the operator
we would like to have. However,
the $N|\Gamma|$-dimensional space of Chan-Paton indices
can be decomposed into the direct sum of two vector spaces $V_{\pm}$ of
dimension $N|\Gamma|/2$  where  the element 
$-\id\in SU(2)_1$ of the global symmetry acts as $\pm1$
and that  $A^i$ is block-off-diagonal with respect to this decomposition.
Furthermore, this decomposition is compatible with the action
of the gauge group of the quiver, $G=\prod_s SU(d_sN)$, since
$G\subset SU(|\Gamma|N)$ is defined as the subgroup which commutes
with the action of $\Gamma$, see \eqref{X}.
Therefore the dibaryon  \eqref{bigbaryon} is a product of two 
gauge-invariant baryonic operators,
\begin{equation}
\det(\lambda_i A^i)=\cA_{+\to -}(\lambda_i) \cA_{-\to +}(\lambda_i),
\label{Adibaryon}
\end{equation} each of weight $|\Gamma|/2$. 
Here, $\cA_{\pm\to\mp}$ is the determinant of the part of $\lambda_i A^i$
which maps $V_\pm$ to $V_\mp$.
They give $O(N)$ distinct operators each. The precise number will be determined 
in Sec.~\ref{Abranch}.

Next let us consider what happens when the vector $\lambda_i$ 
is at an orbifold point of $\CP^1/\Gamma$, i.e.~when $\lambda_i$ 
is an eigenvector of $g\in \Gamma\subset SU(2)$ generating $\bZ_{2k}$.
We can assume $\lambda^i$ has the eigenvalue $\alpha=\exp(2\pi i/(2k))$,
and  decompose the spaces $V_{\pm}$ into the eigenspaces $V_{i}$ of $g$ 
where $g$ acts as a scalar multiplication by $\alpha^i$. We then have
\begin{equation}
V_{+}=V_{2}\oplus V_{4}\oplus\cdots\oplus V_{{2k}},\quad
V_{-}=V_{1}\oplus V_{3}\oplus\cdots\oplus V_{{2k-1}},
\end{equation} and the dimension of $V_{i}$ is uniformly $|\Gamma|/(2k)$
since $g$ acts as the permutation of the basis vectors of the regular representation.

Now the condition \eqref{AB} means the matrix $\lambda_iA^i$ 
maps the block $V_{i}$ to $V_{{i+1}}$;  the dibaryon \eqref{Adibaryon}
further decomposes into the product \begin{align}
\begin{array}{l}
\cA_{+\to -}=\cA_{0\to1}\ \cA_{2\to3}\cdots
\cA_{{-2}\to{-1}}, \\
\cA_{-\to +}=\cA_{1\to2}\ \cA_{3\to4}\cdots
\cA_{{-1}\to{0}} 
\end{array} \label{decomposition_at_orbifoldpoint}
\end{align} where $\cA_{i\to{i+1}}$ is the determinant
of the part of $\lambda_i A^i$ mapping $V_{i}$ to $V_{{i+1}}$.
Each $\cA_{i\to{i+1}}$ has weight $|\Gamma|/(2k)$, and there are $2k$ of them.
In this way we constructed the baryonic operators
which realizes the expectation from the AdS/CFT correspondence;
we call these operators ``fractional dibaryons.''

However, the analysis of the geometry of $T^{1,1}/\Gamma$
predicts not just the existence of the operators with the prescribed weight;
it also predicts they are the only baryonic operators.
We continue our discussion in the new Section because, as we will see,
the confirmation of the prediction requires a quite lengthy analysis.

\section{$\ccA$-type baryons on $T^{1,1}/\Gamma$: direct analysis}\label{direct}
In the previous section we found that there are $\ccA$-type baryonic operators
of the alternating Dynkin quivers with weight predicted by the geometry.
The aim of this section is to construct such operators explicitly 
and to show that any $\ccA$-type baryon can be written as a polynomial of them.
To this aim, we put geometric  intuition aside for a while, and
consider the problem purely from the viewpoint of the quiver gauge theory.
Sec.~\ref{direct} and Sec.~\ref{indirect} provide two mostly independent
methods of classifying these baryonic operators, and can be read 
in either order according to the reader's taste.

\subsection{Some examples}
Let us first construct some baryonic operators on the alternating Dynkin quiver.
We use the most interesting icosahedral group $\hat\cI=E_8$ as the example.
The analysis of the geometry of $S^2/\cI$ told us that the 
operator with the smallest weight is 
the one with weight $12=120/(2\cdot 5)$. What does it look like?

\FIGURE{
\includegraphics[width=.4\textwidth]{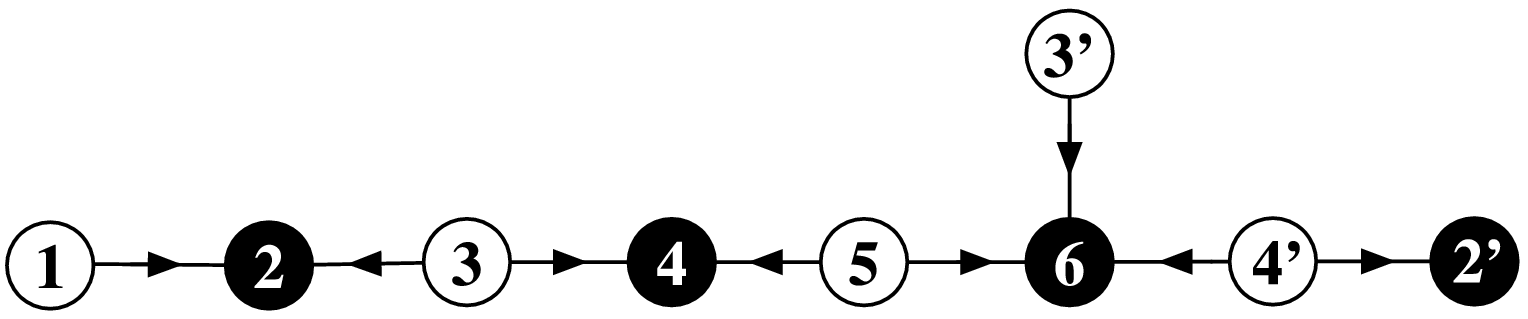}
\caption{Alternating quiver for $E_8$.\label{e8quiver}}
}

The subquiver we are concerned is depicted in Fig.~\ref{e8quiver}.
Let us try to imitate the construction of the dibaryons.   We take
the $N$-th power of  the field $A_{1\to 2}$ and contract the $N$ indices
of $SU(N)$ by the epsilon symbol $\epsilon_{(1)}$ of $SU(N)$. Now we have
an operator of the form \begin{equation}
\epsilon_{(1)}^{i_1\cdots i_N}A_{1\to 2}{}_{i_1}^{j_1}\cdots A_{1\to2}{}_{i_N}^{j_N}
\end{equation}
The indices $j_1$ to $j_N$ for $SU(2N)$ need to be contracted using the 
epsilon symbol $\epsilon^{(2)}$ for $SU(2N)$, 
which has $2N$ indices. Thus, contrary to the case
of the dibaryons, the bifundamental field $A_{1\to 2}$ alone cannot make
a gauge-invariant operator.  We need to contract the extra $N$ indices
by using $N$ of the field $A_{3\to 2}$; now we have 
an operator of the form \begin{equation}
\epsilon_{(1)}^{i_1\cdots i_N}A_{1\to 2}{}_{i_1}^{j_1}\cdots A_{1\to2}{}_{i_N}^{j_N}
\epsilon^{(2)}_{j_1\cdots j_Nj_{N+1}j_{2N}}A_{3\to 2}{}^{j_{N+1}}_{k_1}\cdots
A_{3\to 2}{}^{j_{2N}}_{k_N}
\end{equation} It is not yet gauge invariant, and we need to continue this procedure.
The expression for the operator becomes increasingly  cumbersome, so we abbreviate 
it as \begin{equation}
\epsilon_{(1)} (A_{1\to2})^N \epsilon^{(2)} (A_{3\to2})^N,
\end{equation}where the gauge indices are suppressed and the contraction
against epsilon symbols are understood. 

Now we see that it becomes gauge-invariant at the stage \begin{multline}
\cP_1=\epsilon_{(1)} (A_{1\to2})^N \epsilon^{(2)} (A_{3\to2})^N
\epsilon_{(3)} (A_{3\to 4})^{2N} \times\\
\epsilon^{(4)} (A_{5\to 4})^{2N}
\epsilon_{(5)} (A_{5\to 6})^{3N} \epsilon^{(6)} (A_{3'\to 6})^{3N}
\epsilon_{(3')}\label{baryon1}
\end{multline}  which has weight $12$, as was predicted from the geometry!
We can also construct a baryonic operator  starting from the $A_{3\to 2}$ field,
which results in the operator \begin{equation}
\cP_2=\epsilon^{(2)} (A_{3\to2})^{2N}
\epsilon_{(3)} (A_{3\to 4})^{N} \epsilon^{(4)} (A_{5\to 4})^{3N}
\epsilon_{(5)} (A_{5\to 6})^{2N} \epsilon^{(6)} (A_{4'\to 6})^{4N}
\epsilon_{(4')}.\label{baryon2}
\end{equation} Its weight is also $12$.  To verify the prediction 
from the geometry, we need to find three other operators with weight $12$,
and to check that they are the only ones.

Let us introduce another notation for the baryons thus constructed: 
we introduce a vector $(m_i)_{i=1,\ldots,s}$ where
$m_i$ is the number of the epsilon symbols used 
for the gauge group $SU(d_i N)$. 
We call it the `dimension vector' of the operator
for the reason we will see in Sec.~\ref{indirect}.
When we write down the dimension vector explicitly,
we put the numbers in the form of the Dynkin diagram, 
e.g.~for the operators \eqref{baryon1} and \eqref{baryon2}, 
they are \begin{equation}
v_1=\Ee101011111,\quad
v_2=\Ee000111111.\label{examples}
\end{equation}  
If we hypothetically enlarge the $i$-th gauge group from $SU(d_i N)$ to
$U(d_i N)$, the vector $(N m_i)$ gives the charge vector of the baryonic operator
under the $U(1)$ parts of the gauge groups of the nodes.  These $U(1)$ symmetries
are anomalous, yet useful in classifying the baryonic operators
of the theory\footnote{The importance of  anomalous baryonic symmetries
was pointed out to the authors by A. Hanany.}.

Now it is easy to see that $\cP_1$ and $\cP_2$ are the only operators
with dimension vector $v_1$ and $v_2$, respectively. The dimension vector
of $(\cP_1)^2$ is $2v_1$, and the analysis of the geometry of $T^{1,1}/\Gamma$
predicts that it is the only operator with this dimension vector. 
It is not obvious from the point of view of the gauge theory.
Indeed, an operator with dimension vector $2v_1$
has the form \begin{equation}
\epsilon_{(1)}\epsilon_{(1)} (A_{1\to2})^{2N} 
\epsilon^{(2)}\epsilon^{(2)} (A_{3\to2})^{2N}\cdots,
\end{equation} and there seems to be the choice of how many
$A_{1\to2}$ fields contract the indices of the first $\epsilon_{(1)}$ and
of the first $\epsilon^{(2)}$, and of the first $\epsilon_{(1)}$ and 
of the second $\epsilon  ^{(2)}$, etc. If the prediction from the AdS/CFT 
correspondence is true, these multitude of operators should be proportional
to each other. Similar choices in the contraction of indices
arise for each bifundamental fields, and in total there are milliards of possibilities
in the way of contraction. 
Therefore we need  powerful methods to `untangle' these
complicated contraction of indices of epsilon symbols, which we develop
in Sec.~\ref{sec:untangling} and \ref{duality}.

\subsection{Untangling procedure}\label{sec:untangling}
\FIGURE{
\includegraphics[width=.4\textwidth]{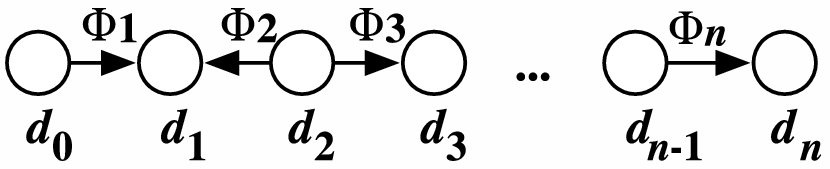}
\caption{Linear alternating quiver.\label{linear}}
}
\noindent In this section we explain how one can transform
baryonic operators of the linear alternating quiver, see Fig.~\ref{linear},
to a standard `untangled' form. 
There, the gauge group of the $i$-th node is $SU(d_i)$, 
and there is a bifundamental $\Phi_i$ between $SU(d_{i-1})$ and $SU(d_i)$
with the index structure specified by the direction of the arrow, i.e.
all the indices of $SU(d_{\text{even}})$ transforms as anti-fundamental,
and those of $SU(d_{\text{odd}})$ as fundamental.
As a further technical assumption,
we demand that $d_i \le d_{i+1}$ is satisfied for arbitrary $i$.

Suppose the baryonic operator contains $m_i$
epsilon symbols for $SU(d_i)$. In order to specify the contraction
of indices, we denote the $\alpha$-th epsilon symbol of $SU(d_i)$ by
$\epsilon_i^{(\alpha)}$, $\alpha=1,\ldots,m_i$.
The standard `untangled' form  of the operator
is such that any bifundamental field $\Phi_i$ is contracted
against $\epsilon_{i-1}^{(\alpha)}$ and $\epsilon_{i}^{(\alpha)}$ with the same $\alpha$,
see Fig.~\ref{untangled}. 
The arrows in this figure represent the contraction of gauge indices.
\FIGURE{
\includegraphics[width=.4\textwidth]{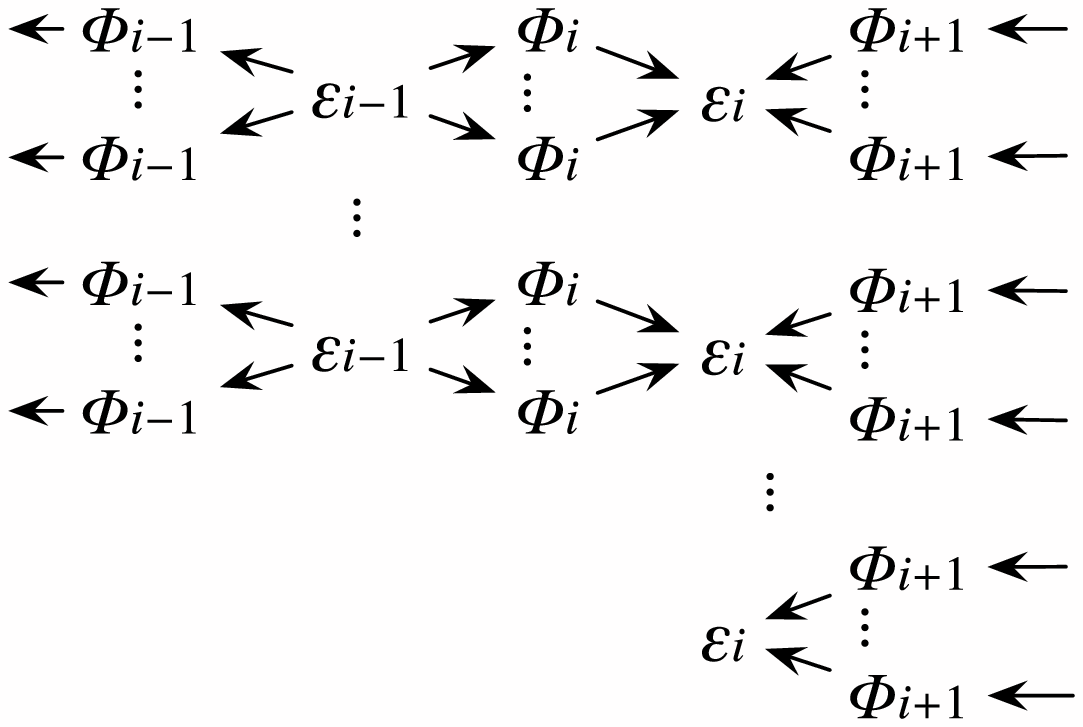}\vspace*{-1em}
\caption{Untangled operator .\label{untangled}}
}

First, for any given baryonic operator,
all the $SU(d_{i-1})$ and $SU(d_i)$ indices of $\Phi_i$'s in the operator
are contracted against
invariant tensors $\varepsilon_{i-1}$ and $\varepsilon_{i}$ respectively,
and the remaining indices of $\varepsilon_{i-1}$ and $\varepsilon_{i}$
are contracted with those of 
$\Phi_{i-1}$ and $\Phi_{i+1}$ as the left hand side of Fig.~\ref{procedure}.
The region with the question mark `?' in this figure
 is an arbitrary permutation of arrows.
We can simplify this by using the identity 
\begin{eqnarray}
\varepsilon^{a_1\cdots a_k c_1\cdots c_{d-k}}
 \overline\varepsilon_{b_1\cdots b_k c_1\cdots c_{d-k}}
\propto \delta^{[a_1}_{[b_1} \cdots \delta^{a_k]}_{b_k]}, \label{ee}
\end{eqnarray}where $\epsilon$ is the epsilon symbol which can be used
to contract the bifundamentals in the quiver,
and $\bar\epsilon$ is the one which cannot be.
Suppose $k$ of $\Phi_i$s are contracted with 
$\varepsilon_{i-1}^{(\alpha)}$.
Since the $SU(d_{i})$ indices of these $\Phi_i$  are then completely antisymmetric,
we can insert $ \delta^{[a_1}_{[b_1} \cdots \delta^{a_k]}_{b_k]}$ 
for $SU(d_{i})$ there, 
and we replace the Kronecker deltas by two epsilon symbols
using \eqref{ee}.
That is, we rewrite the relevant part as follows:
\begin{multline}
(\varepsilon_{i-1}^{(\alpha)}) ^{a_1 \cdots a_k} (\Phi_i) _{a_1}^{i_1} \cdots 
(\Phi_i) _{a_k}^{i_k}
\propto \\
(\varepsilon_{i-1}^{(\alpha)}) ^{a_1 \cdots a_k} (\Phi_i) _{a_1}^{j_1} \cdots 
(\Phi_i) _{a_k}^{j_k} 
\varepsilon_{j_1\cdots j_k c_1\cdots c_{d-k}}
 \overline\varepsilon^{i_1\cdots i_k c_1\cdots c_{d-k}}
\label{part_op},
\end{multline} 
 see the right hand side of Fig~\ref{procedure}.
 
If $m_{i-1}\le m_i$,
 we insert $m_{i-1}$ pairs of $\epsilon_i$ and $\overline\epsilon_i$ for
all $\epsilon_{i-1}^{(\alpha)}$, $\alpha=1,\ldots, m_i$.
Then, we use (\ref{ee}) again to eliminate the newly-introduced $\overline\epsilon_i$ 
against $\varepsilon_{i}$ 
which is originally 
in the baryonic operators from the beginning. The elimination turns them
into antisymmetrized product of Kronecker deltas, 
and at this stage, the indices of $SU(d_{i})$ of 
the $\Phi_i$s which are contracted with $\varepsilon_{i-1}^{(\alpha)}$ 
are contracted with one and the same $\varepsilon_{i}$
as Fig.~\ref{untangled}. 
We rename this $\varepsilon_{i}$ as $\varepsilon_{i}^{(\alpha)}$.
We eliminate all of $\overline\epsilon$'s in this way,
and the result is now in the standard, `untangled' form.
We call this procedure as the `untangling of $\Phi_i$ at $SU(d_i)$'.

If $m_{i-1}>m_i$, the operator vanishes if $d_{i-2}<d_{i-1}$, or
contains $\det\Phi_{i-1}$  as a factor if $d_{i-2}=d_{i-1}$.
Indeed, we can exchange the roles of $SU(d_{i-1})$ and $SU(d_{i})$
in the discussion above and  introduce $m_i$ pairs
of $\overline\epsilon_{i-1}\epsilon_{i-1}$ for each $\epsilon_{i}^{(\alpha)}$.
After the elimination of newly introduced $\overline\epsilon_{i-1}$,
$m_i-m_{i-1}$ of  $\epsilon_{i-1}$ have their indices all contracted
against $\Phi_{i-1}$, not against any $\Phi_{i}$.
If $d_{i-2}<d_{i-1}$, such an operator vanishes from the rank condition 
because $\Phi_{i-1}$ is a $d_{i-2}\times d_{i-1}$ matrix.
If $d_{i-2}=d_{i-1}$, we can show that it contains $\det\Phi_{i-1}$ as a factor
by untangling of $\Phi_i$ at $SU(d_i)$.

Similarly, if $d_{i}=d_{i+1}$ and $m_{i-1}<m_i$, the untangling of 
$\Phi_i$ at $SU(d_i)$  makes all the indices of $m_i-m_{i-1}$ $\epsilon_{i}$
to contract against $\Phi_{i+1}$. Then, untangling of 
$\Phi_{i+1}$ at $SU(d_{i+1})$ produces $m_i-m_{i-1}$ factors of $\det\Phi_{i+1}$
from the original operator.

\FIGURE{
\includegraphics[width=.95\textwidth]{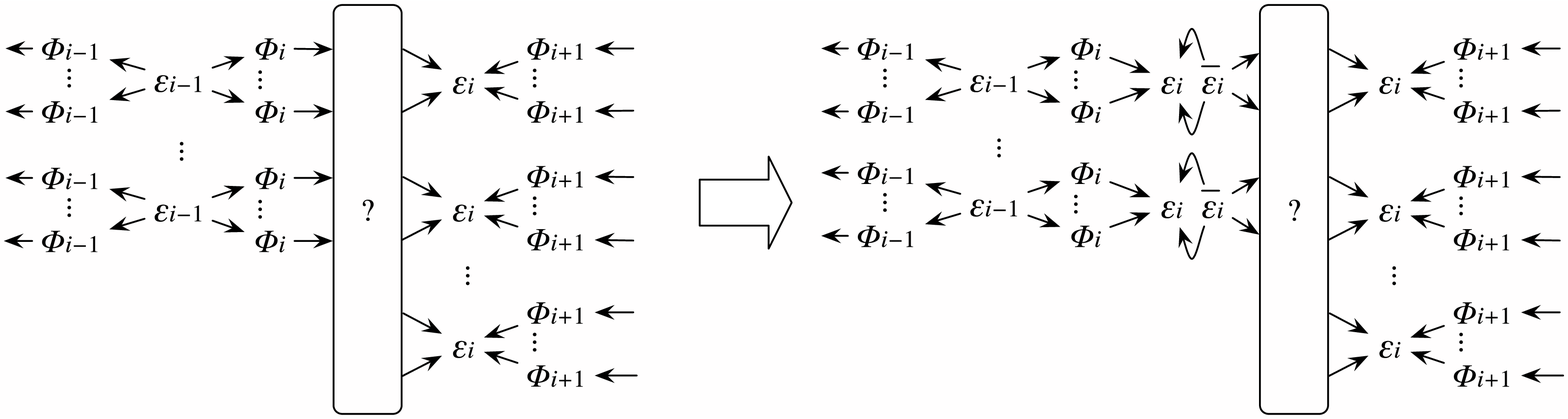}
\caption{Untangling procedure.\label{procedure}}
}

\subsection{Action of the Seiberg duality}\label{duality}
In this section we study the action of the Seiberg duality \cite{seiberg} 
on our quiver theory\footnote{
The authors would like to thank I. Klebanov for his suggestion to study 
the Seiberg duality acting on the baryonic operators.}.
Let us first consider the application of the duality  at one of the nodes,
see Fig.~\ref{seiberg}.  
We use a slightly general setup where $M\ne N$.
The subtheory we consider has $SU(N)$ as the gauge group,
and we treat $SU(M_{1,2})$ and $SU(M)$ as the global symmetry group.
We name the bifundamental fields as in the figure.
We assume $2M=M_1+M_2$ so that the gauge anomaly vanishes. Then
the $SU(N)$ gauge group has effectively $2M$ flavors, and 
the dual theory has $SU(N')$ with $N'=2M-N$ as the gauge group.
The arrows are reversed, since the representation of the dual quarks under
the global symmetry is the complex conjugate of the original ones.
There are extra meson fields in the dual theory,
but we are more interested in the baryons, so let us discuss
them first. 
\FIGURE{
\includegraphics[width=.6\textwidth]{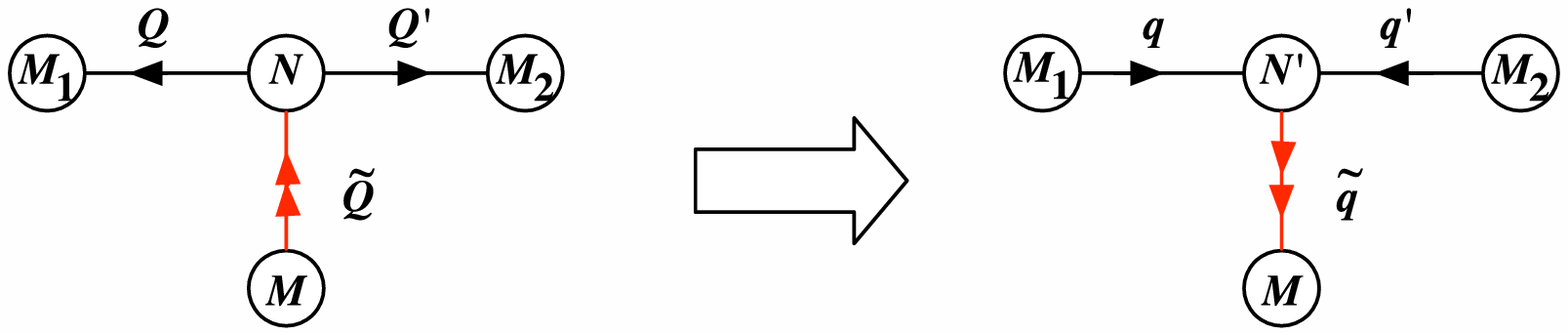}
\caption{Seiberg duality at one of the nodes.\label{seiberg}}
}

It is known that the baryons in the $SU(N_c)$ theory with $N_f$ flavors
$Q_i$, $\tilde Q_i$ and the dual  $SU(N'_c=N_f-N_c)$ theory with $q^i$, $\tilde q^i$
 are related by the rule \begin{equation}
Q_{i_1}\cdots Q_{i_{N_c}}\epsilon_{(N_c)}
=
\epsilon_{i_1i_2\cdots i_{N_f}}
q^{i_{N_c+1}}\cdots q^{i_{N_f}}\epsilon^{(N_c')},
\end{equation} where $\epsilon_{(X)}$ and $\epsilon^{(X)}$ are the epsilon
symbols of $SU(X)$ theory with all indices up or down, respectively, and
we omitted the gauge indices for $SU(N_c)$ and $SU(N_c')$ for simplicity.
Then, if we decompose the global symmetry $SU(N_f)$ to $SU(M_1)\times SU(M_2)$,
we find the correspondence \begin{equation}
Q^k \epsilon^{(N)} Q'{}^{N-k}  = 
\epsilon^{(M_1)} q^{M_1-k} \epsilon_{(N')} q'{}^{M_2+k-N} 
\epsilon^{(M_2)}\label{baryondual}.
\end{equation} This equivalence of baryons of two quivers with opposite
orientation of arrows 
has been known in mathematical literature for twenty years \cite{Kac1}.

Now let us perform the Seiberg duality on our quiver gauge theory,
simultaneously for all of the white nodes.
Then we arrive at the quiver with all arrows reversed compared to the original theory.
In our conformal case  we have $N=M$, so that $N'=N$.
It is easy to see that the meson fields can be integrated out to give back the
purely quartic superpotential formed by the dual quarks, as was the case
in the conifold theory.   Note that it is a matter of convention
which of the two $N$-dimensional representations of $SU(N)$ one calls the fundamental
representation, so we are back at the original theory.  
Under this self-duality, the baryons are transformed non-trivially.
Indeed, using \eqref{baryondual} repeatedly, 
the dimension vector $m'_i$ of the transformed baryon is given by
the dimension vector $m_i$ of the original one via the relation
\begin{align}
m'_i &= m_i  &\text{for white nodes},\\
m'_i &= -m_i +\sum_{j\ \text{connected to}\ i} m_j  & \text{for black nodes}.
\end{align} We denote this action by $m'=W(m)$ where we chose the letter
 $W$ to remind us that we performed the duality for white nodes. 
There is a similar transformation
$m'=B(m)$ performed by the Seiberg duality of the black nodes.

It is instructive to calculate explicitly the action of $B$ and $W$ to the known
baryons constructed in the last sections, e.g.~the ones in \eqref{examples}.
For definiteness we consider the Dynkin subquiver where the node
corresponding to the trivial representation is a white node.
We obtain the following actions : \begin{equation}
\begin{array}{ccccccccc}
v_1=\Ee101011111& \stackrel{\hbox{$B$}}{\longleftrightarrow}&
v_2=\Ee000111111&\stackrel{\hbox{$W$}}{\longleftrightarrow}&
v_3=\Ee010111110\\
\mathop{\vcenter{\hbox{\rotatebox{180}{$\curvearrowright$}}}}\limits_{\hbox{$W$}}
&&&&B \ \vcenter{\hbox{\rotatebox{90}{$\longleftrightarrow$}}}\\
&B \ \vcenter{\hbox{\rotatebox{90}{$\curvearrowright$}}}&v_5=\Ee001121000 &\stackrel{\hbox{$W$}}{\longleftrightarrow}& 
v_4=\Ee011111100
\end{array}\label{seibergactiononfractionals}
\end{equation} We obviously have $W^2=B^2=\mathop{\mathrm{id}}$, but
we have a surprising result that $B$ and $W$ do not commute. We would like to
understand its interpretation from the bulk AdS side, but it is beyond the scope
of the present paper. 

We utilize these actions of $W$ and $B$ in a more
practical way. Recall that the Seiberg duality acts not just on the dimension
vectors but on the individual operators, and there is a one-to-one
mapping between the operators. Thus, if two dimension vectors $v$ and $v'$
are connected by the action of the Seiberg duality,
we are guaranteed to have the same 
number of independent baryonic operators with dimension vector $v$ and $v'$.
For example, to count the number of operators with dimension vector $v_5$,
we apply the dualities $B$ and $W$
to map the dimension vector to $v_3$.
Now it is easy to see that there is one and only one 
baryonic operator with dimension vector $v_3$, by untangling
the operator from both ends.
Thus we also have
one and only one baryonic operator with dimension vector $v_5$.

\subsection{Classification of $\ccA$-type baryons}\label{direct-classification}
With these preparations,
we begin our direct analysis of $\ccA$-type  baryons.
We heavily 
utilize the untangling procedure
and the Seiberg duality discussed  above.
As we will see, the analysis is now straightforward but tedious, so we
split some part of the exposition in the Appendix~\ref{continued}.
In this subsection we enumerate baryons  for $\Gamma$ being
 the cyclic groups and dihedral groups, with weight not more than $|\Gamma|/2$.
In the Appendix~\ref{continued} we show that the basic operators
we find in this subsection generate the whole $\ccA$-type baryons 
for $\Gamma=\bZ_{2n}$ and $\hat\cD_n$; we also see there
how we can analyze the polyhedral cases.
We will see in Sec.~\ref{indirect} how the theory of quiver representation
can give an indirect but efficient  way to analyze the baryons.

\subsubsection{Cyclic groups $\bZ_{2n}$}\label{A-direct}
Let us label the gauge groups as $SU(N)_0$, $SU(N)_1$,\ldots $SU(N)_{2n-1}$,
 and the bifundamental connecting $SU(N)_i$ 
and $SU(N)_{i+1}$ as $\Phi_i$. We identify the index $i$ of the gauge group
modulo $2n$. See the example in Fig.~\ref{z6quiver} for $\Gamma=\bZ_6$.
Here, the circle with a number $i$ in it denotes the  gauge group $SU(N)_i$.
\FIGURE{
\includegraphics[width=.2\textwidth]{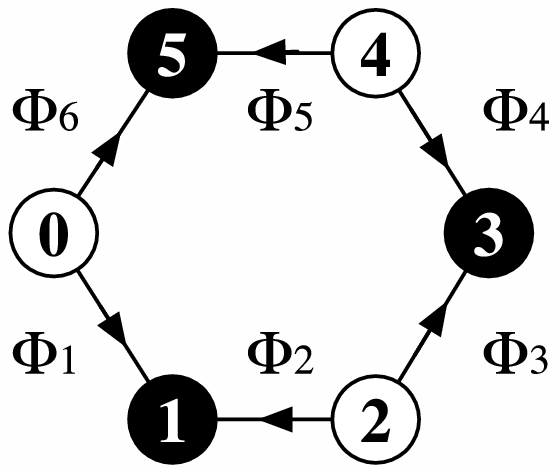}
\caption{Alternating quiver for $\bZ_6$.\label{z6quiver}}
}

Suppose we have an operator with dimension vector $m_i$, i.e.
a gauge-invariant operator which 
is constructed by contracting the gauge indices of bifundamental fields
by $m_i$ epsilon symbols $\epsilon_i$ for $SU(N)_i$. 
We first show that  the operator contains a factor of a dibaryon $\det\Phi_i$
if not all of $m_i$ are equal. 
Indeed,  then,  there is an integer $j$ such that $m_{j-1} < m_j$. 
Untangling $\Phi_{j}$ at $SU(N)_j$,  we find
$m_j-m_{j-1}$ epsilon symbols are contracted with $\Phi_{j+1}$ only,
not at all with $\Phi_{j}$.  
Untangling $\Phi_{j+1}$  at $SU(N)_{j+1}$ then makes
 $(\det\Phi_{j+1})^{m_j-m_{j-1}}$ factored out of the original operator.

Thus we first find $2n$ dibaryons $\det\Phi_i$, $i=1,2,\ldots,2n$ 
as a part of the generators of the baryonic operators.  The first candidates of
operators which cannot be  written as their polynomial 
should have $m_1=m_2=\cdots=m_{2n}=1$ from the discussion above.
We can write down $N-1$ kinds of such gauge-invariant operators  as follows:
\begin{equation}
\cO_k=\epsilon_0 (\Phi_1)^k \epsilon_1 (\Phi_2)^{N-k} \epsilon_2 (\Phi_3)^k
\cdots (\Phi_{2n-1})^k \epsilon_{2n-1} (\Phi_{2n})^{N-k},
\end{equation}where $k=1,2,\ldots,N-1$. They are all of weight $n=|\Gamma|/2$.

Combining the results so far, we have found $2n$ generators of weight 1
and $N-1$ generators of weight $n$, which precisely matches with the
prediction from the geometry of $T^{1,1}/\bZ_{2n}$ we discussed in Sec.~\ref{geom-A}.
We can show that any operator of the higher weight can be written as a polynomial
of these generators using a careful application of untangling,
which we will  discuss in Appendix~\ref{direct-A}.
We will derive the same fact using the theory
of quiver representations in Sec.~\ref{application-of-kac}.

\subsubsection{Dihedral groups $\hat\cD_n$}\label{D-direct}
\FIGURE{
\includegraphics[width=.25\textwidth]{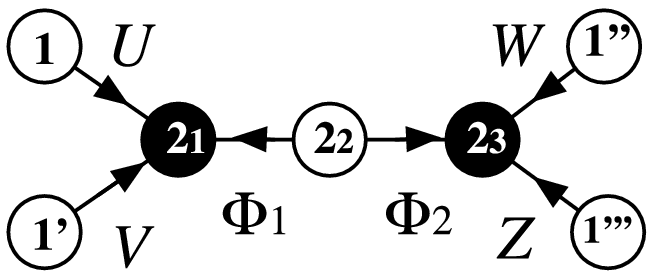}
\caption{Alternating quiver for $\hat\cD_4$.\label{d4quiver}}
}

We call the gauge groups as $SU(N)$, $SU(N)'$, $SU(N)''$,
$SU(N)'''$, $SU(2N)_1$, $SU(2N)_2$,\ldots, and $SU(2N)_{n-1}$.
We have bifundamentals 
$U$ connecting $SU(N)$ and $SU(2N)_1$,
$V$ connecting $SU(N)'$ and $SU(2N)_1$,
$W$ connecting $SU(N)''$ and $SU(2N)_{n-1}$,
$Z$ connecting $SU(N)'''$ and $SU(2N)_{n-1}$;
we also have bifundamentals $\Phi_i$ connecting
$SU(N)_i$ and $SU(N)_{i+1}$ for $i=1,2,\ldots,n-2$. See the example $\hat\cD_4$
depicted in Fig.~\ref{d4quiver}. 
There, a circle with $1'$ in it denotes the $SU(N)'$ gauge group, 
 a circle with $2_2$ in it  the gauge group $SU(2N)_2$, and so on.

One can construct $n$ gauge-invariant  operators $\cP_{1,\ldots,n}$
of weight $2$ as follows: \begin{equation}
\cP_1=\epsilon_N U^N \epsilon_{2N,1} V^N\epsilon_{N'},\quad
\cP_{i}=\det \Phi_{i-1},\quad
\cP_n=\epsilon_{N''} W^N \epsilon_{2N,n-1} Z^N\epsilon_{N'''},
\end{equation} and $i=2,\ldots,n-1$.
Here, $\epsilon_{2N,i}$ is the epsilon symbol of $SU(2N)_i$
and $\epsilon_{N'}$ is that for $SU(N)'$, etc. We omitted the gauge indices
for brevity, but it is clear that there is only one way of contracting the indices.

Let us take a gauge-invariant operator $\cO$ with dimension vector 
($m$, $m'$, $m''$, $m'''$; $\{m_i\}$) where $m'$ is the number of the epsilon symbols
for $SU(N)'$ etc., and $m_i$ is the number of the epsilon symbols for $SU(2N)_i$.
We first show that   some of $\cP_i$ can be factored out of $\cO$
unless $m_i$ is all equal. Indeed, if $m_1>m_2$, we untangle
 $\Phi_1$ at $SU(2N)_1$ to find that the indices of 
at least $m_1-m_2$ epsilon symbols are contracted against only $U$ or $V$,
which inevitably leads to $m_1-m_2$ factors of $\cP_1$. Similarly, 
we find $m_{n-1}-m_{n-2}$ factors of $\cP_n$ if $m_{n-1}>m_{n-2}$.
If neither is the case, there should be $i$ in the range $1<i<n-1$
such that  and $m_i>m_{i+1}$  or $m_{i-1}<m_{i}$. The untangling 
then yields $m_{i}-m_{i+1}$ factors of $\det \Phi_{i-1}$ or
$m_i-m_{i-1}$ factors of  $\det\Phi_{i+1}$, respectively.

In the following we assume $m_i=\mu$ for all $1\le i\le n-2$.
Next we note that if an operator $\cO$ is decomposable
if $m+m'\ne \mu$.
It can be proved by taking the Seiberg dual at $SU(N)$, $SU(N)'$ and the nodes
with the same color. Then the resulting operator $\cO'$ has $m_1\ne m_2$ 
because
\begin{equation}
m_1(\cO')=m+m'+m_2-m_1\ne \mu,\quad
m_2(\cO')=m_2(\cO)=\mu.
\end{equation} Thus we can apply the preceding argument to show
it decomposes.

Now let us analyze the operator with $m_i=1$ for all $1\le i\le n-1$.
It is indecomposable if we have $(m,m,m',m'')=(1,0,1,0)$,
$(1,0,0,1)$, $(0,1,1,0)$ or $(0,1,0,1)$, which lead to four operators \begin{align}
\cQ_1&=U^N (\Phi_1)^N \cdots (\Phi_{n-2})^N  W^N,&
\cQ_2&= V^N (\Phi_1)^N \cdots (\Phi_{n-2})^N  Z^N\\
\cR_1&=U^N  (\Phi_1)^N \cdots (\Phi_{n-2})^N  Z^N,&
\cR_2&= V^N (\Phi_1)^N \cdots (\Phi_{n-2})^N  W^N
\end{align} where the contraction of the indices against one epsilon symbol for each gauge group is understood. They all have weight $n$.

Next let us analyze the operator with $m_i=2$ for all $1\le i\le n-1$.
We first show that it decomposes if $(m,m')=(2,0)$.
Indeed, we can apply the untangling procedure repeatedly,
starting by $W$ and $Z$ at $SU(2N)_{n-1}$, then
for $\Phi_{n-2}$ at $SU(2N)_{n-2}$, \ldots and finally for $U$ at $SU(N)_1$,
which makes the operator proportional to either $\cQ_1^2$ or $\cQ_1\cR_1$.
Similar arguments can be made for the case $(m,m')=(0,2)$, etc.
Therefore, to be indecomposable, we need to have $m=m'=m''=m'''=1$. 

These operators are automatically of weight $2n$.
Now we apply the untangling procedure, starting from $Z$ and $W$ at $SU(2N)_{n-1}$,
then for $\Phi_{n-2}$ at $SU(2N)_{n-2}$, all the way to $\Phi_1$ at $SU(2N)_1$.
Then they are combined into the following parts which only have
$SU(2N)_1$ indices:
\begin{align}
\cW^{a_1\cdots a_N}&=\epsilon^{a_1\cdots a_N c_1\cdots c_N}
(\Phi_1)^N_{c_1\cdots c_N} \epsilon_2 (\Phi_2)^N 
\cdots (\Phi_{n-2})^N \epsilon_{n-2}W^N\epsilon_{N''},\\
\cZ^{a_1\cdots a_N}&=\epsilon^{a_1\cdots a_N c_1\cdots c_N}
(\Phi_1)^N_{c_1\cdots c_N} \epsilon_2 (\Phi_2)^N 
\cdots (\Phi_{n-2})^N \epsilon_{n-2}Z^N\epsilon_{N'''}.
\end{align} 
Therefore the baryonic  operator is of the form \begin{equation}
\cO_k=\cU_{a_1\cdots a_k b_1\cdots b_{N-k}} \cV_{c_1\cdots c_{N-k} d_1\cdots d_k}
\cW^{a_1\cdots a_k c_1\cdots c_{N-k}}
\cZ^{b_1\cdots b_{N-k} d_1\cdots d_k},\label{Dgeneral}
\end{equation} with $k=1,2,\ldots,N-1$, where \begin{equation}
\cU_{a_1\cdots a_N}=\epsilon_{N}(U^N)_{a_1\cdots a_N},\qquad
\cV_{a_1\cdots a_N}=\epsilon_{N'}(V^N)_{a_1\cdots a_N}.
\end{equation} We omitted the gauge indices other than that of $SU(2N)_1$ 
to reduce the clutter. 
We will show in Appendix \ref{non-linear} 
that a certain linear combination of them is decomposable for $N > 1$.
We will also see that the remaining $N-2$ of them are linearly independent in Sec.~\ref{Abranch}.

Summarizing, we found that there are $n$ operators $\cP_i$ with weight two 
and four operators $\cQ_{1,2}$ and $\cR_{1,2}$ with weight $n$.
We additionally found order $N$ of operators  with weight $2n$.
This spectrum is as it should be from the analysis of the geometry of $T^{1,1}/\cD_n$.

We can show that any gauge-invariant operator can be written
as a polynomial of the operators found above using the untangling procedure,
for the detail see Appendix~\ref{direct-D}.
We also see the same result can be derived using the structure of quiver 
representations in the next section.

\section{Baryonic operators and quiver representations}\label{indirect}

In the last section we performed a direct analysis of the baryonic
operators of the alternating Dynkin quiver, by the technique
of the untangling of epsilons and by the application of the 
Seiberg duality.  We studied the operators for $A$- and $D$-type
subgroups, but the classification became quite formidable
for other cases.  In this section we will take an indirect approach
utilizing  the  mathematical theory of  quiver representations. 
Our general strategy is the following. We first show that the baryonic operators
are spanned by the generalized determinants, defined in Sec.~\ref{determinants}.
Then in Sec.~\ref{quiver-rep} we will see that each generalized determinant operator
can be associated with a representation of the quiver. It reduces the enumeration
of baryonic operators to the study of stably indecomposable representations of the quiver.
We quote the theorem of Kac in Sec.~\ref{kac} which accomplishes the task
for the extended Dynkin quivers.  We apply the theorem to our gauge theory
in Sec.~\ref{application-of-kac} to confirm the prediction of the 
number of baryonic operators from the geometry of $T^{1,1}/\Gamma$.

\subsection{Generalized determinants}
\label{determinants}
\FIGURE{
\includegraphics[width=.3\textwidth]{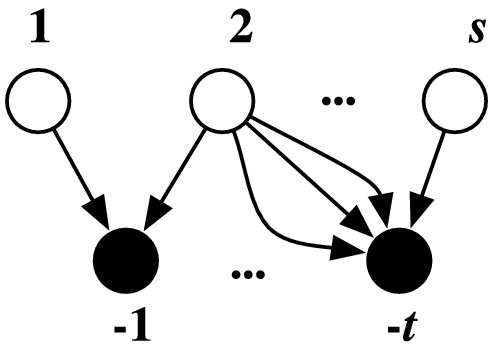}
\caption{Bipartite quiver.\label{bipartite}}
}

We first describe the baryonic operators which can be defined
as the determinant of a big matrix constructed from the bifundamental fields.
They can be defined for arbitrary bipartite quivers, i.e. quivers whose nodes 
can be divided into two classes, say white and black, and all the arrows
are from a white node to  a black node, see Fig.~\ref{bipartite}.
Let us label the white nodes by $i=1,\ldots,s$ and the black nodes by 
$i=-1,\ldots,-t$. Let us label  the arrows by $a=1,\ldots,u$ and 
denote the nodes of the tail and the head of the $i$-th arrow by
$t(a)$ and $h(a)$, respectively.

Now let us assign gauge groups $SU(d_i)$ to the nodes and bifundamental fields
$\Phi_a$ to the arrows $a=1,\ldots,u$. 
$\Phi_a$ has the index structure $\Phi_a{}^\alpha_\beta$
where $SU(d_{h(a)})$ and $SU(d_{t(a)})$ act on the indices
$\alpha=1,\ldots,d_{h(a)}$ and $\beta=1,\ldots,d_{t(a)}$
as the fundamental and as the anti-fundamental representation, respectively.

The fundamental theorem of the classical invariant theory states
that the only way of making gauge-invariant operators out of a monomial
of the bifundamental fields
\begin{equation}
\prod_a \Phi_a {}^{n_a}
\end{equation} is  to contract their indices against the epsilon
tensors of the gauge groups. 
Let $m_i$ be the number of the epsilon symbols
used for the $i$-th gauge group, which should satisfy \begin{equation}
m_i d_i = \sum_{t(a)=i} n_a \label{def-m}
\end{equation} for each $i>0$, and  a similar expression for each $i<0$.
It follows that \begin{equation}
\sum_{i>0} m_i d_i= \sum_{i<0} m_i d_i = \sum_a n_a \equiv w.\label{w}
\end{equation} 

Any gauge invariant operator with prescribed $n_a$
is a linear combination of the operators \begin{equation}
\underbrace{\epsilon_1 \cdots \epsilon_1}_{m_1}
\cdots
\underbrace{\epsilon_s \cdots \epsilon_s}_{m_s}
\underbrace{\epsilon_{-1} \cdots \epsilon_{-1}}_{m_{-1}}
\cdots
\underbrace{\epsilon_{-t} \cdots \epsilon_{-t}}_{m_{-t}}
\prod_a \Phi_a ^{n_a}
\end{equation} with various ways of contracting indices.
Here $\epsilon_i$ is the epsilon tensor of the $i$-th gauge group.
We call the vector $(m_i)$ of the number of the epsilon symbols
the dimension vector of the operator. The origin of the somewhat unnatural
name will be explained later.

To facilitate the specification of the way of contraction,
let us label each of $m_i$ epsilon tensors of $SU(d_i)$ as
$\epsilon_i^{(k)}$ with $k=1,\ldots,m_i$. Then the contraction
is fully specified by giving for each arrow $a$ the numbers $n_a{}^k_l$
of bifundamentals $\Phi_a$ connecting $\epsilon_{h(a)}^{(k)}$ and 
$\epsilon_{t(a)}^{(l)}$.  We denote the operator as \begin{equation}
\cO(\Phi_a,n_a{}^k_l).\label{basis1}
\end{equation}
Different sets of numbers $n_a{}^k_l$ may
correspond to linearly-dependent operators, but it is obvious
they give an over-complete set of gauge-invariant operators with given $n_a$.

It is still formidable to obtain the linearly independent basis of the operators
from the set \eqref{basis1}.  Let us now introduce another set of operators 
for the given number $m_i$ of epsilon symbols.  They are parametrized
by specifying for each arrow $a=1,\ldots,u$ 
a complex matrix $\lambda_a{}^k_l$ with indices $k=1,\ldots,m_{h(a)}$ and
$l=1,\ldots,m_{t(a)}$. Then we form a matrix $M(\Phi,\lambda)$
with  blocks  $M_{(i,j)}$, $i=1,\ldots,s$ and $j=-1,\ldots,-t$,
which is
a $m_i d_i\times m_{-j} d_{-j}$ matrix
\begin{equation}
M_{(i,j)}{}^{k\alpha}_{l\beta}
=\sum_{a\ \text{with}\ t(a)=i,\ h(a)=j}  \lambda_a{}^k_l \Phi_a{}^\alpha_\beta
\label{blockdef}
\end{equation} where $k=1,\ldots,m_j$, $l=1,\ldots,m_i$, $\alpha=1,\ldots,d_j$
and $\beta=1,\ldots,d_i$. 
These blocks form a $w$-by-$w$ matrix $M(\Phi,\lambda)$
thanks to the relation \eqref{w}.
Thus we can take its determinant \begin{equation}
D(\Phi,\lambda)\equiv \det M(\Phi,\lambda)
\end{equation} to get a gauge-invariant operator. We call them the 
generalized determinants.  

To the authors' knowledge, operators of this type were
first used  in string theory literature by
\cite{HerzogWalcher} in the study of baryonic operators
for the quiver gauge theory dual to the complex cones over  del Pezzos,
and they seem to have been known to mathematicians for decades.
A crucial observation by \cite{DZ} made in this century is that $D(\Phi,\lambda)$
also forms an over-complete basis of gauge invariant operators.
The only  thing to be shown is that the operator \eqref{basis1} 
can be obtained as the linear combination of operators $D(\Phi,\lambda)$.
It can be achieved by averaging $D(\Phi,\lambda)$ over $\lambda$: 
\begin{equation}
\cO(\Phi,n_a{}^k_l) \propto
\left(\prod_{a,k,l} \oint \frac{d\lambda_a{}^k_l}{ (\lambda_a{}^k_l)^{1+n_a{}^k_l}}\right)
D(\Phi,\lambda)\label{averaging}
\end{equation} where $\oint d\lambda$ is a contour integral along the unit circle
$|\lambda|=1$. Indeed, the averaging above picks the term proportional
to $\prod_{a,k,l}\left(\lambda_a{}^k_l\right)^{n_a{}^k_l}$ in $D(\Phi,\lambda)$,
which is seen to be $\cO(\Phi,n_a{}^k_l)$ by some mental gymnastics.

One immediate application is to the baryons of the conifold gauge theory, recall 
Fig.~\ref{KWquiver}.
The preceding theorem says that a baryon constructed from $A_{1,2}$ 
using one epsilon symbol for each gauge group is given by \begin{equation}
D(A,\lambda)=\det(\lambda_i A^i),
\end{equation} as it should be. Thus the analysis presented here 
can be thought of as a generalization of this well-known fact to general
bipartite quiver gauge theories.

\subsection{Relation to quiver representations}
\label{quiver-rep}
The blocks of $M(\Phi,\lambda)$ in \eqref{blockdef}
is defined symmetrically with the exchange of $\Phi$ and $\lambda$. 
Thus,  we can define the action of $g_i\in  GL(m_i)$ on $\lambda_a{}^k_l$,
which we schematically denote as $\lambda\to g \lambda g^{-1}$.
The generalized determinant then transforms as \begin{equation}
D(\Phi,g \lambda g^{-1}) = \left( \prod_{i>0}(\det g_i)^{d_i }
\prod_{j<0}(\det g_j)^{-d_j } \right) D(\Phi,\lambda), 
\end{equation}  i.e.~$D(\Phi,\lambda)$ and  $D(\Phi,g \lambda g^{-1})$
determine the same operator.  The important point for us
is that the equivalence classes
of matrices $\lambda$ under the action of $GL(m_i)$ is a well-studied
and beautiful branch of mathematics called the theory of  representations of quivers.

Let us introduce some terminologies.  (We drop the bipartite assumption
for the time being.)
A quiver  $Q$
is now a set of nodes $i=1,\ldots,s$ and arrows $a=1,\ldots,u$,
which connect  the node $t(a)$ to the node $h(a)$. 
A representation $\lambda$
of $Q$ is the assignment of vector spaces $\Lambda_i$ to the nodes,
and linear maps $\lambda_a:\Lambda_{t(a)}\to \Lambda_{h(a)}$ to the arrows.
The set of numbers $\dim \lambda = (\dim \Lambda_i)_{i=1,\ldots, s}$ is called
the dimension vector of $\lambda$.
Two representations $\lambda$, $\lambda'$ is called isomorphic if
$\dim \lambda =\dim \lambda'$ and moreover there is the choice
of invertible matrices $g_i\in GL(\dim \Lambda_i)$ acting on $\Lambda_i$ such that
$\lambda'_a= g_{h(a)} \lambda_a g_{t(a)}^{-1}$  for all arrows $a$.
The representation theory of quivers has been utilized in string theory,
see e.g.~\cite{He,CKV,BerensteinDouglas}. Previous usage of quiver representations
viewed $\Phi$ as the representation, whereas we mainly study the `dual'
quiver representation defined by $\lambda$ in the expression above.

What we showed above can be rephrased as the fact that
$D(\Phi,\lambda)$ and $D(\Phi,\lambda')$ define
the same operator if $\lambda$ and $\lambda'$ are isomorphic,
and that the dimension vector of $D(\Phi,\lambda)$ as defined in
the previous section, i.e.~the vector $(m_i)$ where
$m_i$ is the number of epsilon symbols
used for the $i$-th gauge group, is the dimension vector $\dim\lambda$,
which explains our terminology.

Another concept is the direct sum $\lambda\oplus \lambda'$ 
of two representations: it is defined as the assignment of
$\Lambda_i\oplus \Lambda_i'$ 
to the nodes and of $\lambda_a\oplus \lambda_a'$ to the arrows. 
A representation which can be written as a direct sum is called decomposable,
and if not,  indecomposable.
An indecomposable representation is called stably indecomposable if no infinitesimal
deformation makes the representation decomposable.

\FIGURE{
\includegraphics[width=.15\textwidth]{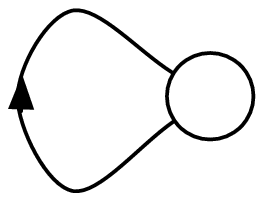}
\caption{A quiver with one node.\label{1pt-quiver}}
}
For an example, consider a quiver with one node and a loop attached to it,
see Fig.~\ref{1pt-quiver}.
A representation of this quiver with the dimension vector $(N)$ is just a
$N\times N$ matrix, and the classification of the representations is
just that of  square matrices up to conjugation. It is easy to see that
an indecomposable representation is a Jordan block, but a Jordan block with more than
one row  becomes diagonalizable i.e.~decomposable by a small
perturbation. Therefore stably indecomposable representations are
1-by-1 matrices.

The usefulness of the concept of indecomposability lies in the fact
 that the matrix $M(\Phi,\lambda\oplus\lambda')$ 
is just the block diagonal sum of the matrices $M(\Phi,\lambda)$
and $M(\Phi,\lambda')$. As their determinants,
$D(\Phi,\lambda)$ then satisfies the relation \begin{equation}
D(\Phi,\lambda\oplus\lambda')=D(\Phi,\lambda) D(\Phi,\lambda').
\label{decomposition-of-D}
\end{equation}
Thus, the generalized determinant for a decomposable $\lambda$ 
decomposes as the product of gauge-invariant operators.
One word of caution is necessary, because $D(\Phi,\lambda)$ can be
decomposable even when $\lambda$ is indecomposable. It often vanishes completely,
e.g.~for a basic indecomposable representation $e_i$ of the quiver $Q$ which assigns
$\Lambda_i=\bC$ and $\Lambda_j$ zero-dimensional for $i\ne j$, 
the maps $\lambda_a$ are automatically zero
so that $D(\Phi,\lambda)$ is also zero. 
A more subtle example is when $\lambda$ is indecomposable but not stably indecomposable.
Take a sequence $\lambda_i$ which converges to $\lambda$.
By assumption $\lambda_i=g_i (\lambda_i'\oplus \lambda_i'') g_i^{-1}$ 
and \begin{equation}
D(\Phi,\lambda_i)=D(\Phi,\lambda_i')D(\Phi,\lambda_i'').\label{lambda-i}
\end{equation} Let the limits of $\lambda_i'$ and $\lambda_i''$ be 
respectively $\lambda'$
and $\lambda''$ ; $\lambda$ is not isomorphic to $\lambda'\oplus \lambda''$
because the limit of $g_i$ does not exist. Still we can take the limit of the relation
\eqref{lambda-i} and  we have \begin{equation}
D(\Phi,\lambda)=D(\Phi,\lambda')D(\Phi,\lambda''),
\end{equation} i.e.~$D(\Phi,\lambda)$ decomposes
if $\lambda$ is not stably indecomposable.
Even $D(\Phi,\lambda)$ for a stably indecomposable $\lambda$ sometimes
decomposes as we will see, but the
preceding arguments tell us that 
 we only need to consider stably indecomposable representations
to find the generators of the baryonic operators.

\subsection{Theorems of Gabriel and Kac}
\label{kac}
The discussion above is extremely useful because
much is known about the representations of the quivers,
in particular for which the underlying diagram 
is one of extended or non-extended Dynkin diagram.

Now let us naively count the number of the parameters 
of the gauge equivalence class of 
a representation $\lambda$ with the dimension vector $\alpha=\dim \lambda$.
It has $\sum_a \alpha_{t(a)}\alpha_{h(a)}$ components,
and $G=\prod_i GL(\alpha_i)$ which has $\sum_i \alpha_i{}^2$ parameters
acts on it. The diagonal $GL(1)$ of $G$ does not act on the matrices, so the
naive number $\mu_\alpha$ of the parameters is \begin{equation}
\mu_\alpha=\sum_a \alpha_{t(a)}\alpha_{h(a)}-\sum_i\alpha_i{}^2+1
=1-\langle \alpha,\alpha\rangle/2,\label{number}
\end{equation} where  the pairing $\langle \alpha,\beta\rangle$
is by the Cartan matrix of the graph, \begin{equation}
\langle\alpha,\beta\rangle =
2\sum_i \alpha_i \beta_i -\sum_a \alpha_{t(a)}\beta_{h(a)}
-\sum_a \alpha_{h(a)}\beta_{t(a)}.
\end{equation}
Appearance of the Cartan matrix in the counting of the parameters
makes  it very natural to identify the dimension vector $\alpha$
as an element $\sum \alpha_i e_i$  
of the root lattice associated to the quiver, where $e_i$ is the
simple root corresponding to the $i$-th node.

Recall that the pairing $\langle \alpha,\beta\rangle$ is positive definite
if and only if the quiver is one of the non-extended Dynkin diagram,
and is positive semi-definite if and only if it is one of the extended Dynkin diagram.
A vector $\alpha$ is called real 
if $\langle\alpha,\alpha\rangle>0$ 
and imaginary if $\langle r,r\rangle\le 0$.
Then the relation \eqref{number} tells us that, naively speaking,
we can expect a discrete number of indecomposable representations 
if $\alpha$ is a real root, i.e. $\langle\alpha,\alpha\rangle=2$,
and a $\mu_\alpha$-parameter family if $\alpha$ is imaginary. 

The statement is made mathematically precise by Gabriel,
who introduced the terminology ``quiver'' in the first place \cite{Gabriel}:
\begin{itemize}
\item The number of indecomposable representations of a quiver $Q$
is finite if  and only if the underlying diagram of $Q$ is a non-extended Dynkin diagram.
Thus, such quivers are classified by $A$, $D$ and $E$.
\item When $Q$ is Dynkin, 
 a representation $\lambda$ is indecomposable if and only if
$\dim\lambda$ is one of the positive root.
\item There is one and  only one indecomposable representation for each positive root.
\end{itemize}
The proof  utilize the so-called reflection functor which implements
the Weyl reflection by the simple roots at the level of the representation of the quiver.
With  this tool, the proof goes almost the same as the classification
of the simply-laced root systems. 
The reflection functor also acts on the baryonic operators
through the generalized determinants $D(\Phi,\lambda)$,
and it is the mathematical realization of the Seiberg duality
discussed in Sec.~\ref{duality}.


After Gabriel's work, many people studied the extension to more general quivers,
and one culmination is the result by Kac \cite{Kac1}:
\begin{itemize}
\item A representation $\lambda$ is indecomposable if and only if
$\dim\lambda$ is one of the positive root of the associated Kac-Moody algebra.
\item There is one and  only one indecomposable representation for each positive real root.
\item  For each positive imaginary root $\alpha$,
there is a $\mu_{\alpha}$-parameter family of
 indecomposable representations $\lambda$ 
 with $\alpha=\dim\lambda$.
\end{itemize}
If the quiver is one of the extended Dynkin diagram, the associated
Kac-Moody algebra is the untwisted
current algebra for the corresponding simply-laced group,
and  the structure of the roots are well-known, which can be readily utilized
to the analysis of the $\ccA$-type baryons of our theory.
For a readable account of the theorem, 
see \cite{KraftRiedtmann}; 
indecomposable representations of extended Dynkin quivers
are explicitly listed in \cite{DlabRingel}, although the direction of the arrows
are not the same as ours.

The theorems above classify  indecomposable representations, but as argued in the last
subsection, stably indecomposable representations are more relevant for our purposes.
Fortunately they were also  classified  for the extended Dynkin quivers \cite{Kac1}.
To state the theorem, let us introduce the Ringel pairing $R(\alpha,\beta)$
which is not necessarily symmetric and depends on the orientation
of the arrows: \begin{equation}
R(\alpha,\beta)=\sum_{i}\alpha_i \beta_i -\sum_{a} \alpha_{t(a)}\beta_{h(a)}.
\end{equation} It is related to the Cartan pairing via $\langle \alpha,\beta\rangle
=R(\alpha,\beta)+R(\beta,\alpha)$.
Next, let us recall the set of the positive roots of 
untwisted simply-laced affine Lie algebras $\hat\frakg$, which can be
described as follows \cite{KacText}:  let us relabel the nodes so that $0$-th node
corresponds to the extending node of the extended Dynkin diagram.
We denote the simple roots as $e_i$, $i=0,1,\ldots, r$, and identify
the subspace generated by $e_1,\ldots,e_r$ with the root lattice
of the corresponding finite dimensional Lie algebra $\frakg$.
Then, the set of positive imaginary roots are $\{k \delta\}$ for $k$ a positive integer,
with \begin{equation}
\delta=\sum_{i=0}^r d_i e_i
\end{equation} where $d_i$ is the $i$-th Coxeter label, i.e.~the dimension of the 
indecomposable representation of the corresponding discrete subgroup of $SU(2)$.
The set $\hat\Delta_+$ of positive real roots  are given by \begin{equation}
\hat \Delta_+=\Delta_+ \cup \{k\delta\pm r\ | \ r\in \Delta_+,\ k\in \bZ_{>0}\},
\end{equation}where $\Delta_+$ is the set of positive roots of $\frakg$.  
Now the theorem states that the dimension vector of a stably indecomposable
representation is in either of the two sets \begin{align}
\hat\Delta_{+,0}&=\{\delta\}\cup\{\alpha,\ \delta-\alpha\ | \ \alpha\in\Delta_+\ 
\text{and}\  R(\delta,\alpha)=0\} ,\label{delta+0}\\
\hat\Delta_{+,1}&=\{\alpha\in\hat\Delta_+\ | \ R(\delta,\alpha)\ne0\}.
\end{align}

\subsection{Application to the study of $\ccA$-type baryons}\label{application-of-kac}
Let us apply the mathematical theory reviewed in
the previous sections 
to the classification of the baryons of our theory. 
The inspection of the quiver diagram reveals that the $\ccA$-type bifundamental fields
form two disjoint sets of extended Dynkin diagrams, and thus
any gauge-invariant operator is constructed by fields coming from
only one of the two Dynkin diagrams.  In the following we only 
consider one set of bifundamentals forming a Dynkin diagram.

Its nodes are colored in white and black, and all of the arrows are from 
white to black, and therefore from the argument in Sec.~\ref{determinants}
the only gauge invariant operators are the generalized determinants
$D(\Phi,\lambda)$. As explained in Sec.~\ref{quiver-rep}, they decompose
as the product of two gauge-invariant operators if $\lambda$ is not
stably indecomposable.
We abuse the notation and often identifies the dimension vector 
$\dim \lambda$ and the indecomposable representation $\lambda$  itself
if $\dim\lambda$ is a positive real root, since no confusion should arise.

As cautioned in Sec.~\ref{quiver-rep}, it is not that
all of the stably indecomposable representations of the quiver correspond to an
indecomposable baryonic operators. But the set above gives the only possibility
for such operators.  The set can further be constrained, because 
$\dim \lambda=(\lambda_i)$,
which is the number of the epsilon symbols for $SU(d_iN)$ 
one use to construct the operator,
needs to satisfy the relations \eqref{def-m} and \eqref{w} for some
set of non-negative integers $n_a$ assigned to the bifundamental fields.
One immediate consequence is that \begin{equation}
\sum_{i\ \text{white}}d_i \lambda_i 
=\sum_{i\ \text{black}}d_i \lambda_i,
\end{equation}  which is equivalent to the condition \begin{equation}
R(\delta,\dim\lambda)=0
\end{equation} using the relation \eqref{eigenvector}.
Thus we only need to study the set $\hat\Delta_{+,0}$ \eqref{delta+0},
and given the condition above, 
the weight of the operator $D(\Phi,\lambda)$ is \begin{equation}
w(D(\Phi,\lambda))=\frac12\sum_i d_i \lambda_i.\label{w-formula}
\end{equation}

\subsubsection*{Bayons of weight $|\Gamma|/2$}
Let us first study the baryons with $\dim\lambda=\delta$.
The weight of such operators is $|\Gamma|/2$ from \eqref{w-formula}.
There is a one-parameter family of stably indecomposable representation $\lambda$
with  $\dim\lambda=\delta$, and
it is known  that the moduli space of such $\lambda$
is
$\CP^1/\Gamma$ with the orbifold points removed\footnote{%
Strictly speaking,
it is imprecise to refer the moduli space as
$\CP^1/\Gamma$ with orbifold points removed,
because the classification of the indecomposable representation
of the quiver is done in the sense of algebraic geometry,
and $\CP^1/\Gamma$ with orbifold points removed is
isomorphic to $\CP^1$ with three points removed.}  \cite{Kac1}, 
and it nicely matches with the moduli of the $\ccA$-type brane 
we found in Sec.~\ref{geom-A}. We will return to the problem
of counting the number of baryons of weight $|\Gamma|/2$ in
Sec.~\ref{Abranch}.

\subsubsection*{Baryons with weight less than $|\Gamma|/2$}
The positive real roots in the set $\hat\Delta_{+,0}$ give operators with weight
less than $|\Gamma|/2$. There is at most only one baryonic
operator for each of such dimension vectors.
It is from the fact that 
the integrand in the formula \eqref{averaging}
gives the same baryonic operator for almost all $\lambda$
because
there is only one stably indecomposable representation for each dimension vector.
Still, $D(\Phi,\lambda)$  might be decomposable to the
products of two baryons of lower weight.
For example,  if a real root $w$ in $\hat\Delta_{+,0}$ is the sum of two
real root  $v_{1,2}\in\hat\Delta_{+,0}$
 and if $D(\Phi,v_1)$ and $D(\Phi,v_2)$ is non-zero,
then \begin{equation}
D(\Phi,w)\propto D(\Phi,v_1)D(\Phi,v_2).
\end{equation} The reason is that any baryonic operator
with dimension vector $w$ is of the form $D(\Phi,w)$ 
as argued above, and the non-zero operator
$D(\Phi,v_1)D(\Phi,v_2)$ has the dimension vector $w=v_1+v_2$.
Therefore, to make a baryonic operator which is not a product of 
gauge-invariant operators,
 we need to take a positive real root vector in $\hat\Delta_{+,0}$ which cannot be
written as the sum of vectors in $\hat\Delta_{+,0}$. We call such a vector
indecomposable.

Let us classify $\ccA$-type baryons of weight less than $|\Gamma|/2$
using the strategy outlined above.
We analyze  cyclic groups, dihedral groups and the
polyhedral groups  in turn.
\subsubsection*{Cyclic groups, $A_{2n-1}$}
It is easy to check that only indecomposable vector
in $\hat\Delta_{+,0}$ is of the form $e_i+e_{i+1}$, $i=0,\ldots,n-1$
where $e_n$ is identified with $e_0$. Then  the corresponding
baryonic operator is just the dibaryons $\det A_{i\to i+1}$.
The result agrees with the previous direct analysis in Sec.~\ref{direct-classification}.
\subsubsection*{Binary dihedral groups, $D_{n+2}$}
It is straightforward to list all of elements of $\hat\Delta_{+,0}$, of which 
the indecomposable vectors are $n$ of weight two \begin{gather}
\Dn111{0\, \cdots \,0}000,\quad
\Dn001{1\,\,\,0\,\cdots\, 0}000 \quad\ldots,\quad
\Dn000{\cdots \, 0 \,\,\, 1 \,\,\, 1 \,\,\, 0 \,\cdots}000, \nonumber \\
\ldots,\quad \Dn000{0 \,\cdots\, 0\,\,\,1}010,\quad
\Dn000{0\cdots \,0}111,
\end{gather} and four of weight $n$: \begin{gather}
\Dn011{1\cdots1}011,\quad
\Dn101{1\cdots1}110, \\
\Dn011{1\cdots1}110,\quad
\Dn101{1\cdots1}011. \label{Dn-n}
\end{gather}

The list above matches the expectation from the analysis of the geometry.
Indeed, since the corresponding Platonic triple is now $p=n$, $q=2$ and $r=2$,
we expected $n$ operators $\cP_1,\ldots,\cP_n$ of weight 2
and four operators $\cQ_{1,2}$ and $\cR_{1,2}$ of weight $n$. 
Furthermore, the dimension vector of $\cP_1\cdots\cP_n$,
$\cQ_1\cQ_2$ and $\cR_1\cR_2$ are all equal to $\delta$.
This corresponds to the decomposition (\ref{decomposition_at_orbifoldpoint}) 
of the dibaryon at the orbifold point of $\CP^1/\Gamma$.

\subsubsection*{Binary tetrahedral group, $E_6$}
The vectors in $\hat\Delta_{+,0}$ are
six of weight four\begin{gather}
v_1=\Esix1011100, \quad
v_2=\Esix0101110, \quad
v_3=\Esix0010111,\\ 
v_4=\Esix0001111, \quad
v_5=\Esix1010110, \quad
v_6=\Esix0111100;
\end{gather}  two of weight six, \begin{gather}
z_1=\Esix0011210,\quad
z_2=\Esix1111111;
\end{gather} and six of weight eight \begin{gather}
w_1=\Esix0111221,\quad
w_2=\Esix1021211,\quad
w_3=\Esix1112210,\\
w_4=\Esix1121210,\quad
w_5=\Esix0112211,\quad
w_6=\Esix1011221.
\end{gather} 

It is easy to check any of the weight-eight vectors
is the sum of two weight-four vector, e.g. $w_1=v_2+v_3.$
Therefore, as argued previously,
$D(\Phi,w_1)\propto D(\Phi,v_2)D(\Phi,v_3)$.
Thus, all the baryonic operators with weight less than $|\Gamma|/2=12$
is generated by six operators of weight four, and two of weight six.
We have $v_1+v_2+v_3=v_4+v_5+v_6=z_1+z_2=\delta$. 
Hence it seems reasonable to identify the operator with dimension vector
$v_{1,2,3}$ as $\cP_i$, $v_{4,5,6}$ as $\cQ_i$ and $z_{1,2}$ as $\cR_i$.

\subsubsection*{Binary octahedral group, $E_7$}
$\hat\Delta_{+,0}$ consists of four vectors of weight 6 \begin{gather}
v_1=\Eseven01011100,\qquad
v_2=\Eseven00011110, \nonumber \\
v_3=\Eseven00101111,\qquad
v_4=\Eseven11111000;
\end{gather}
 three vectors of weight 8 \begin{gather}
w_1=\Eseven00112100,\qquad
w_2=\Eseven01111110,\qquad
w_3=\Eseven11011111;
\end{gather} six of weight 12, three of weight 16, and four of weight 18.
Of the six weight-12 vectors, four can be written as the sum of two weight-four vectors.
The indecomposable ones are then two remaining ones : \begin{equation}
z_1=\Eseven01122111,\qquad
z_2=\Eseven11112210.
\end{equation} All of weight-16 and weight-18 vectors are decomposable.
We find $v_1+v_2+v_3+v_4=w_1+w_2+w_3=z_1+z_2=\delta$.

\subsubsection*{Binary icosahedral group, $E_8$}
The vectors in $\hat\Delta_{+,0}$ are  five of weight 12 \begin{gather}
v_1=\Ee101011111,\qquad 
v_2=\Ee000111111,\qquad 
v_3=\Ee010111110, \nonumber \\
v_4=\Ee011111100,\qquad 
v_5=\Ee001121000;
\end{gather} three of weight 20, \begin{equation}
w_1=\Ee001122210,\qquad
w_2=\Ee011122111,\qquad
w_3=\Ee111221111;
\end{equation} two of weight 30, \begin{equation}
z_1=\Ee012232221,\qquad
z_2=\Ee111233211;
\end{equation} and five of weight 24, five of weight 36,
three of weight 40 and five of weight 48.
The vectors with weight 24, 36, 40 or 48 are all decomposable.
We also find $\sum v_i=\sum w_i=\sum z_i=\delta$.

\subsubsection*{Summary}
For all the cases, we found $k$ operators of weight $|\Gamma|/(2k)$ for each
of the orbifold points of $S^2/\Gamma$,
 as predicted by the geometry of the orbifold of $T^{1,1}$.
We also found that together with operators of dimension vector $\delta$,
they generate the whole $\ccA$-type baryons, which matches the prediction
of the AdS/CFT correspondence from the analysis of the bulk side.
We have also constructed operators of dimension $|\Gamma|/(2k)$ 
in Sec.~\ref{A-gauge}. We check that those operators
have the dimension vectors listed above  in Appendix~\ref{fractional-as-generalized}.

\section{Dimension of the $\ccA$-type baryonic branch}\label{Abranch}
As a final exercise, let us count the dimension of the moduli space
of alternating Dynkin quivers and compare with
the number of the generator of the $\ccA$-type baryonic operators.
We will find that for $N>1$ there is no non-linear relation
among the generators of the baryonic operators. In a more mathematical
parlance, it means that the chiral ring  of $\ccA$-type baryons of our theory
is just a polynomial ring.

The moduli space we study in this section is not the full moduli space
of the gauge theory, but presumably it will describe the branch where
all of the vacuum expectation values for $\ccB$-type baryonic fields vanish.
Then, the remaining fields are just $\ccA$-type bifundamentals, and therefore
the branch will be just two copies of the moduli space we study.

\subsection{Dimension of the moduli space}
The moduli space in question is that of the $\ccA$-type fields
which form one alternating Dynkin diagram, i.e.~we have
gauge groups $SU(d_sN)$ at the nodes and the bifundamentals are specified by the
arrows.  
We denote by $n_\Gamma$ 
 the number of the nodes of  the extended Dynkin
diagram of type $\Gamma$.
As always, we complexify the gauge groups to $SL(d_sN)$ instead of
imposing the D-term conditions. If we further enlarge the gauge groups
to $GL(d_sN)$, the study of the moduli space is exactly equivalent to the
study of the indecomposable representation of the same quiver with
dimension vector $N\delta$. 

As we quoted in Sec.~\ref{kac},  generic points in the moduli are the direct sum
of $N$ indecomposable representations of dimension vector $\delta$, each of which has
one parameter.  Thus it has $N$ complex parameters.  But it is for the gauge group
$U(d_sN)$, and we need the result for the gauge group $SU(d_sN)$.
Of the $U(1)^{n_\Gamma}$ in the difference, only $n_\Gamma-1$ act
non-trivially on the bifundamentals, because
 the simultaneous  $U(1)$ rotation for each of the $U(d_sN)$ gauge groups
 does not change the bifundamental fields at all. 
Thus, there are 
$n_\Gamma-1$  extra degrees of freedom in the moduli space
for the gauge group $SU(d_sN)$ in addition to the moduli space for the gauge group
$U(d_sN)$.
 Therefore the number of the parameters is \begin{equation}
N+n_\Gamma-1.\label{dim-of-A-moduli}
\end{equation}

\subsection{Number of generators}\label{num-gen}
We counted the number of generators with weight less than $|\Gamma|/2$
in Sec.~\ref{application-of-kac}. We already
saw for $\bZ_{2n}$ we have $2n$ operators
of weight $2$, and for non-Abelian $\Gamma$ with the associated
Platonic triple $(p,q,r)$ we had $p$ operators $\cP_{1,\ldots,p}$,
$q$ operators $\cQ_{1,\ldots,q}$ and $r$ operators $\cR_{1,\ldots,r}$.

Hence we only need to find the number of the operators with weight $|\Gamma|/2$.
As we discussed, they are spanned by the generalized determinants $D(A,\lambda)$
with
the number of the epsilon symbols dictated by the dimension vector $\delta$.
From the theorem of Kac, we know that 
$\lambda$ has one complex parameter up to gauge equivalence.
We also discussed the operator $\cA_{+\to -}$ and $\cA_{-\to +}$
of weight $|\Gamma|/2$ in Sec.~\ref{A-gauge},
where it appeared as the decomposition of the dibaryon $\det(\lambda_i A^i)$ 
in the unorbifolded theory.  We saw that $\lambda_i$ and $\lambda_i'$ related
by the action of $\Gamma$ give the same baryonic operator thanks
to the condition \eqref{AB}, \begin{equation}
A^i{}=\rho_\two(h)^i_j \rho_r(h) A^j{} \rho_r(h)^{-1}, 
\end{equation}and that the moduli space of $\lambda_i$ can be identified
with $S^2/\Gamma$.  Thus what needs to be studied is the number of 
linearly-independent
operators obtained from $\cA_{+\to -}(\lambda)$.

Let us recall 
$\cA_{+\to -}(\lambda_i)$ is the determinant of a block $\lambda_iA^i_{+\to -}$
of the matrix $\lambda_iA^i$ which maps
$V_+$ to $V_-$, where $V_\pm$ is the eigenspace of
$-1\in SU(2)$ with eigenvalue $\pm1$ of  the regular representation $\rho_r$. 
Let us say $\rho_r=\rho_+\oplus\rho_-$ under this decomposition.
Then
the orbifold projection is now \begin{equation}
A^i_{+\to -}=\rho_\two(h)^i_j \rho_-(h) A^j_{+\to -} \rho_+(h)^{-1}. 
\end{equation} Therefore the determinant
$\cA_{+\to -}(\lambda_i)$  satisfies the relation \begin{align}
\cA_{+\to -}(\lambda)&=\left(\det\rho_+(h) \det\rho_-(h)^{-1}\right)^N
\cA_{+\to -}(\rho_\two(h)\lambda)\\
&=\left(\det\rho_r(h)\right)^N \cA_{+\to -}(\rho_\two(h)\lambda),
\label{Arelation}
\end{align}where we used the facts that there are $N$ copies
of the regular representation, and $\det\rho_\pm=\pm 1$.

$\cA_{+\to-}$ is a polynomial of $\lambda_{1,2}$ of pure degree $N|\Gamma|/2$.
Then, the relation above \eqref{Arelation} means that $\Gamma$
acts on the polynomial as the representation \begin{equation}
(\det\rho_r)^N\otimes\mathrm{Sym}^{N|\Gamma|/2}(\rho_\two).\label{hage}
\end{equation}  If a polynomial is not invariant under $\Gamma$,
the operator vanishes by averaging over $\Gamma$.
Thus the number of
linearly independent operators is  at most the number of invariant vectors 
in the representation \eqref{hage}.

The one-dimensional representation
$\det\rho_r$ is found to be non-trivial only for $D_{\text{odd}}$,
where it is equal to the representation $1'$ in the table \eqref{1prime}
in the Appendix~\ref{bigtable}.  Then a straightforward application of the 
orthogonality of the characters or  Molien's formula shows
there are $N+1$ invariant vectors, irrespective of  $\Gamma$.
This number $N+1$ includes the decomposable baryons 
formed by the fractional baryons.

\subsubsection*{Abelian $\Gamma$}
It is the case when $\Gamma=\bZ_{2n}$.
It is easy to see  that there are two ways of constructing baryons
of dimension vector $\delta$ from the baryons of weight $2$.
Thus we expect $N-1$ independent operators with weight $|\Gamma|/2=n$,
which precisely agrees with what we found in Sec.~\ref{A-direct}.
In total, we have $N+2n-1$ operators, which is equal to  \eqref{dim-of-A-moduli}.

\subsubsection*{Non-Abelian $\Gamma$}
As we saw, $\prod \cP_i$, $\prod \cQ_i$ and $\prod \cR_i$ 
give three baryonic operators with dimension vector $\delta$.
Since we have $N+1$ independent operators with this dimension vector,
we believe for $N>1$ these three products
 are linearly-independent, and the rest of $N+1$ operators
gives $N-2$ linearly independent baryons. 
 In total, we have \begin{equation}
N-2 + p + q + r\label{num-of-generators}
\end{equation} generators of baryons.
Now, an interesting fact is that
we have the relation \begin{equation}
p+q+r=n_\Gamma+1
\end{equation} for Dynkin diagrams of $D$ and $E$ type.
Thus \eqref{dim-of-A-moduli} and \eqref{num-of-generators}  give the same number.

Now we have shown that the number of generators of the $\ccA$-type baryonic operators
is equal to the dimension of the moduli space of the $\ccA$-type baryons. 
Therefore, there can be no non-linear relation among the generators
obtained thus far, and we have a surprisingly simple result
that  the chiral ring of $\ccA$-type baryons is just a polynomial ring for $N>1$.
Indeed, it agrees with the result of \cite{SkowronskiWeyman} which was obtained
in a different method.
The case $N=1$
is further discussed in  Appendix \ref{non-linear}.

\section{Summary and discussion}\label{summary}
Let us summarize what we have obtained so far.
We considered the AdS/CFT duality between Type IIB string theory
on $T^{1,1}/\Gamma$ and the corresponding gauge theory,
especially the mapping between the wrapped D3-branes and the
baryonic operator of the quiver gauge theory.
We first started by constructing the gauge theory by applying the
prescription of Douglas and Moore to the theory of Klebanov and Witten.
The geometry of $T^{1,1}/\Gamma$ told us that
the number of the baryonic operators in the gauge theory
 is dictated by the structure of the action 
of the group $\Gamma$ on $S^2$.   We found the expected number
of the baryonic operators by decomposing the dibaryons of 
un-orbifolded theory.  

The rest of the paper was devoted to show that the baryons
thus discovered exhaust the set of indecomposable baryons.
It was with the help of 
the untangling procedure, the Seiberg duality and the theory of
quiver representations that we accomplished the task.
Moreover, we found that there is no non-linear relation
among the generators of the baryonic operators.
We believe the technique we developed and/or imported from the
mathematics of quiver representation can be utilized in the study
of the baryons of non-toric and/or non-conformal quiver gauge theory,
where the ranks of the gauge groups are in general different from each other,
as they were in our case.

An immediate generalization will be the study of the baryonic operators
of the dual gauge theory of other non-Abelian orbifolds of
Sasaki-Einstein spaces.  One natural candidate is $S^5/\Gamma$,
where $\Gamma$ is a non-Abelian finite subgroup of $SU(3)$.
The main difficulty is that the moduli space of the wrapped D3-branes 
is much more intricate in the geometry side, and that the quiver does not
nicely split into alternating Dynkin diagrams in the gauge theory side.

Another candidate will be the study of the orbifold of $Y^{p,q}$.
Here $Y^{p,q}$ spaces are the infinite series of explicit Sasaki-Einstein
spaces in five dimensions found in \cite{Ypq} with the isometry $SU(2)\times U(1)^2$,
and the corresponding quiver was constructed in \cite{YpqQuiver}.
We can take a non-Abelian subgroup $\Gamma$ of $SU(2)$ isometry and
consider the space $Y^{p,q}/\Gamma$, which has $U(1)^2$ 
as the isometry group and not more.  The quiver for $Y^{p,q}/\Gamma$
can be constructed exactly as in Sec.~\ref{construction-of-quiver}
and is nicely described using the alternating Dynkin quiver.
The analyses of the geometry and of the gauge theory 
in Sec.~\ref{baryons} carry through mostly unchanged also in these cases.
There might still be a new phenomena  in these examples.

Non-conformal deformation of our quiver gauge theory might also be interesting;
the Klebanov-Strassler solution \cite{KS}, 
and the baryonic deformation of it \cite{BZ,DKS} breaks
the $U(1)_R$ symmetry but respects $SU(2)\times SU(2)$ symmetry.
Thus the non-conformal version of our quiver should have a moduli space
of stable supersymmetric vacua.  It might have some interesting properties
which are not directly inherited from the un-orbifolded cases. 
We hope to revisit these questions in a future publication.

\acknowledgments
The authors would like to thank Dario Martelli for various stimulating discussions.
They also appreciate helpful comments by Amihay Hanany,
Yoshiyuki Kimura, Igor Klebanov, and Johannes Walcher.
The work of FY is supported by Japan Society for the Promotion of Science
Research Fellowships for Young Scientists.
 YT is supported by the United States  DOE Grant DE-FG02-90ER40542.

\appendix
\section{Tables of representations of discrete subgroups of $SU(2)$}\label{bigtable}
In this appendix we list the irreducible representations
of discrete subgroups of $SU(2)$.  We list the explicit
representation matrices for  cyclic and binary dihedral groups,
and we present  only the character tables for binary tetra-, octa- and
icosahedral groups.
\subsection*{Cyclic groups}
The group is $\bZ_{n}$ generated by $g$,
 and there are $n$ irreducible one-dimensional representations
$\rho_i$ on which $g$ is represented by $\alpha^i$.
Here, $\alpha$ is $\exp(2\pi i/n)$.

\subsection*{Binary dihedral groups}
It is denoted by $\hat\cD_n$, and
 has $4n$ elements, and is generated by elements $a$, $b$ and
 $z$ with the relations  $a^n=b^2=(ab)^2=z$, $z^2=1$.
There are $n-1$ two-dimensional irreducible representations $\rho_{2,k}$,
$k=1,\ldots,n-1$ where $a$ and $b$ are represented by \begin{equation}
 a=\begin{pmatrix}
 \alpha^k& \\ & \alpha^{-k}
\end{pmatrix},\quad
b=\begin{pmatrix}
&i^k\\  i^k&
\end{pmatrix}
\end{equation} where $\alpha=\exp(\pi i/n)$.
$\rho_{2,1}$ is the fundamental two-dimensional representation
$\rho_\two$ which is defined through the embedding $\Gamma\subset SU(2)$.
We can similarly define the representations $\rho_{2,0}$ and $\rho_{2,n}$,
but each of them decomposes as the sum of two one-dimensional representations,
\begin{equation}
\rho_{2,0}=\rho_1 \oplus \rho_1',\qquad
\rho_{2,n}=\rho_1'' \oplus \rho_1'''
\end{equation} where $a$ and $b$ are represented by the following
scalar multiplication:\begin{equation}
\begin{array}{c|cccc}
&1&1'&1''&1'''\\
\hline
a&1&1&-1&-1\\
b&1&-1&i^n&-i^n
\end{array}.\label{1prime}
\end{equation}
 
 \arraycolsep = .6em
 
\subsection*{Binary tetrahedral group, $E_6$}
 The group has $24$ elements, and is generated by elements $a$, $b$, $c$ and
 $z$ with the relations  $a^3=b^3=c^2=z$, $c=ab$, $z^2=1$.
 Irreducible representations are three of dimension one
 which we call $1$, $1'$, $1''$; three of dimension two $2$, $2'$, $2''$
 and one three-dimensional representation $3$. The character table follows:
 \begin{equation} 
 \begin{array}{l|rrrrrrrr}
 &e&z&c&a&a^2&b&b^2\\
 \hline
 1 & 1& 1& 1& 1&1 &1 &1 & \\
 1' & 1& 1& 1& \omega & \omega^2 & \omega^2& \omega & \\
 1'' & 1& 1& 1& \omega^2& \omega & \omega& \omega^2& \\
 2 & 2& -2& 0& 1& -1& 1& -1& \\
 2' & 2& -2& 0& \omega& -\omega^2& \omega^2&-\omega & \\
 2'' & 2& -2& 0& \omega^2& -\omega & \omega& -\omega^2 & \\
 3 & 3& 3 & -1& 0& 0& 0& 0& 
\end{array}
\end{equation} where $\omega=\exp(2\pi i/3)$.
 
\subsection*{Binary octahedral group, $E_7$}
 The group has $48$ elements, and is generated by elements $a$, $b$, $c$ and
 $z$ with the relations  $a^4=b^3=c^2=z$, $c=ab$, $z^2=1$.
 Irreducible representations are $1$, $2$, $3$, $4$, $3'$, $2'$,
 $1'$ and $2''$, where the number denotes the respective dimension
 and the prime distinguishes different irreducible representations of the same
 dimension. The character table is the following :
 \begin{equation}
 \begin{array}{l|rrrrrrrr}
 &e&z&c&a&a^2&a^3 &b&b^2\\
\hline
1  &1&1&1&1&1&1&1&1\\
1' &1&1&-1& -1 &1&-1&1&1\\
2  &2&-2&0&\sqrt2&0&-\sqrt2&1&-1\\
2' &2&-2&0&-\sqrt2&0&\sqrt2&1&-1\\
3  &3&3&-1&1&-1&1&0&0\\
3' &3&3&1&-1&-1&-1&0&0\\
4  &4&-4&0&0&0&0&-1&1\\
2'' &2&2&0&0&2&0&-1&-1\\
\end{array}.
\end{equation}

\subsection*{Binary icosahedral group, $E_8$}\label{E8charactertable}
 The group has $120$ elements, and is generated by elements $a$, $b$, $c$ and
 $z$ with the relations  $a^5=b^3=c^2=z$, $c=ab$, $z^2=1$.
Irreducible representations are $1$, $2$, $3$, $4$, $5$,
 $6$, $4'$, $2'$, and $3''$, where the notation is as before.
The character table is given by \begin{equation}
\begin{array}{l|rrrrrrrrr}
 &e&z&c&a&a^2&a^3&a^4 &b&b^2\\
\hline
1&1&1&1&1&1&1&1&1&1\\
2&2&-2&0&\varphi&\varphi^{-1}&-\varphi^{-1}&-\varphi& 1 & -1\\
3&3&3&-1&\varphi&-\varphi^{-1}&-\varphi^{-1}&\varphi&0&0\\
4&4&-4&0&1&-1&1&-1&-1&1\\
5&5&5&1&0&0&0&0&-1&-1\\
6&6&-6&0&-1&1&-1&1&0&0\\
4'&4&4&0&-1&-1&-1&-1&1&1\\
2'&2&-2&0&-\varphi^{-1}&-\varphi&\varphi&\varphi^{-1}&1&-1\\
3''&3&3&-1&-\varphi^{-1}&\varphi&\varphi&-\varphi^{-1}&0&0
\end{array},
\end{equation} where $\varphi$ is the golden ratio $(1+\sqrt5)/2$.
As is well known, $\varphi$ and $-\varphi^{-1}$ solve $x^2=x+1$.

\section{Fractional dibaryons as generalized determinants}
\label{fractional-as-generalized}
In this section
 we study the relation of 
the fractional dibaryons constructed geometrically in Sec.~\ref{A-gauge}
and
 the baryonic operators constructed in Sec.~\ref{application-of-kac}.

To see this, let us calculate the dimension vector of the fractional dibaryon
$\cA_{i\to i+1}$. Recall that  it is defined as the determinant of the block
which maps $V_i$ to $V_{i+1}$, where $V_i$ is the eigenspace of 
$g\in\bZ_{2k}\subset\Gamma$ with eigenvalue $\alpha^i$
acting on the $N$ copies of the regular representation $\rho_\Gamma$ of $\Gamma$.
 Here, $\alpha$ is $\exp(\pi i/k)$.
 
To translate the operator  to the language of the gauge theory,
we change the basis of $\rho_\Gamma$ to the one as
the direct sum of irreducible representations, see \eqref{regulardecomposition}.
As explained in Sec.~\ref{construction-of-quiver}, $N$ copies
of $\rho_\Gamma$ decomposes
as \begin{equation}
\rho_\Gamma^{\oplus N}=\bigoplus_s \bC^{Nd_s}\otimes \rho_s,
\end{equation} where the gauge groups $SU(N d_s)_{1,2}$  act
on the factor $\bC^{Nd_s}$ and $\Gamma$ acts on $\rho_s$.
Thus the eigenspaces $V_i$ is given by \begin{equation}
V_i=\bigoplus \bC^{Nd_s}\otimes \rho_{s,i}
\end{equation} where $\rho_{s,i}$ is the eigenspaces of $g$ acting
on $\rho_s$ with the eigenvalue $\alpha^i$.
Thus, the fractional dibaryon $\cA_{i\to i+1}$ uses
$\dim\rho_{s,i}$ epsilon symbols of $SU(d_sN)_1$ and 
$\dim\rho_{s,i+1}$ epsilon symbols of $SU(d_sN)_2$.
The dimension of the eigenspaces $\dim\rho_{s,i}$ can be determined
from the data summarized in the Appendix \ref{bigtable}.

Let us for example consider the fractional dibaryon for the icosahedral group
with $p=5$.
The element $a$ is represented in each of the irreducible representations
\begin{gather}
\rho_1(a)=1,\ 
\rho_2(a)=\diag(\alpha,\alpha^{-1}),\ 
\rho_3(a)=\diag(\alpha^2,1,\alpha^{-2}),\\
\rho_4(a)=\diag(\alpha^3,\alpha,\alpha^{-1},\alpha^{-3}),\ 
\rho_5(a)=\diag(\alpha^4,\alpha^2,1,\alpha^{-2},\alpha^{-4}),\\
\rho_6(a)=\diag(-1,\alpha^3,\alpha,\alpha^{-1},\alpha^{-3},-1),\\
\rho_{4'}(a)=\diag(\alpha^{4},\alpha^2,\alpha^{-2},\alpha^{-4}),\\
\rho_{2'}(a)=\diag(\alpha^3,\alpha^{-3}),\ 
\rho_{3''}(a)=\diag(\alpha^4,1,\alpha^{-4}),
\end{gather} as can be inferred from the table
in Appendix~\ref{E8charactertable}.
Here  $\alpha=\exp(\pi i/5)$.
Then, the dimension vectors of $\cA_{0\to 1}$, for example,
 is found to be \begin{equation}
v_1=\Ee101011111,
\end{equation}  which is defined in \eqref{seibergactiononfractionals}.
Thus $\cA_{0\to 1}$ is the operator $\cP_1$ constructed in
\eqref{baryon1}.  We also find that the dimension vectors
of $\cA_{2\to 3}$,  $\cA_{4\to 5}$,  $\cA_{6\to 7}$, $\cA_{8\to 9}$
respectively 
to be $v_3$, $v_5$, $v_4$, $v_2$, which are
 defined also in \eqref{seibergactiononfractionals}.
In a similar manner, we can calculate the dimension
vectors of  the fractional dibaryons for other 
orbifolding group $\Gamma$.
They give exactly the dimension vectors tabulated in Sec.~\ref{application-of-kac}.

\section{Direct analysis  of $\ccA$-type baryons, continued.}\label{continued}
\subsection*{Cyclic groups, $\Gamma=\bZ_{2n}$}\label{direct-A}
We will continue the discussion of Sec.~\ref{A-direct},
using  the same notation.  There, we found that any indecomposable
operator other than the dibaryon $\det\Phi_i$ has $m_1=m_2=\cdots=m$,
and we also have found $N-1$ operators $\cO_k$, $k=1,\ldots,N-1$ with $m=1$.
In this section we show that any operator with $m>1$
can be rewritten as a polynomial in the gauge-invariant operators just mentioned.

As a preparation, we apply the untangling procedure repeatedly,
starting from $\Phi_1$ at $SU(N)_1$, then for $\Phi_2$ at $SU(N)_2$, all the way
to $\Phi_{2n-1}$ at $SU(N)_{2n-1}$. Then the operator is now some complicated
contraction by $m$ epsilon symbols of $SU(N)_0$ of the following
operators $O_k$ 
which are gauge-invariant under $SU(N)_{i}$ for $1\le i\le 2n-1$: \begin{equation}
O_k=(\Phi_1)^k \epsilon_1 (\Phi_2)^{N-k} \epsilon_2 (\Phi_3)^k
\cdots (\Phi_{2n-1})^k \epsilon_{2n-1} (\Phi_{0})^{N-k}.
\end{equation}  For $m=1$, the only way to make it gauge invariant
is to contract $k$ indices of $(\Phi_1)^k$  and $N-k$ indices of $(\Phi_{2n})^{N-k}$,
which gives the operators $\cO_k$ we found in Sec.~\ref{A-direct}.

For $m>1$, the operator is now of the form \begin{equation}
\epsilon^{(1)} O_{k_1} 
\epsilon^{(2)} O_{k_2} 
\cdots
\epsilon^{(m)} O_{k_m} \label{AAA}
\end{equation} 
where $k_i$ of $\Phi_1$ are all contracted against the epsilon symbol 
$\epsilon^{(i)}$ of $SU(N)_0$. We can assume $k_1\le k_2\le \cdots\le k_m$
without loss of generality. Now the remaining choice in the contraction
is how $\Phi_{2n}$ in $O_{k_i}$ are contracted against $\epsilon^{(j)}$.
Let $\ell_1$ be the number of $\Phi_{0}$ in $O_{k_1}$ contracted
against $\epsilon^{(1)}$.

We use a double mathematical induction in $k_1$ and $\ell_1$ to show
that it can be reduced to a polynomial in $\det\Phi_i$ and $\cO_k$.
First, it contains $\det \Phi_{0}$ as a factor if $k_1=0$.
Second, it contains $\cO_k$ as a factor if $\ell_1=N-k_1$.
Third, let us assume that any operator can be decomposed if
$k_1<k$ or 
$k_1=k$,  $\ell_1>\ell$, and consider an operator with $k_1=k$ and $\ell_1=\ell$.
Now $k$ of $\Phi_1$, $\ell$ of $\Phi_{0}$  contract
against  $\epsilon^{(1)}$. Since  $k+\ell<N$, 
 we have at least one $\Phi_{0}$ in $O_{k_1}$ contracting
against $\epsilon^{(i)}$, $i>1$.  Now we apply the Pl\"ucker relation
\begin{equation}
\epsilon_{[a_1a_2\cdots a_N} \epsilon_{a_{N+1}]b_2\cdots b_N} =0
\end{equation} to $\epsilon^{(1)}\epsilon^{(i)}$, with the index
contracted to $\Phi_{0}$ in $O_{k_1}$ as $a_{N+1}$. 
Then the terms in the resulting expression either have
$\ell_1=\ell+1$, $k_1=k$ or $\ell_1=\ell$, $k_1=k-1$, see Fig.~\ref{plucker}.
\FIGURE{
\includegraphics[width=.7\textwidth]{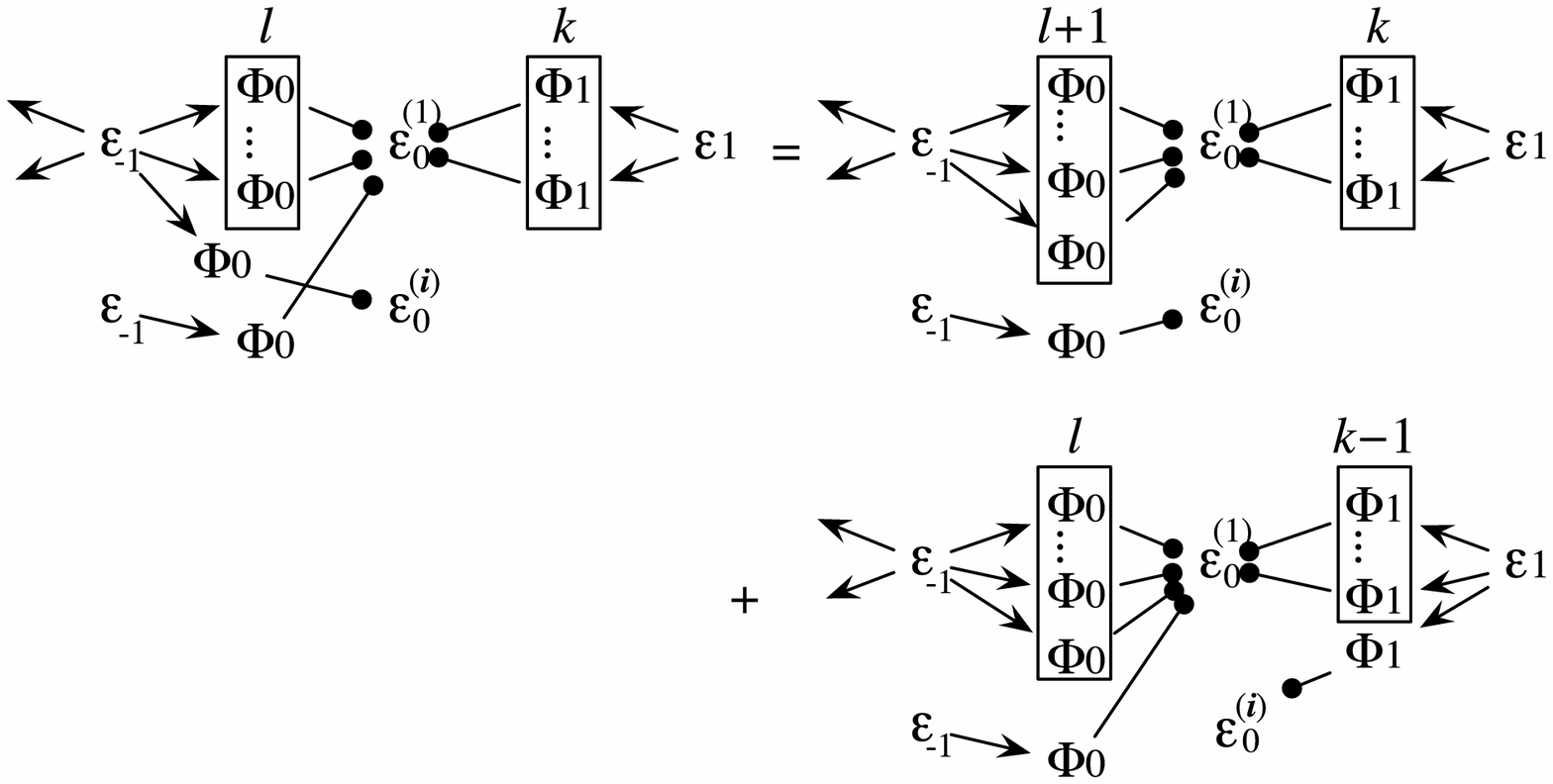}
\caption{Application of the Pl\"ucker relation. Black blobs are the indices
$a_1,\ldots,a_{N+1}$.\label{plucker}}
} The terms with $k_1=k-1$ is not exactly of the form in \eqref{AAA},
but they can be made so by untangling $\Phi_2$, $\Phi_3$, \ldots, $\Phi_{2n}$.
Then the mathematical induction implies it can be decomposed into
a polynomial of $\det\Phi_i$ and $\cO_k$.

\subsection*{Binary dihedral groups, $\Gamma=\hat\cD_n$}\label{direct-D}
Here we show that any baryonic operator of the $D$-type alternating Dynkin quiver
can be written as the polynomial of the basic operator which we obtained
in Sec.~\ref{D-direct}. We continue to use the notation in that section.
We denoted the dimension vector as $(m,m',m'',m''';m_i)$,
and we showed that the operator decomposes unless $m+m'=m''+m'''=m_i$.
Let us set $\mu=m_i$ for brevity. 
We show any operator with $\mu\ge 3$ is a product
of operators with $\mu=1,2$.

We first perform the untangling procedure beginning from $W$ and $Z$ 
at $SU(2N)_{n-2}$, repeatedly to $\Phi_1$ at $SU(2N)_1$.  Then the
bifundamental fields other than $U$ and $V$ 
are combined to the parts 
\begin{align}
\cW^{a_1\cdots a_N}&=\epsilon^{a_1\cdots a_N c_1\cdots c_N}
(\Phi_1)^N_{c_1\cdots c_N} \epsilon_2 (\Phi_2)^N 
\cdots (\Phi_{n-2})^N \epsilon_{n-1}W^N\epsilon_{N''},\\
\cZ^{a_1\cdots a_N}&=\epsilon^{a_1\cdots a_N c_1\cdots c_N}
(\Phi_1)^N_{c_1\cdots c_N} \epsilon_2 (\Phi_2)^N 
\cdots (\Phi_{n-2})^N \epsilon_{n-1}Z^N\epsilon_{N'''}.
\end{align}
 Here,  $\epsilon^{a_1\cdots a_{2N}}$ is the epsilon
symbol for $SU(2N)_1$, 
$\epsilon_i$ is the one for $SU(2N)_i$,
and we suppressed the indices of the gauge groups other than
$SU(2N)_1$ for the sake of simplicity.
As a byproduct of the untangling procedure above, we have the relations \begin{equation}
\cW^{[a_1\cdots a_N}\cW^{b_1]\cdots b_N}=
\cZ^{[a_1\cdots a_N}\cZ^{b_1]\cdots b_N}=0 \label{mock-plucker}.
\end{equation}

Then we untangle $U_a$ and $V_a$ connected to $\cW$'s at
$SU(N)$ and $SU(N)'$. 
It makes $U$, $V$ and $\cW$ to combine into polynomials of
 $\cU$, $\cV$ and $O_k$ ($k=0,\ldots,N$) 
 defined as follows:
\begin{align}
\cU_{a_1\cdots a_N}&=\epsilon_{N}(U^N)_{a_1\cdots a_N},\\
\cV_{a_1\cdots a_N}&=\epsilon_{N'}(V^N)_{a_1\cdots a_N},\\
O_k{}_{[a_1\cdots a_{k}][b_1\cdots b_{N-k}]}
&= \cU_{ a_1\cdots a_{k}c_1\cdots c_{N-k}}
\cV_{b_1\cdots b_{N-k}d_1\cdots d_k}
\cW^{c_1\cdots c_{N-k} d_1\cdots d_{k}}.
\end{align}
Thus, the problem is now reduced to the study of the contraction
of operators $\cU$, $\cV$, $O_k$ 
to the operators $\cZ$.  The important point here is
that $\cZ$ satisfies  the Pl\"ucker-like 
relation \eqref{mock-plucker}. 

First, when a $\cU$ is contracted to a product of 
several $\cZ$, repeated application of the Pl\"ucker-like relation
can make all of the indices of $\cU$ to contract against one $\cZ$.
Thus it contains $\cU\cZ$ as a factor.  One can make the same argument
for $\cV$. 

Then the remaining case to analyze is a baryonic operator where 
the product of $\mu$ of $O_{k}$'s  is contracted against
$\mu$ of $\cZ$, which  we distinguish as $\cZ_{(i)}$, $(i=1,\ldots,\mu)$.
Suppose there are $k_i$ of $U$ fields contracted against $\cZ_{(i)}$. 
Application of the untangling procedure for bifundamentals in $O_k$'s 
in the order $U$, $V$, $\Phi_1$, $\Phi_2$,\ldots, we see that such an operator
can be expressed as  \begin{equation}
\prod_i O_{k_i} \cZ_{(i)}
\end{equation} where $k_i$ indices of  $U$ inside $O_{k_i}$
are all contracted against $\cZ_{(i)}$. Let $\ell_i$ be the number of 
indices of $\cV$ inside $O_{k_i}$ contracted against $\cZ_{(i)}$.
Now we can apply the same mathematical induction for $l_1$ and $k_1$
as in the case of cyclic groups treated in the previous subsection,
and we find that the operator can be decomposed as a polynomial
of $O_{k,a_1\cdots a_N} \cZ^{a_1\cdots a_N}$. This is what we wanted to prove.

\subsection*{Binary icosahedral group, $\cI$}
Let us study  the icosahedral case to exemplify how we can enumerate
baryons of alternating Dynkin quiver of exceptional type. The quiver
was already depicted in Fig.~\ref{e8quiver}.  
Suppose we are given a baryonic operator.
We first apply the untangling procedure to the bifundamentals
repeatedly, from the endpoint of three legs of the extended quiver to the junction of them.
Then the bifundamentals of
 each leg are organized into the following combinations: \begin{equation}
 \begin{array}{l@{\,}l@{\,}l@{}l@{}l@{\,}l@{\,}l}
\cU_{3N}&=&\epsilon_{(1)}\cA_{1\to2}{}^N
&\epsilon_{(2)}\cA_{3\to 2}{}^N
&\epsilon_{(3)}\cA_{3\to 4}{}^{2N}
&\epsilon_{(4)}\cA_{5\to 4}{}^{2N}
&\epsilon_{(5)}\cA_{5\to 6}{}^{3N},\\
\cU_{2N}&=&&\epsilon_{(2)}\cA_{3\to 2}{}^{2N}
&\epsilon_{(3)}\cA_{3\to 4}{}^{N}
&\epsilon_{(4)}\cA_{5\to 4}{}^{3N}
&\epsilon_{(5)}\cA_{5\to 6}{}^{2N},\\
\cU_{4N}&=&&
&\epsilon_{(3)}\cA_{3\to 4}{}^{3N}
&\epsilon_{(4)}\cA_{5\to 4}{}^{N}
&\epsilon_{(5)}\cA_{5\to 6}{}^{4N},\\
\cU_{N}&=&&&
&\epsilon_{(4)}\cA_{5\to 4}{}^{4N}
&\epsilon_{(5)}\cA_{5\to 6}{}^{N},\\
\cU_{5N}&=&&&&
&\epsilon_{(5)}\cA_{5\to 6}{}^{5N};\\
\cV_{2N}&=&\epsilon_{(2')}\cA_{4'\to2'}{}^{2N} &\epsilon_{(4')}\cA_{4'\to 6}{}^{2N},\\
\cV_{4N}&=&&\epsilon_{(4')}\cA_{4'\to 6}{}^{4N}; \\
\cW_{3N}&=&\epsilon_{(3')}\cA_{3'\to 6}{}^{3N}.
\end{array}
\end{equation}
Here, $\cA_{a\to b}$ stands for the bifundamental field connecting
$SU(aN)$ and $SU(bN)$ gauge groups, $\epsilon_{(i)}$ the epsilon symbol
for $SU(iN)$, and contraction of the gauge indices other than those of $SU(6N)$ 
should be understood. The subscripts of $\cU$, $\cV$ and $\cW$
denote the number of anti-symmetric indices of $SU(6N)$.
The remaining task is to combine these operators with as many
epsilon symbols for $SU(6N)$ as necessary.

Now we can enumerate the baryons according to the number $m_6$ 
of the epsilon symbols used for $SU(6N)$, but it becomes more and more
cumbersome as $m_6$ increases. Let us content ourselves
by showing that the operators of lowest weight is of weight $12$,
and there are five of them.

For $m_6=1$, we find the following four operators: \begin{equation}
\begin{array}{lc}
\text{operator}&\text{dim. vector}\\
\hline
\epsilon_{(6)}\cU_{3N}\cW_{3N}, & \Ee101011111,\\
\epsilon_{(6)}\cU_{2N} \cV_{4N},& \Ee000111111,\\
\epsilon_{(6)}\cU_{4N}\cV_{2N}, & \Ee010111110,\\
\epsilon_{(6)}\cU_{N}\cV_{2N}\cW_{3N}, & \Ee011111100.
\end{array}\label{FUBAR}
\end{equation}They are all of weight $12N$.  Other combinations automatically vanish.
For example, $\epsilon_{(6)}\cU_{4N} \cU_{2N}$ antisymmetrizes
$6N$ of the bifundamental $\cA_{5\to 6}$, which is a $5N\times 6N$
matrix. Therefore it vanishes from the consideration of the rank.
For $m_6=2$, the only possibility with weight not more than $12$ is \begin{equation}
\epsilon_{(6)}\epsilon_{(6)} \cU_{5N} \cV_{4N} \cW_{3N},
\end{equation}with dimension vector  \begin{equation}
\Ee001121000.
\end{equation} The fact that we have two epsilon symbols means that
there are various way of contracting indices, with many possibly linearly-independent
operators. However, as was discussed in Sec.~\ref{duality},
the Seiberg duality turns it to one of the operators in \eqref{FUBAR},
which is guaranteed to have one unique operator for one dimension vector.
Thus we only have one independent operator of weight $12$ with $m_6=2$.
Finally we can check there are no operator of weight not more than $12N$ 
with $m_6>2$. This completes the enumeration of lowest-weight operators,
that is, there are five operators with weight $12$.

\section{Non-linear relations among baryonic generators for $N=1$}\label{non-linear}
As discussed in Sec.~\ref{num-gen}, we have $N+1$ linearly independent operators
with dimension vector $\delta$, while it is easy to see that three operators
$\cP_1\cdots \cP_p$, $\cQ_1\cdots \cQ_q$ and $\cR_1\cdots \cR_r$
share the same dimension vector $\delta$. Therefore, there should be
one non-linear relation among the generators $\cP_i$, $\cQ_i$ and $\cR_i$
for  $N=1$.
We report here how such a relation can be derived for $\Gamma=\hat\cD_{n}$.

The gauge groups are now $SU(2)_1\times\cdots\times SU(2)_{n-1}$,
As matter superfields,
we have bifundamentals $\Phi_i$ connecting $SU(2)_i$ and $SU(2)_{i+1}$,
and in addition two fundamentals $U$, $V$  for $SU(2)_1$
and $W$, $Z$ for $SU(2)_{n-1}$.  We need not distinguish fundamental
and anti-fundamental representation, since any gauge group is $SU(2)$.
Thus we can assume any contraction is done by $\epsilon_{ab}$.
A fundamental identity is the Pl\"ucker relation\begin{equation}
\epsilon_{ab}\epsilon_{cd}=
\epsilon_{ac}\epsilon_{bd}-\epsilon_{ad}\epsilon_{bc},
\end{equation}which can be depicted as \begin{equation}
{}^a_b{|\ |}{}^c_d = 
{}^a_b{\stacks{-}{-}}{}^c_d -
{}^a_b{\times}{}^c_d \label{magic}
\end{equation} where a line connecting two indices stands for the epsilon symbol.
Now the product of $n$ operators of type $\cP_i$ is \begin{equation}
\stack{U}{V}
{|\ |}
\stack{\Phi_1}{\Phi_1}
{|\ |}
\cdots
{|\ |}
\stack{\Phi_{n-2}}{\Phi_{n-2}} 
{|\ |}
\stack{W}{Z}.
\end{equation} Applying \eqref{magic}, we have \begin{align}
&=\stack{U}{V}
(\stacks{-}{-} - {\times})
\stack{\Phi_1}{\Phi_1}
(\stacks{-}{-} - {\times})
\cdots
(\stacks{-}{-} - {\times})
\stack{\Phi_{n-2}}{\Phi_{n-2}} 
(\stacks{-}{-} - {\times})
\stack{W}{Z}\\
&\propto
(U\Phi_1\cdots\Phi_{n-2}W)(V\Phi_1\cdots\Phi_{n-2}Z)
-(U\Phi_1\cdots\Phi_{n-2}Z)(V\Phi_1\cdots\Phi_{n-2}W)\\
&=\cQ_1\cQ_2-\cR_1\cR_2,
\end{align}
which was what to be shown.  

Similar analysis for $N>1$ expresses 
that a certain linear combination $\sum a_k \cO_k$ 
of the operators defined in \eqref{Dgeneral}
as a linear combination of 
$\prod \cP_i$, $\cQ_1\cQ_2$,  $\cR_1\cR_2$. 
Therefore it  just eliminates one of $\cO_k$ from the set of the generators,
and it does not introduce non-linear relation among true generators.
Direct derivation of similar relations  for $E_{6,7,8}$ seems to be much more difficult.

\end{document}